\documentclass[useAMS,usenatbib]{mnras}
\usepackage{epsfig,appendix,array,rotating,pdflscape,array,longtable,colortbl,appendix}
\usepackage{multirow,url}
\usepackage{booktabs}
\usepackage{lscape, amsmath} 
\usepackage{paralist}
\usepackage{relsize}
\usepackage{amssymb}
\usepackage{xtab}

\newcommand{\hi}{H\,{\sc i}}
\newcommand{\hii}{H\,{\sc ii}}
\newcommand{\prim}{$^{\prime}$}
\newcommand{\prin}{$^{\prime\prime}$}
\newcommand{\aprox}{${\sim}$}

\newcommand{\km}{km\,s$^{-1}$}
\newcommand{\degree}{$^{\circ}$}
\newcommand{\halpha}{H${\alpha}$}
\newcommand{\cg}{CGCG\,097-}
\newcommand{\msolar}{M$_{\odot}$}

\newcommand{\mhi}{$M({\mathrm {HI}})$}

\newcommand{\hdos}{H$_{2}$}

\newcommand {\apgt} {\ {\raise-.5ex\hbox{$\buildrel>\over\sim$}}\ }
\newcommand {\aplt} {\ {\raise-.5ex\hbox{$\buildrel<\over\sim$}}\ }

\newcommand{\af}{$A_\mathit{flux}$}
\setlongtables

\title[]{Abell\,1367: a high  fraction of  late--type galaxies displaying \hi\ morphological and kinematic perturbations  }
\author[]{T. C. Scott$^{1,2}$\thanks{E-mail: tom.scott@astro.up.pt (TCS)}, E. Brinks$^{2}$, L. Cortese$^{3}$, A. Boselli$^{4}$, and H. Bravo--Alfaro$^{5}$  \\
$^{1}$Institute of Astrophysics and Space Sciences (IA), Rua das Estrelas, 4150--762 Porto, Portugal\\
$^{2}$Centre for Astrophysics Research, University of Hertfordshire, College Lane, Hatfield, AL10 9AB, UK \\
$^{3}$International Centre for Radio Astronomy Research, The University of Western Australia, 7 Fairway, 6009, Crawley WA, Australia\\
$^{4}$Aix Marseille Universit\'e, CNRS, LAM (Laboratoire d'Astrophysique de Marseille) UMR 7326, 13388, Marseille, France\\
$^{5}$Departamento de Astronom\'\i a, Universidad de Guanajuato, Apdo.\ Postal 144, Guanajuato 36000, Mexico}

\begin{document}
\date{\ref{•}. Received ; in original form }


\maketitle

\label{firstpage}

\begin{abstract}
To investigate the effects the cluster environment has on Late-Type Galaxies (LTGs) we studied HI perturbation signatures for all Abell\,1367 LTGs with HI detections. We used new VLA \hi\  observations combined with AGES single dish blind survey data. Our study indicates that the asymmetry between the high--and low--velocity wings of the characteristic double--horn integrated \hi\  spectrum as measured by the asymmetry parameter, \af, can be a useful diagnostic for ongoing and/or recent HI stripping. 26\% of A1367 LTGs have an \af\  ratio, more asymmetrical than 3 times the 1$\sigma$ spread in the \af\  ratio distribution \textcolor{black}{of an undisturbed sample of isolated galaxies}  (2\%) and samples from \textcolor{black}{other} denser environments (10\% to 20\%). Over half of the A\,1367 LTGs, which are members of groups or pairs, have an \textcolor{black}{\af\ } ratio larger than twice the 1 $\sigma$  spread found in the \textcolor{black}{isolated}  sample. This suggests inter-group/pair interactions could be making a significant contribution to the LTGs displaying such \af\  ratios.
The study also demonstrates that the definition of the \hi\ offset from the optical centre of LTGs is resolution dependent, suggesting that unresolved AGES \hi\  offsets that are significantly larger than the pointing uncertainties ($>$ 2 $\sigma$) reflect interactions which have asymmetrically displaced significant masses of lower density \hi, while having minimal  impact on the location of the highest density HI  in resolved maps.  \textcolor{black}{ The distribution of \af\ from a comparable sample of Virgo galaxies  provides a clear indication that the frequency of \hi\ \textcolor{black}{profile}  perturbations  is lower than in  A\,1367.}

\end{abstract}


\section{INTRODUCTION}
\label{into}

Since at least  \textit{z} $\simeq$ 0.5 mechanisms have operated within the galaxy cluster environment to transform late--type  galaxies (LTGs) into S0 Hubble types \citep{fasa00,goto03}. A  number of different mechanisms have been proposed with the principle  \textcolor{black}{ones being stripping of the Interstellar Medium (ISM) by interaction with the diffuse intra--cluster  medium (ICM), as originally proposed by Gunn \& Gott (1972; primarily ram pressure stripping)\nocite{gun72}} and  tidal interactions. In clusters tidal interactions divide into two main types:  repeated low impact high velocity ``harassment" interactions near the cluster core  \citep[][]{moore96} and higher impact lower velocity ``preprocessing''  interactions amongst members of  infalling groups \citep[][]{dress04}. To date it is unclear whether the same mechanism is dominant at all cluster radii, in all cluster types and at all redshifts \citep{doming01,dress04,bosel06a,poggianti09,bosel14}. 

An important aspect of these questions is the  mechanism  removing cold neutral gas (\hi\ and molecular gas)  from cluster LTGs.  At low  \textit{z} ($\lesssim$ 0.1) the fraction of  \hi\ deficient galaxies\footnote{A galaxy's \hi\ deficiency is defined as the log of the ratio of the expected to observed gas mass \citep{hayn84}. Negative values indicate \textcolor{black}{ an \hi\ } excess over  the expected content. } in galaxy clusters  rises toward the cluster centres   \citep{sola01,vgork04}.  \textcolor{black}{\cite{koop04,bosel06b}} and \cite{cort12} showed that the \hi\ deficiencies in cluster  LTGs correlate with the truncation of their star forming and dust disks. This implies   progressive outside--in \hi\ disk truncation is responsible for  a corresponding  quenching  of star formation. X--ray emission from the ICM in clusters typically shows that  ICM densities rise steeply toward cluster cores and models predict that ram pressure \textcolor{black}{stripping} will reach its maximum efficiency \textcolor{black}{when  a galaxy transits the high density ICM in  the cluster  core region }    \citep{roed05,tonn07}. A number of low redshift (\textit{z }$\lesssim$ \textcolor{black}{0.2}) resolved \hi\ studies explain  truncated \hi\  disks and \hi\ deficiencies of LTGs projected within  $\sim$ 1 Mpc of nearby cluster  ICM cores  as arising from  ram pressure stripping \textcolor{black}{\cite[e.g.][]{bravo00,voll08,chung09,jaffe15,yoon2017}. }  \textcolor{black}{Deep  narrow band \halpha\ imaging studies revealed  long \halpha\ tails near the cores of several nearby clusters \citep[e.g.][]{gava01b,bosel14, bosel16,fossati16,yagi17} and these are usually also attributed to ram pressure stripping.}  The typical absence of evidence of tidally perturbed old stellar disks in these and other studies have led, along with other circumstantial evidence, to a  consensus view that ram pressure stripping is the dominant mechanism accelerating the evolution of LTGs in nearby clusters \citep{bosel06a,bosel14}. However, the density of galaxies also increases toward the  clusters cores so ``harassment" \citep{moore96} might also be involved in depleting  and truncating the \hi\ disks. \hi\ observational evidence for harassment is quite limited \textcolor{black}{\citep[one example being][]{hayn07}}, but the concentration of  intra--cluster light near the centres of clusters is consistent with such a tidal mechanism \citep[e.g.][]{montes2014}. An alternative explanation is that the \hi\ removal and morphology transformations occur in the cluster outskirts as \textcolor{black}{a} result of group ``preprocessing" tidal interactions \citep{dress04},   prior to the LTGs arriving in the core region of clusters\textcolor{black}{, e.g.,} the Blue Infalling Group \citep{cort06}, and in the outskirts of Abell\,1367, RSCG 42 \textcolor{black}{ \citep{scott12}. Studies of group galaxies confirm that their \hi\ content and distribution are impacted by the group environment \citep{hess2013,brown17}.}    It is already clear that multiple \hi\ removal mechanisms can  operate \textcolor{black}{simultaneously}  in individual galaxies and clusters. But, particularly beyond the virial radius,   the relative importance of each type of mechanism remains controversial.  
For example modeling by \citep{tonn07} claims ram pressure  effects can extend to 3 virial radii, whereas modelling by \cite{bekki14} indicates that a Milky Way mass spiral might not be fully stripped of its \hi\ \textcolor{black}{during }  a transit though a Virgo mass ICM core. 

Abell 1367 (\textit{z} = 0.02) is a dynamically young spiral--rich galaxy cluster with an X--ray luminosity 1.25  $\times$  10$^{44}$  ergs s$^{-1}$\citep[][]{plionis09}. The cluster's total mass is $\sim 6.9 \times 10^{14}$\,\msolar\  \citep{bosel06a}, about half that of  Coma. Its two subclusters (SE and NW) are in the early stages of an approximately equal mass merger, about 200 Myr from core contact \citep{don98}. \cite{cort04} reported the cluster velocity  as  6484 $\pm$81 \km\  and established that  each subcluster has its own  substructure.  The majority  of a sample  of bright  LTGs  projected within a radius of 1  Mpc from the cluster centre and two groups near the virial radius have  \hi\  and/or  molecular gas signatures indicating that they are suffering recent or  on--going interactions, which have strongly impacted  their ISM \textcolor{black}{\citep{gava87,gava89}}. A  common signature being a larger than expected offset between \hi\  and optical intensity maxima, referred to hereafter as an \hi\ offset,  \textcolor{black}{\citep[][papers I, II, III and IV respectively from \textcolor{black}{here onwards}]{scott10,scott12,scott13,scott15}.}

In this paper we report on observations with the \textcolor{black}{ NRAO\footnote{The National Radio Astronomy Observatory is a facility of the National Science Foundation operated under cooperative agreement by Associated Universities, Inc.} Karl G. Jansky Very Large Array (VLA)}  in its C configuration ($\sim$ 15 arcsec resolution) mapping  15 A\,1367 LTGs and an extragalactic \hii\ region (referred to from now on as the C--array detections).  From the cumulatively available \hi\  data  for A\,1367 LTGs we  investigate \textcolor{black}{four} \hi\ disk perturbation signatures   (resolved and unresolved \hi\ offsets, \textcolor{black}{the asymmetry in the  \hi\ spectral profiles as well as optical/\hi\ velocity offsets}) and  their relation to selected  galaxy and cluster properties. \textcolor{black}{An important aspect was to improve the understanding of the relation between the  perturbation signatures revealed by single dish and resolved \hi\ observations.} \textcolor{black}{Disturbances to a LTGs virialised and regularly rotating  \hi\  disk can produce asymmetric signatures  in its 
 1D \hi\ profile and/or 2D \hi\ velocity field.  Either of the signatures provides evidence of  perturbed \hi. }  The paper also addresses the question of whether or not the  cumulative \hi\ data now available supports the earlier indications, from  \textcolor{black}{Paper I }  that the frequency of LTGs with   perturbed \hi\ is abnormally high in A\,1367.  

We utilise Arecibo single dish (AGES\footnote{Arecibo Galaxy Environment Survey} \textcolor{black}{and ALFALFA\footnote{\textcolor{black}{The Arecibo Legacy Fast ALFA survey}}}) and our own VLA \hi\ data together with optical data from NED\footnote{NASA/IPAC Extragalactic Database} and Hyperleda\footnote{ \textcolor{black}{ Hyperleda \citep[][http://leda.univ-lyon1.fr]{makarov14}}} as well as SDSS\footnote{\textcolor{black}{Sloan Digital Sky Survey }}  and \textcolor{black}{archived} X--ray data from \textit{ROSAT} and \textit{XMM Newton} for A\,1367 \textcolor{black}{ from The High Energy Astrophysics Science Archive Research Center (HEASARC)}. We also used \textit{Spitzer}, \textit{WISE}\footnote{Wide-field Infrared Survey Explorer}, GOLDMine\footnote{ Galaxy Online Database Milano Network,\citep[][http://goldmine.mib.infn.it/]{gava03b} } and 2MASS\footnote{Two Micron All Sky Survey} near infrared (NIR) as well as  \textit{GALEX}\footnote{\textcolor{black}{Galaxy Evolution Explorer}} UV archive  data. Section \ref{obs} gives details of the VLA C--array observations with the VLA results \textcolor{black}{ given in section}  \ref{results}. Section  \ref{hioff} describes the \hi\  data used in the \hi\ offset  and spectral profile analysis set out  in section \ref{offsets}.   A discussion follows in section \ref{discuss} with concluding remarks in section \ref{concl}. Based on a redshift to A\,1367  of 0.022 and assuming $\Omega_M =0.3$, $\Omega_\Lambda =0.7$, and  $H_o  =72$\,\km\,Mpc$^{-1}$ \citep{sperg07} the distance to the cluster is 92\,Mpc and the   angular scale is 1\,arcmin $\sim$ 24.8\,kpc. All $\alpha$ and $\delta$ positions referred to throughout this paper are  J2000.0. 

\section{OBSERVATIONS}
\label{obs}
\begin{figure}
\begin{center}
\includegraphics[scale=0.38]{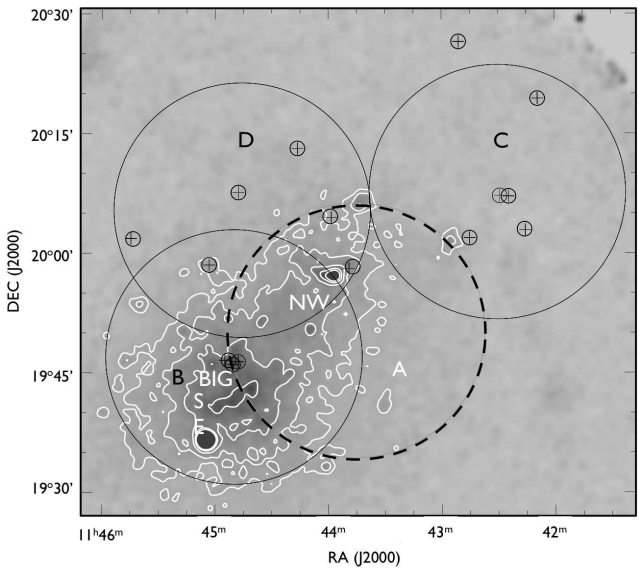}
\caption{\textbf{A\,1367:} The FWHP (32 arcmin) primary beams for the three VLA fields (B, C and D) are indicated with large black circles. The primary beam of the VLA D--array observations (Field A) reported in Paper I is shown \textcolor{black}{with a dashed circle} for reference. The \hi\ positions of our VLA C--array \hi\ detections from fields B, C  and D are shown with small black circles, the positions of their optical counterparts shown with black  crosses. The grayscale is a \textit{ROSAT} X--ray image with white contours overlaid. The projected positions of the SE, NW subclusters and the BIG compact group are labelled.}
\label{bigrosat} 
\end{center}
\end{figure}

We observed \hi\ in three fields using the  VLA  in C--array configuration. The full width half power (FWHP) primary beams ($\sim$ 32 arcmin) of all three fields are projected well within the 104 arcmin (2.58 Mpc) virial radius of the cluster \citep{moss06}. The VLA FWHP beams and velocity ranges are fully within the A\,1367  volume  of the AGES blind single dish \hi\ survey \citep{cort08}. The   observations were made in  correlator mode 4 which gave two independent intermediate frequencies (IFs), each with dual polarization (right and left--hand circular). With the exception of field B, the  IFs  for each field's observations were set up so that each IF generated a position--position--velocity cube ($\alpha,  \delta,  $velocity) with a \aprox\ 500 \km\  velocity range, with the velocity ranges of the two cubes set  adjacent to each other  with a small velocity  overlap. Where \hi\ for the same target was detected in more than one  IF,    cubes with adjacent contiguous  velocities were combined, using the AIPS task  {\sc mcube},  to \textcolor{black}{produce} a  cube for the field with a velocity range of \aprox 1000 \km. The observations and final cubes had a velocity resolution of  $\sim$ 11 \km. Figure \ref{bigrosat} indicates the FWHP primary beams for the three pointings with large circles. The figure also shows the intensity of the X--ray emission \textcolor{black}{\citep[\textit{ROSAT}; ][]{don98}}  from the cluster's ICM (white contours) and the positions of the VLA C--array \hi\ detections (small black circles enclosing crosses). Observational parameters for the  three fields are listed in Table \ref{table1}, including the velocity resolution, central velocities, and
the rms noise per channel.

Sensitivity is non--uniform between the  VLA fields because of their differing  integration times. Additionally, for all fields, there is a sharp drop in sensitivity beyond the primary beam FWHP radius. We applied the AIPS task  {\sc PBCOR}  to the cubes to correct for this. Consequently  the \hi\ flux density for  SDSS J114250.97+202631.6, which projected beyond the FWHP of the Field C primary beam, is more uncertain than \textcolor{black}{for} the   other C--array detections.

Our observations were carried out under dynamic scheduling   between 14 April 2008 and 30 May  2008. All of the observations were made during the period in which the VLA antennas were being systematically upgraded to \textcolor{black}{``EVLA--standard"} antennas. During the observations  the array consisted of approximately equal numbers VLA and EVLA antennas. As a result the observations  were significantly  impacted by transition effects. The data were calibrated and imaged with the AIPS software package, but the standard calibration and reduction procedures had to be modified to overcome a number of issues associated with the VLA--EVLA transition. In particular all EVLA--EVLA baselines were discarded and the band pass calibration was carried out as the first step of the calibration rather than at the end as is usual. The observations were made without on--line Doppler tracking or Hanning smoothing. This required post observation correction for  changes in Doppler shifts during the observations and Hanning smoothing using the AIPS task {\sc cvel}. For all fields self--calibration was carried out to mitigate the effects of side lobes from strong continuum sources, with  side lobes from 3C264 presenting a particular problem. Self--calibration  improved on the standard complex gain calibration and in all cases the solutions converged on the first iteration (phase only).   Our final data cubes were produced with robust weighting ({\sc robust = 0}) and have a spatial resolution of $\sim$ 15 arcsec. The {\sc robust } option corrects the weights of the visibilities in the Fourier transform for the fact that there is a much higher density of measured visibilities in the inner part of the {\em uv--}plane compared to the outer regions \citep{briggs95}. This comes at a cost of a slightly increased noise compared to using what is known as `natural weights'. \textcolor{black}{ Natural weighting leads to a conversion to brightness temperature, T$_\mathrm{B}$, of 1\,mJy\,beam$^{-1} = 3$\,K}.   Appendix \ref{vlafields} gives further details of the observational set up for  the individual VLA fields.

It was found that continuum subtraction was a critical step in the data reduction. Ordinarily one would search for line--free channels in each
cube and then use, e.g., {\sc uvlin} to subtract the continuum. The problem with our observations was that  in several cubes some \hi\ emission was detected in almost every channel.  This required a more elaborate approach consisting of deriving a continuum map made up of the average of line free channels at each spatial position in the cube. This map was subsequently subtracted  from the line + continuum data to produce cubes containing only line emission.  In practice  this required the following procedure. First, emission from continuum sources with  flux density greater than the estimated peak \hi\ emission, normally $\sim$ 5 mJy/beam, was modelled and removed in the  \textit{uv}--plane with  AIPS task {\sc  uvsub}. This resulted in cubes containing {\it line + residual continuum}  sources. In the second stage AIPS tasks {\sc blank} and {\sc sqash} were used to create an image of the average residual continuum in the line free channels at each spatial position by masking regions of the cube clearly containing \hi\ emission. The average continuum image  was subtracted from the line plus residual continuum cube, producing a cube which more clearly delineated  the extent  of \hi\ emission. In the third stage the previous cube was smoothed with AIPS tasks {\sc convl} and a second  average continuum image was made but this time masking the \hi\  from cube produced in the previous stage. This second version of the residual continuum map was then subtracted from original   {\it line + residual  continuum} data to produce the final cubes containing only line emission and  much reduced rms noise compared to using {\sc uvlin or \textcolor{black}{uvlsf}}. Table \ref{table1} shows the approximate rms per channel for  the final cubes. For consistency we applied \textcolor{black}{the above continuum subtraction method} to all of the cubes.

\begin{table*}
\footnotesize
\begin{minipage}{150mm}
\caption{VLA C--array observational parameters}
\label{table1}
\begin{tabular}{@{}cccccrrrrrrr@{}}
\hline

Field \footnote{VLA Project ID: AB1285. }&Day\footnote{\textcolor{black}{Details of each \textcolor{black}{day's} observations are given in Appendix \ref{vlafields}.}}&IF&
${\alpha}_{2000}$ & ${\delta}_{2000}$ & Int.& Beam\footnote{For robust 0 cubes.}  & Central & rms noise  &Detection N\hi\footnote{\textcolor{black}{3$\sigma$ in 2 consecutive channels  of 11 \km\ each. }} &Detection\footnote{\textcolor{black}{3$\sigma$ in 2 consecutive channels  of 11 \km\ each, with distance = 92 Mpc. }} \\
& &&& &  time &size & Velocity  &per channel &limit $\times$ 10$^{19}$&limit $\times$ 10$^7$ \\
&&& [$^h$ $^m$ $^s$] & [\degree\ \prim\ \prin\ ] &  [hours]& [\prin]& [\km ] & [mJy] &[atoms cm$^{-2}$]&[\msolar]\\ \hline 
 B &5&1&11 44 50.0 & 19 47 00 &  9.3 & 21$\times$18  & 8200 & 0.28&\textcolor{black}{1.8}& \textcolor{black}{5.5}\\
 B &5&2&11 44 50.0 & 19 47 00 &  9.3 & 21$\times$18  & 5100 & 0.29&\textcolor{black}{1.9} &\textcolor{black}{5.7}\\
 C &1&1& 11 42 30.4 & 20 07 41 & 16.7& 19$\times$18 & 5800 & 0.22  &\textcolor{black}{1.6}&\textcolor{black}{4.4}\\
 C &1&2& 11 42 30.4 & 20 07 41 & 16.7& 19$\times$18  & 6300 & 0.22 &\textcolor{black}{1.6}&\textcolor{black}{4.4} \\
 C &2&1& 11 42 30.4 & 20 07 41 & 2.0& 20$\times$18 & 7100 & 0.70 &\textcolor{black}{4.7}&\textcolor{black}{13.9}\\
 C &2&2& 11 42 30.4 & 20 07 41 & 2.0& 20$\times$18 & 7600 & 0.70 &\textcolor{black}{4.7}&\textcolor{black}{13.9}\\
 D &3&1& 11 44 45.6& 20 05 24& 8.9& 14$\times$15  & 7400 & 0.34  &\textcolor{black}{3.9}&\textcolor{black}{6.7}\\
 D &3&2& 11 44 45.6& 20 05 24& 8.9& 14$\times$15  & 5100 & 0.35 &\textcolor{black}{4.0}&\textcolor{black}{6.9} \\
 D &4&1& 11 44 45.6& 20 05 24& 9.3& 14$\times$15  & 6500 & 0.25 &\textcolor{black}{2.9}&\textcolor{black}{5.0}\\
 D &4&2& 11 44 45.6& 20 05 24& 9.5& 14$\times$15  & 7000 & 0.25 &\textcolor{black}{2.9}&\textcolor{black}{5.0}\\
 \hline
\end{tabular}
\end{minipage}
\end{table*}

\section{VLA C--ARRAY RESULTS  }
\label{results}

Velocity integrated \hi\ maps \textcolor{black}{and velocity fields} for each   \textcolor{black}{of  the 15  \hi\ detections from our C--array observations and \cg062 from  NRAO VLA archival data (16 VLA C--array  detections in total)} are  presented in Appendix \textcolor{black}{\ref{appenda}}. A brief analysis for each C--array detection is also included in the Appendix. \textcolor{black}{Basic  properties for each of the 16 C--array detections, \textcolor{black}{including} \cg062, are given in Table \ref{table2} with a summary of their \hi\ properties given in Table  \ref{table3}.  } The column labelled ``\textcolor{black}{Flux} VLA/AGES"  in Table \ref{table3} gives the fraction of  \hi\ \textcolor{black}{flux} recovered by the C--array observations in comparison to that from the AGES  Arecibo single dish \hi\ survey \citep{cort08}. For all but three  galaxies the C--array recovered $>$ 70\% of the AGES \hi\ \textcolor{black}{flux}. The two poorest C--array \textcolor{black}{flux}  recoveries were for \cg087 (0.47) and the \cg102 pair (0.60). 

For \cg087 our previous VLA D--array observations (Paper I) recovered almost 100\% of the AGES \hi\ \textcolor{black}{flux} and revealed  an extensive low surface brightness region NW of the optical disk.  Comparing the VLA C and D--array maps  as well as  the AGES spectrum reveals the bulk of missing C--array flux is higher velocity \hi\ located NW of the optical centre, and in particular in its $\sim$ 70 kpc \hi\ tail reported in Paper 1. A  95\% C--array \hi\ flux recovery for our largest \hi\ diameter ($\sim$ 3 arcmin, 75 kpc) spiral\textcolor{black}{, \cg129W,} lead \textcolor{black}{us} to rule out lack short spacings as the reason for the missing C--array   flux. We therefore conclude  that the interaction \cg087 is undergoing has produced extensive regions of low surface brightness \hi\, which are below the C--array detection threshold of \textcolor{black}{2.9} $\times$ 10$^{19}$ cm$^{−2}$, but above the D--array detection threshold of 1.9 $\times$ 10$^{19}$ cm$^{−2}$. \textcolor{black}{ The \hi\ morphology and velocity field for \cg087 (Figure \ref{97087}) supports the ram pressure plus tidal interaction with \cg087N scenario proposed in \citep{cosolandi17}, see Appendix \ref{appenda}. }   In the case of the \cg102 pair  the C--array \hi\ morphology and kinematics as well as NIR asymmetry analysis  provide evidence indicative of a recent tidal interaction (see appendix \textcolor{black}{ \ref{appenda}}).  As in \cg087, undetected low density \hi\ debris from the interaction could possibly account for the missing C--array \textcolor{black}{flux} in the  \cg102 pair. 

The \hi\ morphologies and/or kinematics of almost all of   the 15 LTGs presented in Appendix \textcolor{black}{\ref{appenda}} show significant signatures  of \textcolor{black}{a} recent or ongoing perturbation, e.g.,  \cg121 (Figure \ref{97121}) which could  be an example of ongoing ram pressure stripping; \cg068 (Figure \ref{97068}) has \textcolor{black}{a} morphology and kinematics which could be explained by a ram pressure and/or a tidal interaction.  \cg062 (Figure \ref{97062}) displays a cometary morphology at both  optical and \hi\ wavelengths.

\textcolor{black}{We tried to fit  rotation curves to  \hi\ emission in the A\,1367 C--array cubes using  $^\mathrm{3D}$B{\scriptsize AROLO}
 software \citep{diTeodoro2015}. Despite experimenting with different parameters, including binning the cube along the velocity axis, the
  $^\mathrm{3D}$B{\scriptsize AROLO} fits either failed or were \textcolor{black}{too} poor to be useful. This appeared to be the result of a an unfortunate combination of the low number of beams across the galaxies (typically 3 or 2), low S/N and in almost all cases evidence from the integrated maps that the \hi\  disks are significantly perturbed.}

\begin{table*}
\centering
\begin{minipage}{140mm}
\caption{VLA C--array \textcolor{black}{detections} \textcolor{black}{\hi: basic properties }}
\label{table2}
\begin{tabular}{@{}lccllr@{}}
\hline

Galaxy ID\footnote{\textcolor{black}{CGCG IDs  are from the Zwicky catalogue of Galaxies and Cluster Galaxies \citep[][CGCG]{zwicky68}. IDs with the prefix J  are from SDSS, J114229.18+200713.7 = SDSS J114229.18+200713.7 and J114250.97+202631.8 = SDSS J114250.97+202631.8. K2 refers to the extragalactic \hii\ region  $[SKV2002] K2$ in the BIG compact group \citep{cort06}.  }}  &RA&DEC& V$_{\mathrm{opt}}$\textcolor{black}{\footnote{\textcolor{black}{V$_{\mathrm{opt}}$ = optical velocity from NED.}}}  & Hubble\footnote{From NED, except $[SKV2002] K2$ which is from \cite{cort06}. }&Major axis\footnote{\textcolor{black}{Major axis diameters are   r--band (SDSS Isophotal) diameters from NED, except for \cg072, CGCG 127032  which are K$_s$ (2MASS ``total") diameters from NED and \cg138, \cg087 which are  25\,mag\,arcsec$^{-2}$ isophotal B--band level diameters from Hyperleda.}}\\
& [$^h$ $^m$ $^s$] & [\degree\ \prim\ \prin\ ] &[\km ] &Type&[arcmin] \\ 
\hline
\cg125&11 44 54.8& +19 46 35&8225&\textcolor{black}{Sa}&\textcolor{black}{0.92}\\
$[SKV2002] K2$ &11 44 50.6& +19 46 02& \textcolor{black}{8152}&HII&--\\
\cg114&11 44 47.8& +19 46 24&8293& Pec&\textcolor{black}{0.48}\\
\cg129W&11 45 03.9& +19 58 25&5091&Sb&\textcolor{black}{2.77} \\
\cg138&11 45 44.7& +20 01 52& 5322&Pec&\textcolor{black}{0.65} \\
\cg068&11 42 24.5& +20 07 09& 5974&Sbc&\textcolor{black}{1.14}\\
\textcolor{black}{J114229.18+200713.7}&11 42 29.2& +20 07 14& 6250&Irr&--\\
\cg062&11 42 14.8& +19 58 36&7815&Pec&\textcolor{black}{0.86}\\
\cg063&11 42 15.6 & +20 02 56&6087&Pec&\textcolor{black}{0.58}\\
\cg072&11 42 45.1& +20 01 56&6338&Sa&\textcolor{black}{1.11}\\
CGCG 127032&11 42 09.1& +20 18 56&5764&S0&\textcolor{black}{1.47} \\
\cg087&11 43 49.1& +19 58 06& 6725&Pec&\textcolor{black}{1.86}\\
\cg121&11 44 47.0 & +20 07 30&6583&Sab&\textcolor{black}{1.25}\\
\cg97102N&11 44 17.2& +20 13 24&6368&Sa&\textcolor{black}{1.04}\\
\cg091&11 43 58.9& +20 04 37&7372&Sa&\textcolor{black}{1.15}\\
\textcolor{black}{J114250.97+202631.8}&11 42 51.0&+20 26 32&5718&S&\textcolor{black}{0.70}\\
 \hline
\end{tabular}
\end{minipage}
\end{table*}

\begin{table*}
\centering
\begin{minipage}{180mm}
\caption{\hi\ properties for \textcolor{black}{the} VLA C--array \hi\ detections }
\label{table3}
\begin{tabular}{@{}lllrrllllr@{}}
\hline

Galaxy ID\footnote{\textcolor{black}{See Table \ref{table2} note a. }} &\textcolor{black}{$V_{\mathrm {HI}}$}\footnote{$V_{\mathrm {HI}}$ calculated using the method  described in  Paper I.}& W$_{20}$\footnote{W$_{20}$ calculated using the method described in  Paper I.} &\hi\  offsets &\hi\  offsets &\textcolor{black}{\af}\footnote{\textcolor{black}{\textcolor{black}{AGES} \af\ as defined in section  \ref{aflux}\textcolor{black}{. In two cases, (\cg125, K2, \cg114) and (\cg068,  \textcolor{black}{SDSS J114229.18+200713.7}), VLA \hi\ detections are confused within the 3.5 arcmin AGES FWHP beam and in these cases the \af\ in the table was measured  from the VLA C--array spectra rather than the AGES spectra. }} }  & Flux\footnote{As noted in footnote d, some \hi\ detections are confused in the AGES FWHP beam\textcolor{black}{. In those cases the brackets in the table  indicate the discrete VLA \hi\ detections which are included in each integrated AGES Flux value and similarly for the AGES \mhi\ and the \hi\ deficiency values.}}& M(HI)\footnote{\textcolor{black}{Based on the \textcolor{black}{flux} from AGES, except for SDSS J114250.97  and  cases were the galaxy is confused in the AGES FWHP beam, i.e., for \cg125, $[SKV2002] K2$, \cg114, \cg068, \textcolor{black}{SDSS J114229.18+200713.7} which are from their VLA C--array spectra. \mhi\ = 2.36 $\times$ 10$^2$ D$^2$ S$_{HI}$ where \mhi\ is in \msolar, D = distance (92 Mpc)  and S$_{HI}$ is the AGES or VLA \textcolor{black}{flux}  in mJy. }    }& Def. \hi\footnote{\textcolor{black}{ \hi\ deficiencies are calculated using the method as described in  Paper I.} } &Sub \footnote{Projected distance from the nearest A\,1367 subcluster ICM core assuming a spatial scale at the cluster distance of  24.8 kpc per arcmin. For \cg114, \cg125 and $[SKV2002] K2$ the nearest subcluster in projection is the SE subcluster and in all other cases it is the NW subcluster.}\\
&VLA--C&VLA--C&VLA--C&AGES&AGES&VLA/AGES&AGES&AGES&cluster\\
&&&$>$ 2 $\sigma$&$>$ 2 $\sigma$&&&&&distance\\
&   [\km ] &  [\km ]&  [arcsec ] &  [arcsec ] && [fraction]&[10$^9$\msolar]&&[kpc]\\ 
\hline
\textcolor{black}{CGCG 97--125}&8277$\pm$18&424$\pm$36&14.4$\pm$2.9&--&1.47$\pm$0.09&$\rceil$ &$\rceil$4.7&$\rceil$-0.1&121\\
\textcolor{black}{$[SKV2002]$ K2}&8152$\pm$10&261$\pm$19&--&--&1.19$\pm$0.04&\,$\mid$ 1.27&\,$\mid$&\,$\mid$&101\\
\textcolor{black}{CGCG 97--114}&8451$\pm$20&54$\pm$40&10.4$\pm$2.9&--&1.13$\pm$0.07&$\rfloor$&$\rfloor$&$\rfloor$&109\\
\textcolor{black}{CGCG 97--129W} &5102$\pm$11&512$\pm$22&13.1$\pm$2.9&--&1.19$\pm$0.02&\,\,\,\,0.95&\,\,\,\,\,7.8&\,\,\,\,-0.1&362\\
\textcolor{black}{CGCG 97--138}&5315$\pm$2&64$\pm$3&8.1$\pm$3.3&--&1.17$\pm$0.03&\,\,\,\,0.65&\,\,\,\,\,2.0&\,\,\,\,-0.1&610 \\
\textcolor{black}{CGCG 97--068}&5969$\pm$11&355$\pm$21&18.6$\pm$2.9&--&1.12$\pm$0.01&$\rceil$ 1.12&$\rceil$ 7.2&$\rceil$-0.3&621\\
\textcolor{black}{J114229.18+200713.7}&6114$\pm$2&86$\pm$5&7.9$\pm$2.9&--&1.52$\pm$0.02&$\rfloor$&$\rfloor$&$\rfloor$&597\\
\textcolor{black}{CGCG 97--062}&7803$\pm$12&228$\pm$25&--&39.6$\pm$15.2&1.90$\pm$0.04&\,\,\,\,0.96&\,\,\,\,\,2.5&\,\,\,\,\,0.3&626\\
\textcolor{black}{CGCG 97--063}&6082$\pm$2&172$\pm$3&--&--&1.22$\pm$0.08&\,\,\,\,0.87&\,\,\,\,\,1.3&\,\,\,\,\,0.0&637\\
\textcolor{black}{CGCG 97--072}&6327$\pm$21&301$\pm$41&8.6$\pm$2.9&--&1.03$\pm$0.05&\,\,\,\,1.09&\,\,\,\,\,0.9&\,\,\,\,\,0.5&464\\
\textcolor{black}{CGCG 127--032}&5738$\pm$11&311$\pm$21&30.8$\pm$2.9&29.5$\pm$14.5&1.13$\pm$0.02&\,\,\,\,\,0.73&\,\,\,\,\,1.9&\,\,\,\,0.3&853 \\
\textcolor{black}{CGCG 97--087}&6729$\pm$11&561$\pm$22&10.9$\pm$2.9&--&1.82$\pm$0.09&\,\,\,\,0.47&\,\,\,\,\,5.9&\,\,\,\,\,0.4&80\\
\textcolor{black}{CGCG 97--121}&6578$\pm$63&388$\pm$126&13.6$\pm$2.9&--&1.92$\pm$0.08&\,\,\,\,1.09&\,\,\,\,\,1.3&\,\,\,\,\,0.4&370\\
\textcolor{black}{CGCG 97--102N}&6373$\pm$11&302$\pm$22&11.1$\pm$2.9&--&1.22$\pm$0.07&\,\,\,\,0.60&\,\,\,\,\,1.0&\,\,\,\,\,0.4&411\\
\textcolor{black}{CGCG 97--091}&7372$\pm$2&260$\pm$4&9.0$\pm$2.9&--&1.17$\pm$0.02&\,\,\,\,0.87&\,\,\,\,\,6.5&\,\,\,\,-0.2&190\\
\textcolor{black}{J114250.97+202631.8}&5705$\pm$3&130$\pm$6&10.6$\pm$2.9&N/A&1.17$\pm$0.07&\,\,\,\,N/A &\,\,\,\,\,2.7&\,\,\,\,-0.5&841\\
 \hline
\end{tabular}
\end{minipage}
\end{table*}

\section{ A\,1367: CUMULATIVE  \hi\  DATA }
\label{hioff} 
 
The  A\,1367 field within the  AGES single dish blind \hi\ survey, carried out  with the 305--m Arecibo telescope using the ALFA\footnote{ The seven--beam Arecibo L--Band Feed Array(ALFA) receiver \textcolor{black}{\citep{giovanelli05a}}.} receiver,  covers a $\sim$ 5\degree\ $\times$ 1\degree\ sky area centred on the core of A\,1367. At   A\,1367's distance  the AGES survey  covers a projected  area of $\sim$ 7.4 Mpc$^2$ and   extends in RA, east and west, beyond the  A\,1367 virial radius (see Figure  \ref{fig2}). \hi\ detections in the  AGES A\,1367 field were  made in the velocity range 1142 \km\ to 19052 \km.   Full details of the AGES survey are given in \cite{cort08}.   For this study we define the  5\degree\ $\times$  1\degree\ sky area of  \textcolor{black}{the} AGES A\,1367 field  with a  more restricted  velocity range of 4600 \km\ to 8600 \km\ as the ``A\,1367 volume". This  more restricted velocity range is approximately \textcolor{black}{4 times} the 891 $\pm$58 \km\ velocity dispersion of the cluster \citep{cort04}. Our study uses the AGES third  data release positions in  the A\,1367 volume as the projected positions of the AGES \hi.  Within the A\,1367 volume there are  47  AGES \hi\ detections with optical  counterparts. Five  of these have severe radio frequency interference (RFI) features in their spectra making their \hi\ positions and spectra unreliable. For one of these galaxies (\cg079) the position and spectrum from the low resolution VLA D--array map was substituted  for the AGES data. \textcolor{black}{For \cg079 the integrated   \hi\ flux from VLA D--array observations was 0.518 Jy \km\ compared to 0.247 Jy \km\ from AGES, which reflects the impact of RFI on the AGES spectrum. We} excluded the other four galaxies from our analysis. For the remaining 43 AGES detections the mean AGES \hi\ positional  uncertainties  are RA: (11.72$\pm$3.21 arcsec) and DEC (11.23$\pm$2.37 arcsec).  We were concerned the AGES positions  might be affected by  systematic pointing errors. In that case  errors in RA would be expected to dominate over DEC errors because of the drift scanning observation method used at Arecibo. Our tests found no indication that the AGES offsets or  their PA 's are correlated with either RA or DEC.

Within the A\,1367 volume there are  29 LTGs and an extragalactic  \hii\ region (30  objects) with  VLA C--array  and/or D--array  resolved  \hi\ maps with optical counterparts \textcolor{black}{(Paper I, Paper II and this paper)}.  The position of the \hi\ column density maximum in each object  was determined from  the maps.  Positional uncertainties  for the \hi\ column density maxima  in the maps are $\sim$ 2 arcsec and $\sim$ 4 arcsec  for the VLA C and D arrays respectively. For 21 LTGs  in the A\,1367 volume  we have both  VLA   \hi\ column density maxima positions and  AGES \hi\ positions. Nine VLA \hi\ detections are not in AGES and  are, with one exception, galaxies resolved in the VLA maps but  confused within the larger $\sim$ 3.5 arcmin Arecibo \textcolor{black}{FWHP} beam. The exception is SDSS J114250.97+202631.8 which was mapped with  the VLA C--array but was not reported in AGES because it was only partially detected at the northern edge of the AGES field.

\begin{table*}
\centering
\begin{minipage}{180mm}
\caption[]{\textcolor{black}{A\,1367:} magnitude of \textcolor{black}{resolved and unresolved} \hi\ offsets, \textcolor{black}{\af\  and \hi/optical velocity difference}\footnote{\textcolor{black}{VLA \hi\ detections \cg114 and J114229.18 in Table \ref{table3} were excluded from this table because  they are  confused in the AGES \textcolor{black}{FWHP} beam with LTGs with much larger \hi\ masses,  which prevented  reliable AGES spectra being  extracted for them.  J114250.97, although detected with the VLA, was excluded from this table because it does not have an AGES spectrum (see section \ref{hioff}.) }  } }
\label{table5a}
\begin{tabular}{@{}lrrrrrrrr@{}}
\hline
Optical ID\footnote{\textcolor{black}{See Table \ref{table3} note a.  }}&	RA &	DEC &	\hi\ offset\footnote{\textcolor{black}{Projected offsets \textcolor{black}{between} the AGES \hi\ and optical positions. Offset uncertainties were calculated in \textcolor{black}{quadrature using both the AGES \hi\ and NED optical velocity uncertainties}. Offsets $>$ 2 $\sigma$ are highlighted with  bold typeface.}} &	\hi\ offset &\textcolor{black}{\hi\ offset}\footnote{\textcolor{black}{Offset angles are the counter clockwise angle in degrees from the  north, with the origin at the optical position, for a vector joining the  optical and AGES \hi\ positions. LTGs with AGES \hi\ position offsets $>$ 2 $\sigma$ are highlighted with  bold typeface.}} 	&	\hi\ offset 	&\af\footnote{\textcolor{black}{For \cg068 and \cg125 the \af\ is from the VLA C--array spectra, see Table \ref{table3} footnote d. LTGs with \af\ $>$ 1.39 are shown in bold typeface.} } &\textcolor{black}{$\Delta$V\footnote{\textcolor{black}{Difference between the AGES \hi\  and optical velocity. Velocity differences greater than their 3 $\sigma$ velocity uncertainties are highlighted with  bold typeface, with the uncertainties calculated in quadrature using  both the AGES \hi\ and NED optical velocity uncertainties. }}}	\\
Source	&	AGES&AGES&	AGES&	AGES&\textcolor{black}{AGES}&VLA&	AGES&\textcolor{black}{AGES --}\\
	&	&&	&	&\textcolor{black}{angle}&&	&\textcolor{black}{optical}\\\
&[$^h$ $^m$ $^s$] & [\degree\ \prim\ \prin\ ] 	&[arcsec]&	[kpc]&[\degree]&[arcsec]&&	[\km]\\
\hline
SDSS J113544.31+195114.0&11 35 45.3&19 51 20&	16$\pm$16.3 &7$\pm$7&	68.0	&	       --	&\textbf{1.71$\pm$0.06}&	0.0$\pm$9.9\\	
\cg027&11 36 54.2   	&20 00 03        	&	       13$\pm$15.2     	&	5$\pm$6 	&	357.6	&17.9$\pm$5.7& 1.04$\pm$0.07&0.0$\pm$8.5\\	
\cg026& 11 36 57.6&19 58 37& \textbf{52.8$\pm$14.5}&\textbf{22$\pm$6 }&\textbf{65.4}&8.3$\pm$5.7&1.3$\pm$0.08 &0.0$\pm$4.2\\	
\cg033&11 37 36.5	&20 09 34 &18.3$\pm$15.2& 8$\pm$6 	&	153.2	&--&	       1.02$\pm$0.04&	0.0$\pm$4.2\\	
AGC 210538&11 38 05.0      	&	       19 51 46        	&	       17.2$\pm$14.6   	&	       7$\pm$6 	&	73.1	&7.6$\pm$5.9&	1.04$\pm$0.04&0.0$\pm$4.2\\	
ASK 627362.0 &	11 38 05.0&20 14 41&\textbf{64.6$\pm$18.6} 	&\textbf{27$\pm$8}&\textbf{82.3}&--&\textbf{1.46$\pm$0.11}&	0.0$\pm$5.7\\	
\cg036&	       11 38 51.3      	&	       19 36 22        	&	       17.4$\pm$34.2   	&	7$\pm$14&15.7	&8.6$\pm$5.7&\textbf{1.44$\pm$0.05}&0.0$\pm$8.5\\	
FGC 1287&11 39 10.8&19 35 58&	       \textbf{52.0$\pm$14.6}&\textbf{22$\pm$6}&\textbf{357.5}&9.6$\pm$5.9&1.38$\pm$0.06&	\textbf{42.0$\pm$7.1}\\	
\cg041&11 39 23.1&19 32 32&33.9$\pm$17.7&14$\pm$7&	322.9	&3.6$\pm$5.7&	1.02$\pm$0.02&\textbf{5.0$\pm$3.6}\\	
ASK 629113.0    	&	       11 40 39.5      	&	       19 54 55        	&	       18.1$\pm$15.2   	& 7$\pm$6 	&	24.8	& --	&1.06$\pm$0.06&	0.0$\pm$4.2\\	
ABELL 1367:[GP82] 1512&11 41 48.5&19 45 38&\textbf{33.1$\pm$14.5 }& \textbf{14$\pm$6}&\textbf{217.0}	&--&1.18$\pm$0.07&	0.0$\pm$8.5\\	
CGCG 127-032&11 42 09.8&20 19 24&\textbf{29.5$\pm$14.5}&\textbf{12$\pm$6}&\textbf{20.7}&30.8$\pm$2.9&1.13$\pm$0.02&0.0$\pm$7.1\\	
\cg062&11 42 12.2&19 58 27&\textbf{39.6$\pm$15.2}&\textbf{16$\pm$6}&\textbf{257.6}&4.1$\pm$2.9&	\textbf{1.90$\pm$0.04}&\textbf{41.0$\pm$6.7}\\	
\cg063&11 42 16.5 &20 03 02        	&	       14.2$\pm$29.5   	&	       6$\pm$12&	62.9	&2.3$\pm$2.9&1.22$\pm$0.08&0.0$\pm$7.1\\	
\cg068&11 42 24.9&20 07 10 	&6.1$\pm$14.5    	&	       3$\pm$6 	&	85.2	&18.6$\pm$2.9&1.12$\pm$0.01&\textbf{5.0$\pm$2.8}\\	
\cg072&11 42 44.3&	       20 01 43        	&	       18.6$\pm$15.2   	&	       8$\pm$6 	&	224.2	&8.6$\pm$2.9&1.03$\pm$0.05&0.0$\pm$4.2\\	
\cg073&11 42 56.4&19 57 34&24.4$\pm$14.5&10$\pm$6&181.9	&4.4$\pm$5.7&1.14$\pm$0.06&	\textbf{11.0$\pm$9.2}\\	
\cg079&	       11 43 12.3      	&	       20 00 32        	&	       22.3$\pm$15.3   	&	 9$\pm$6 	&	313.7	&13.8$\pm$3.2	&\textbf{2.06$\pm$0.08}   	&	0.0     $\pm$   9.9\\	
ABELL 1367:[GP82] 1227&11 43 12.5&	       19 36 55&11.8$\pm$14.9&5$\pm$6&	313.6&5.6$\pm$6.7&1.04$\pm$0.05&8.0$\pm$11.7\\	
\cg087&11 43 47.6&19 58 20&25.9$\pm$14.5&11$\pm$6&301.7	& 10.9$\pm$2.9&\textbf{1.82$\pm$0.09}&\textbf{179.0$\pm$8.2}\\	
\cg091&11 43 59.0      	&	       20 04 30&7.3$\pm$14.5    	&	       3$\pm$6 	&	175.4	&	       9$\pm$2.9&1.17$\pm$0.02&0.0$\pm$2.8\\	
\cg102N 	&11 44 16.9&20 13 09&	       15.6$\pm$17     	&	       6$\pm$7 	&	197.5	&	       11.1$\pm$2.9&\textbf{1.46$\pm$0.02}&2.0$\pm$15.2\\	
\cg121  	&	       11 44 48.2&20 07 37     	&	       18.6$\pm$15.2   	&	       8$\pm$6 	&	69.1	&13.6$\pm$2.9&\textbf{1.92$\pm$0.08}&0.0$\pm$4.2\\	
\cg122  	&	       11 44 52.3      	&	       19 27 23        	&	       7.9$\pm$14.5    	&	       3$\pm$6 	&	6.5	&	--	&	    1.23$\pm$0.03&10.0$\pm$10\\	
\cg125  	&	       11 44 53.4      	&19 46 35&	21.8$\pm$14.5&	       9$\pm$6 	&	269.2	&	       14.4$\pm$2.9&\textbf{1.47$\pm$0.09}&0.0$\pm$11.3\\	
\cg129W 	&	       11 45 03.7      	&19 58 32&	7.3$\pm$14.5&3$\pm$6 	&	337.8	&	13.1$\pm$2.9&1.19$\pm$0.02&0.0$\pm$2.8\\	
\cg138  	&	       11 45 44.0      	&20 01 47&11.1$\pm$14.6& 5$\pm$6 	&	245.4	& 8.1$\pm$3.3     	&1.17$\pm$0.03&0.0$\pm$2.8\\	
ABELL 1367:[GP82] 0328&11 45 46.4&19 42 10&\textbf{44.3$\pm$17.3}&\textbf{18$\pm$7}&\textbf{24.9}&	--&1.25$\pm$0.07&0.0$\pm$9.9\\	
SDSS J114725.07+192331.3&11 47 24.6&19 23 24 &10.3$\pm$14.5&4$\pm$6 	&	224.3	&	--	&1.09$\pm$0.02  &0.0$\pm$5.7\\	
SDSS J114730.39+194859.0&11 47 27.9&19 49 06&	\textbf{38$\pm$16.3}&\textbf{16$\pm$7}&\textbf{280.6}	&	--	&1.06$\pm$0.05&	0.0$\pm$9.9\\	
ABELL 1367:[BO85] 092&11 47 30.5&19 53 53&\textbf{92.7$\pm$18.5}&\textbf{38$\pm$8 }&\textbf{355.4}	&	--	&1.09$\pm$0.04&19.0$\pm$28.4\\	
\cg152&	       11 47 40.2      	&	       19 56 21        	&	       12.8$\pm$14.5   	&	       5$\pm$6 	&	93.3	&	--	&	    1.12$\pm$0.04   	&	0.0$\pm$2.8\\	
SDSS J114737.56+200900.5&11 47 39.7&20 09 24&\textbf{39.6$\pm$14.5}&\textbf{16$\pm$6}&\textbf{53.8}&--	&1.19$\pm$0.03&	0.0$\pm$7.1\\	
SDSS J114825.21+194217.0&11 48 23.6&19 42 44&	\textbf{36.2$\pm$14.5}&\textbf{15$\pm$6 }	&\textbf{318.2}&--	&\textbf{1.69$\pm$0.03}&	0.0$\pm$9.9\\	
LSBC D571-02&11 48 24.8 &	20 23 31&	       26.9$\pm$22     	&	       11$\pm$9        	&	296.4	&	--	&	    \textbf{1.45$\pm$0.07}   	&	0.0$\pm$11.3\\	
MAPS-NGP O-434-31444    	&11 48 29.8      	&	       19 45 29        	&	       22$\pm$20.1     	&9$\pm$8 	&	108.4	&	--	&1.29$\pm$0.04&0.0$\pm$5.7\\	
MAPS-NGP O-434-21239    	&	       11 50 30.5      	&	       20 05 05        	&8.6$\pm$16.7& 4$\pm$7 	&	109.7	&	--	&1.10$\pm$0.06&0.0$\pm$9.9\\	
SDSS J115033.61+194250.1        	&	       11 50 34.9&	       19 43 01        	&22.2$\pm$14.5&9$\pm$6&60.7&	--	&1.31$\pm$0.05&0.0$\pm$9.9\\	
CGCG 127-072 	&	       11 51 02.3      	&	       20 23 55        	&18.1$\pm$14.5   	&	       7$\pm$6 	&	98.2	&	--	&1.04$\pm$0.03&0.0$\pm$2.8\\	
KUG 1149+199    	&	       11 51 56.6      	&	       19 38 38&	       16.1$\pm$14.5   	&	       7$\pm$6 	&	117.2	&	--	&1.2$\pm$0.02&0.0$\pm$2.8\\	
KUG 1149+206    	&	       11 52 24.0      	&	       20 19 43        	&	       15.6$\pm$16.0   	&6$\pm$7 	&	335.1	&	--	&1.06$\pm$0.04&0.0$\pm$7.1\\	
KUG 1150+203    	&	       11 53 18.5      	&	       20 06 22        	&	       7$\pm$14.9      	&3$\pm$6 	&	141.8	&	--	&1.37$\pm$0.04&0.0$\pm$4.2\\	
\cg180  	&	       11 54 14.6      	&	       20 01 38        	&	       9.9$\pm$15.3    	&	       4$\pm$6 	&	95.8	&	--	&	    1.35$\pm$0.06&0.0$\pm$4.2\\	
 \hline
\end{tabular}
\end{minipage}
\end{table*}

\section{Discussion}
\label{discuss}
\textcolor{black}{We investigated the resolved VLA  and unresolved AGES  \hi\ data within  the  A\,1367 volume with the aim of understanding the magnitude and  spatial  distribution  of   LTG \hi\ perturbation signatures.  This involved analysis of four quantifiable \hi\ perturbation signatures: i) the asymmetry between the upper and lower wings of the \hi\ profile; ii) the difference between the optical and \hi\ velocities; iii) the projected  offset between  AGES \hi\ position and the centre of its optical counterpart (AGES \hi\ offset);  iv) the offset between  the position of the \hi\ column density maximum in a galaxy's resolved VLA map and the centre of its optical counterpart (VLA \hi\ offset). Calculations of  these  offsets used the optical  positions from NED, or if unavailable  from SDSS. We assume  that the optical positions trace the deepest part of the galaxies' potentials. The  optical positions typically have uncertainties  of  0.5 arcsec or less.  Table \ref{table5a} gives the AGES \hi\ offset, and where available the VLA \hi\ offset, for the 43 \hi\ AGES detections with optical counterparts in the A\,1367 volume. The table also gives the \hi\ spectral asymmetry ratios (\af) from the AGES spectra  \textcolor{black}{and the offset between the AGES \hi\ and optical velocities}. We considered the relation of these \hi\ perturbation signatures  to selected properties of their  galaxies and the cluster. Additionally to assess the frequency and strength of  \hi\ perturbations in  A\,1367 relative to other nearby clusters  we \textcolor{black}{determined the \af\ distribution for a sample of Virgo cluster galaxies.}}

\subsection{\textbf{\hi\ spectral profiles -- \textcolor{black}{\af}}}
\label{aflux}
 A galaxy's   \af\ ratio is a measure of the asymmetry in its  integrated \hi\ flux density \textcolor{black}{profile}  (within its W$_{20}$ velocity range) at velocities above and below \textcolor{black}{the}  galaxy's systemic velocity, $V_{\mathrm {HI}}$,  \textcolor{black}{see \cite{espada11} including graphical examples in their figures 3 and 4. } 
Even  isolated LTGs display a scatter of \af\ ratio, which is well characterised by a half Gaussian, \textcolor{black}{with its mean equal to 1.0 and a $1 \sigma$ dispersion of 0.13. \textcolor{black}{This }  half  \textcolor{black}{Gaussian} \textcolor{black}{   was obtained from a fit to} the distribution of \af\ values from a sample of AMIGA\footnote{Analysis of the interstellar Medium of Isolated GAlaxies} isolated galaxies  \citep{espada11}. The value of an \textcolor{black}{\af} deviating by $1 \sigma$ from the mean \textcolor{black}{of that distribution} is then 1.13.}  In Figure  \ref{Fig_3nn} \textcolor{black}{(upper panel)} we show \textcolor{black}{a histogram of the distribution} of  AGES \af\ ratios for the  43 \textcolor{black}{A\,1367} AGES \hi\ detected  LTGs   compared to the half Gaussian \textcolor{black}{fit for} \af\  ratios from the AMIGA sample of isolated galaxies:  11 (26\%) of the 43 LTGs  have  an AGES \textcolor{black}{\af} ratio greater \textcolor{black}{than 1.39, the 3 $\sigma$ value}  from the AMIGA sample. \textcolor{black}{The penultimate column of Table \ref{table5a} shows the \af\ values for the A1367 LTGs and  those with \af\  $>$ 1.39 are highlighted in bold typeface. } \textcolor{black}{The 26\%  for A\,1367} is significantly greater than the 2\% for the AMIGA sample and above the 10\% to 20\%  which \cite{espada11} \textcolor{black}{reported} for other samples \textcolor{black}{from environments with higher galaxy densities}. \textcolor{black}{ Assuming the 2\%  of galaxies with \af\ $>$ 3 $\sigma$ (\af\ = 1.39) from the AMIGA study  reflects the  intrinsic \af\ variation  in a galaxy population. It then follows that the presence of a larger fraction of galaxies with \af\ $>$ 1.39 in samples from higher galaxy density  environments can  be attributed to perturbations caused by interactions with the environment. In the field,  pair or group tidal interactions are probably responsible, whereas  in clusters both tidal and ram pressure interactions are the probable  sources of \textcolor{black}{the \hi\ } perturbations. }

\textcolor{black}{Ideally we would search for  statistically significant correlations between \af\ $>$ 3 $\sigma$ and other  \textcolor{black}{LTG}  properties, but the small number of \textcolor{black}{LTGs} matching this criterion makes this impractical.  However, there are} 17 LTGs with  AGES \textcolor{black}{\af}  greater than the 2$\sigma$ AMIGA  value of 1.26 (AGES \textcolor{black}{\af}  $>$2$\sigma$ from now on) \textcolor{black}{ and for these LTGs we carried out tests for statistically significant correlations between their \af\ and their}:  AGES \hi\ offsets, subcluster distance, \hi\ deficiency,  \hi\ mass, SDSS $g - i$ colour, optical disk inclination and mean S/N in the AGES spectrum. Table \ref{table7a} sets out the Pearson and Spearman \textcolor{black}{$r$} and \textcolor{black}{$p$} values from these tests\textcolor{black}{. We emphasize the tests are only for LTGs with   \af\  $>$2$\sigma$}. For the relations between AGES \textcolor{black}{\af}  $>$2$\sigma$ versus subcluster distance and AGES \af  $>$2$\sigma$ versus \hi\ deficiency, the null hypothesis that the variables  are \textcolor{black}{uncorrelated} \textcolor{black}{ is rejected by a one--tail directional test with 99\%  and 97.5 \% confidence, respectively.} \textcolor{black}{However,  while statistically significant these correlations are rather weak \textcolor{black}{and are not found if the sample is restricted to members with \af\  $>$ 3$\sigma$.} In all other cases the null hypothesis of no correlation is accepted with 99\% confidence level.} Figure \ref{Fig4n} shows plots of both of  these relations with a  linear fit  added  for galaxies with an  AGES \textcolor{black}{\af} ratio $>$ 2 $\sigma$ \textcolor{black}{(a fit exclusively to the black symbols above the dashed lines in the figures)}.  Figure \ref{Fig4n} \textcolor{black}{(left)},  shows a  trend for the AGES \textcolor{black}{\af} ratio$>$ 2 $\sigma$  to increase toward the cluster center, although this \textcolor{black}{is} characterised by significantly larger \textcolor{black}{\af}  ratios near the subcluster cores with  smaller ratios at radial distances \textcolor{black}{$>$} 1.5   Mpc.  The round open symbols in the Figure \ref{Fig4n}  represent members of pairs or groups \textcolor{black}{(close companions per NED)}: 5 out of 9 (56\%) these pair or group members have   \textcolor{black}{\af} ratios $>$ 2 $\sigma$ and make up  31\% of the galaxies with  \textcolor{black}{\af} ratios $>$ 2 $\sigma$.  \textcolor{black}{We used NED to define a search volume, using a search radius of 26 arcmin (750 kpc) and $\pm$ 1000 \km, creating a (fairly) complete list of all companions for each LTG, late-- type as well as early--type galaxies. One of our LTGs was deemed to be in a pair or group if a cluster galaxy was present within a projected separation of 5 arcmin (145 kpc) and radial velocity separation of less than $\pm$ 300 \km\ (except for the two principal members of the RSCG 42 which have a velocity separation of 447 \km).  } The  correlation between   \textcolor{black}{\af} ratio $>$ 2 $\sigma$ and  \hi\ deficiency indicates  an elevated  \textcolor{black}{\af} ratio is  associated with ongoing or recent stripping of \hi\  from cluster LTGs. 
We note that both statistically significant relations are observed despite the self evident exclusion of  LTGs without \hi\  detections, referred to in Paper I, which were presumably previously stripped of their \hi.

Our tests (Table \ref{table7a}) did not find a statistically significant correlation between AGES \textcolor{black}{\af} $>$ 2 $\sigma$  and AGES \hi\  offsets $>$ 2 $\sigma$. However, all  galaxies with resolved one sided \hi\ tail morphologies also have \textcolor{black}{\af} $>$ 2 $\sigma$ and the magnitude of their AGES \hi\ offsets are larger than their VLA  \hi\ offsets (Figure \ref{fig_5nn}), consistent with the single dish observations being  sensitive to displaced low density \hi. Moreover there are five cases \textcolor{black}{(\cg026, ASK 627362.0, FGC\,1287, \cg062, SDSS J114825.21+194217.0)} where both the AGES \hi\ \textcolor{black}{offset and \af\ } interaction signatures $>$ 2 $\sigma$ are present. For  three of them we have VLA \hi\ maps  showing morphology and kinematics consistent with strongly perturbed \hi.  \textcolor{black}{Overall we conclude \af\ can be a useful indicator of ongoing and/or recent \hi\ perturbation and the large fraction of A\,1367 LTGs with \af\ $>$ 1.39 adds to the evidence that  such perturbations are  strong and widespread in A\,1367.}

\begin{figure}
\begin{center}
\includegraphics[ angle=0,scale=0.44] {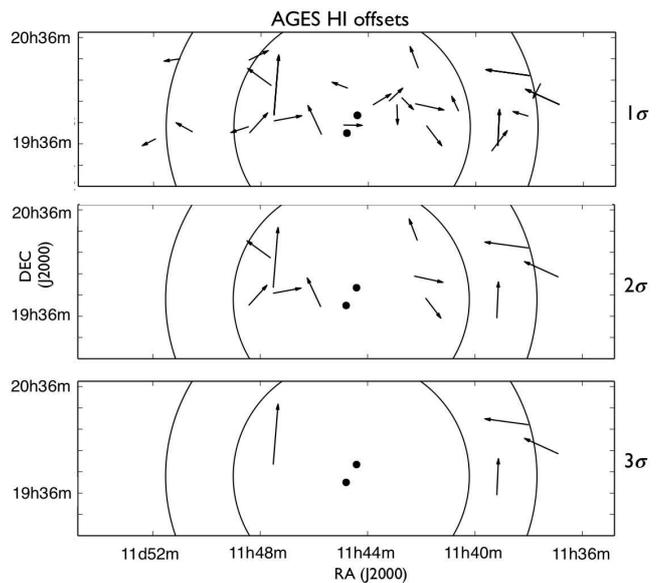}
\vspace{1cm}
\caption{\textbf{Projected AGES \hi\ offsets within  the  $\sim$ 1\degree\ $\times$ 5\degree\   4600 to 8600 \km\ A\,1367 volume:} The arrows indicate the magnitude ($\times$  20) and direction (from the galaxy optical centre) of the \hi\ offsets.   The top, middle and bottom panels show The AGES  \hi\ offsets $>$ 1 $\sigma$, $>$ 2 $>$ and $>$ 3 $\sigma$ pointing \textcolor{black}{uncertainties} respectively.  The centres of the A\,1367 SE and NW subclusters are indicated with filled circles.   The smaller of \textcolor{black}{the} two large circles indicates the $\sim$ 1.64 Mpc (1.1\degree) A\, 1367 $R_{200}$ radius with the larger circle showing the virial radius $\sim$ 2.57 Mpc (1.73\degree).  }
\label{fig2}
\end{center}
\end{figure} 
\begin{figure}
\begin{center}
\includegraphics[ angle=0,scale=.46] {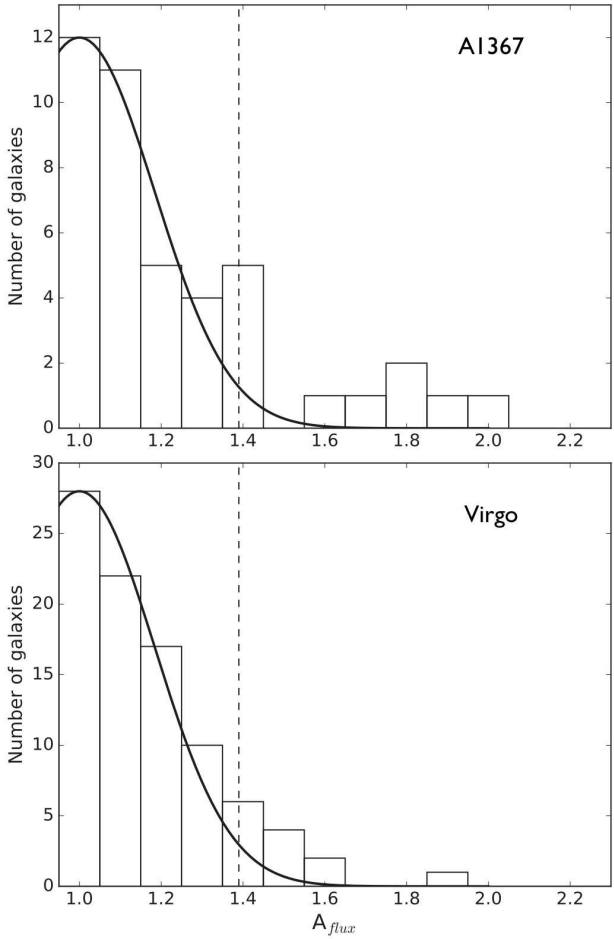}
\vspace{1cm}
\caption{\textcolor{black}{ \textbf{Distribution of \af\  ratios in A\,1367 and Virgo compared to the half  Gaussian fit to the  \af\ distribution from  the AMIGA sample of isolated galaxies. } \textit{\textbf{Top: }} Histogram of \af\ for LTGs within the  A\,1367 volume.  \textit{\textbf{Bottom: }} Histogram of \af\ for galaxies within the  Virgo--ALFALFA  volume. The Virgo sample includes only galaxies with  \mhi\ $ >$ 3 $\times$ 10$^8$ \msolar, i.e., the lowest \hi\ mass LTG in our A\,1367--AGES sample. The half Gaussian fit maximum has been scaled to the  highest value histogram bin.  The dashed line indicates the   3  $\sigma$ value (1.39) from the half   Gaussian fit to the  distribution of \af\ in a sample of AMIGA isolated galaxies.  } }
\label{Fig_3nn}
\end{center}
\end{figure}

\begin{figure*}
\begin{center}
\includegraphics[ angle=0,scale=.52] {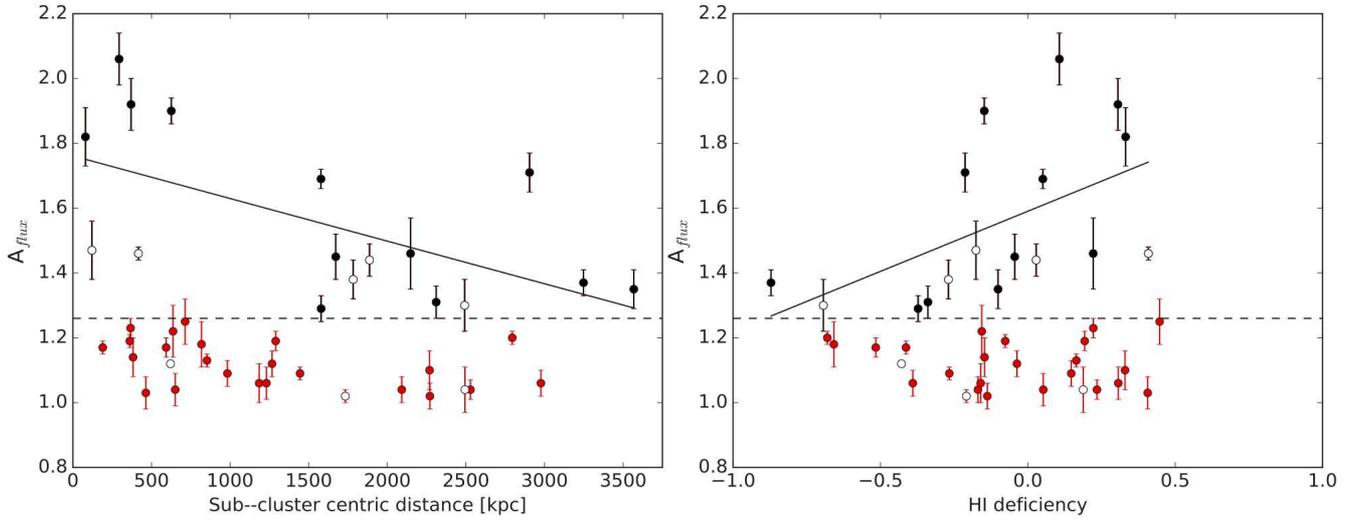}
\vspace{1cm}
\caption{ \textbf{\textcolor{black}{\af\ ratio versus subcluster distance and \hi\ deficiency for LTGs in the A\,1367 volume.}} \textcolor{black}{ LTGs with \af\ $>$ 2 $\sigma$ are shown with black symbols and those with \af\ $<$ 2 $\sigma$ with red symbols.   \textbf{\textit{Left}}: \af\ ratio versus subcluster distance. The solid line shows the least squares fit for LTGs with \af\ $>$ 2 $\sigma$, i.e., a fit exclusively to the black symbols above the dashed line.  }  \textbf{\textit{Right}}: \af\ ratio versus \hi\ deficiency.  \textcolor{black}{ The solid line is a least squares fit for LTGs  with an  \af\  $>$ 2 $\sigma$, i.e., a fit exclusively to the black symbols above the dashed line.  }   The open symbols indicate LTGs which are members of close pairs or groups and the dashed horizontal line indicates the \textcolor{black}{ \af\ } 2 $\sigma $ level from the AMIGA isolated galaxy sample.  }
\label{Fig4n}
\end{center}
\end{figure*}

\begin{figure}
\begin{center}
\includegraphics[ angle=0,scale=.45] {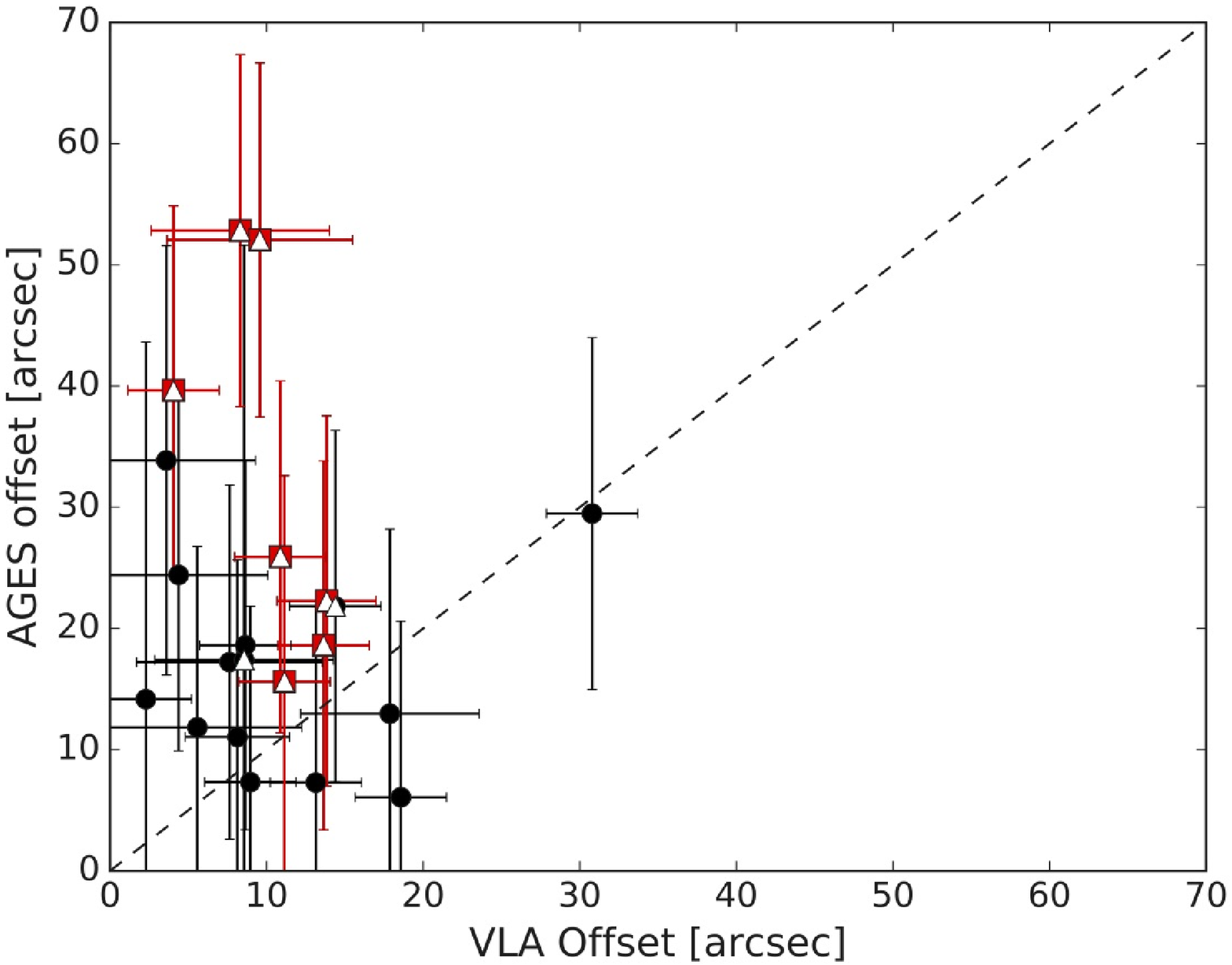}
\vspace{1cm}
\caption[Magnitude of the projected resolved VLA versus AGES \hi\ offsets ]{Magnitude of the resolved VLA versus AGES unresolved  \hi\ offsets  in arcsec.  Galaxies displaying resolved \hi\ morphologies with distinct one sided \hi\ tails are plotted with red square symbols and those with \textcolor{black}{\af} $>$ 2 $\sigma$ \textcolor{black}{are} shown with a white triangle.  The dashed lined indicates where  the 1:1  relationship lies.}
\label{fig_5nn}
\end{center}
\end{figure}

\subsection{\textcolor{black}{\textbf{\hi/optical  -- velocity offsets}}}
\label{voff}
\textcolor{black}{For the AGES detected LTGs (excluding those with  RFI contamination) we  compared  their AGES \hi\  and optical velocities. Six  LTGs had velocity differences greater than their 3 $\sigma$ velocity uncertainties, with the uncertainties calculated in quadrature using  both the AGES \hi\ and \textcolor{black}{NED} optical velocity uncertainties. Those LTGs and the their velocity differences \textcolor{black}{are highlighted with bold typeface in the final column of   Table \ref{table5a}.  It is notable that the three LTGs with a velocity difference $>$ 40 \km\ (FGC 1287,  \cg062  and \cg087)} all have clearly discernible \hi\ tails \textcolor {black}{in resolved maps} and \textcolor{black}{optical disk} inclinations $\geq$ 69\degree. }

\subsection{H\,{\sc i} optical \textcolor{black}{-- spatial} offsets }
\label{offsets}

\subsubsection{Unresolved AGES \hi\ offsets }
\label{offsd}
Table \ref{table4}   summarises the number of AGES \hi\ offsets   $>$1$\sigma$, $>$2$\sigma$  and $>$3$\sigma$ pointing uncertainties.   The  AGES \hi\ offset uncertainties were determined in quadrature using the \textcolor{black}{NED} optical and \hi\ AGES positional uncertainties.    11 (26\%) of the 43  AGES detections with optical counterparts in the A\,1367 volume have AGES \hi\ offsets $>$ 2$\sigma$ pointing uncertainties (referred to as AGES \hi\ offsets $>$ 2$\sigma$ from now on) and they  are \textcolor{black}{ highlighted with bold typeface in the columns headed \hi\ offsets AGES in Table \ref{table5a}.}    Figure \ref{fig2}  shows the  magnitudes (scaled $\times$ 20) of the AGES \hi\ offsets $>$1$\sigma$, $>$2$\sigma$  and $>$3$\sigma$ with arrowheads indicating the projected direction of each offset from the galaxy's optical centre to its AGES \hi\ position. 
We note from Figure \ref{fig2} that the  largest magnitude offsets are not preferentially found near the subcluster cores, with three of the four AGES  \hi\  offsets $>$3$\sigma$  projected beyond the $R_{200}$ radius. Two of these offsets (\cg026 and  FGC\,1287) are known from VLA \hi\ mapping to contain long $\sim$ 160 kpc and $\sim$ 250 kpc \hi\ tails, respectively \textcolor{black}{(Paper II)}.

\begin{table*}
\begin{minipage}[b]{\linewidth}\centering
\caption{\textcolor{black}{AGES} \af\ $>$ 2 $\sigma$  and \textcolor{black}{AGES} \hi\ offset $>$ 2 $\sigma$ relations, \textit{r} and \textit{p}--values}
\label{table7a}
\begin{tabular}{@{}lrrrr@{}}
\hline
& Pearson&&Spearman\\
Relation&\textcolor{black}{$r$}--value\footnote{ \textcolor{black}{Relations in bold typeface are those for which a one--tailed directional test using the Pearson \textit{r} rejects the null hypothesis that the variables are uncorrelated at confidence levels of 95 \% or greater.} \textcolor{black}{The degrees of freedom for the tests is 15, i.e., sample size - 2.} The \hi\ deficiencies and \hi\ mass were  calculated assuming a distance to \textcolor{black}{the cluster of  92 Mpc}.} 	& \textcolor{black}{$p$}--value&\textcolor{black}{$r_s$}&\textcolor{black}{$p$}--value\\
\hline
\textcolor{black}{\af}  $>$ 2 $\sigma$\\
\hline
AGES \hi\ offset &-0.063&0.810&0.071&0.786\\
\textbf{Subcluster distance }&\textbf{-0.598}&\textbf{0.011}&-0.646&0.005\\
\textbf{\hi\ deficiency } &\textbf{0.522}&\textbf{0.031}&0.658&0.004\\
\hi\ mass &-0.240&0.353&-0.210&0.419\\
SDSS $g - i$ &-0.341&0.180&-0.077&0.768\\
Optical disk inclination &0.259&0.315&0.102&0.697\\
AGES spectrum mean S/N &-0.312&0.222&-0.246&0.340\\
\hline
\hline
AGES \hi\ offset  $>$ 2 $\sigma$\\
\hline
Subcluster distance &0.256&0.447&0.391&0.235\\
 \hi\ deficiency &0.099&0.772&0.000&1.000\\
\hi\ mass&0.038&0.910&0.200&0.555\\
SDSS $g - i$&0.333&0.316&0.300&0.370\\
Optical disk inclination &0.012&0.972&0.291&0.385\\
AGES spectrum mean S/N &-0.046&0.894&-0.041&0.905\\
\hline
\end{tabular}
\end{minipage}
\end{table*}

\begin{table}
\centering
\begin{minipage}{140mm}
\caption[]{AGES \hi\ offsets in the A\,1367 volume }
\label{table4}
\begin{tabular}{@{}lrrrr@{}}
\hline
\hi\ offsets from &	Total\footnote{Excludes 4 AGES galaxies with spectra significantly impacted \\ by RFI  } 	&	offset	&	offset 	&offset	\\
AGES \hi\ positions&	 	&	$>$3 $\sigma$ 	&	$>$2 $\sigma$ 	&$>$1 $\sigma$ 	\\
  \hline
Number of galaxies &	43	&	4	&	11&	26\\
Percentage &		&	9	&	26&	60\\
 \hline
\end{tabular}
\end{minipage}
\end{table}

We carried out tests for statistically significant correlations between the AGES \hi\ offsets $>$2$\sigma$ and subcluster distance, \hi\ deficiency, \hi\ mass, SDSS $g - i$ colour, optical disk inclination and mean S/N in the AGES spectrum. Based on the  Pearson correlation coefficient\textcolor{black}{, $r$(\textcolor{black}{df\footnote{\textcolor{black}{Degrees of freedom = sample size - 2.}} = }9), and a one-tailed test}  the \textcolor{black}{null hypothesis that each of these  variables are uncorrelated to AGES \hi\ offsets $>$2$\sigma$ is accepted with a 99.95\% confidence level}.   Table \ref{table7a} sets out the \textcolor{black}{$r$ and $p$} values from these tests.

\subsubsection{Resolved VLA \hi\ offsets }
\label{offvla}
\begin{figure}
\begin{center}
\includegraphics[ angle=0,scale=.45] {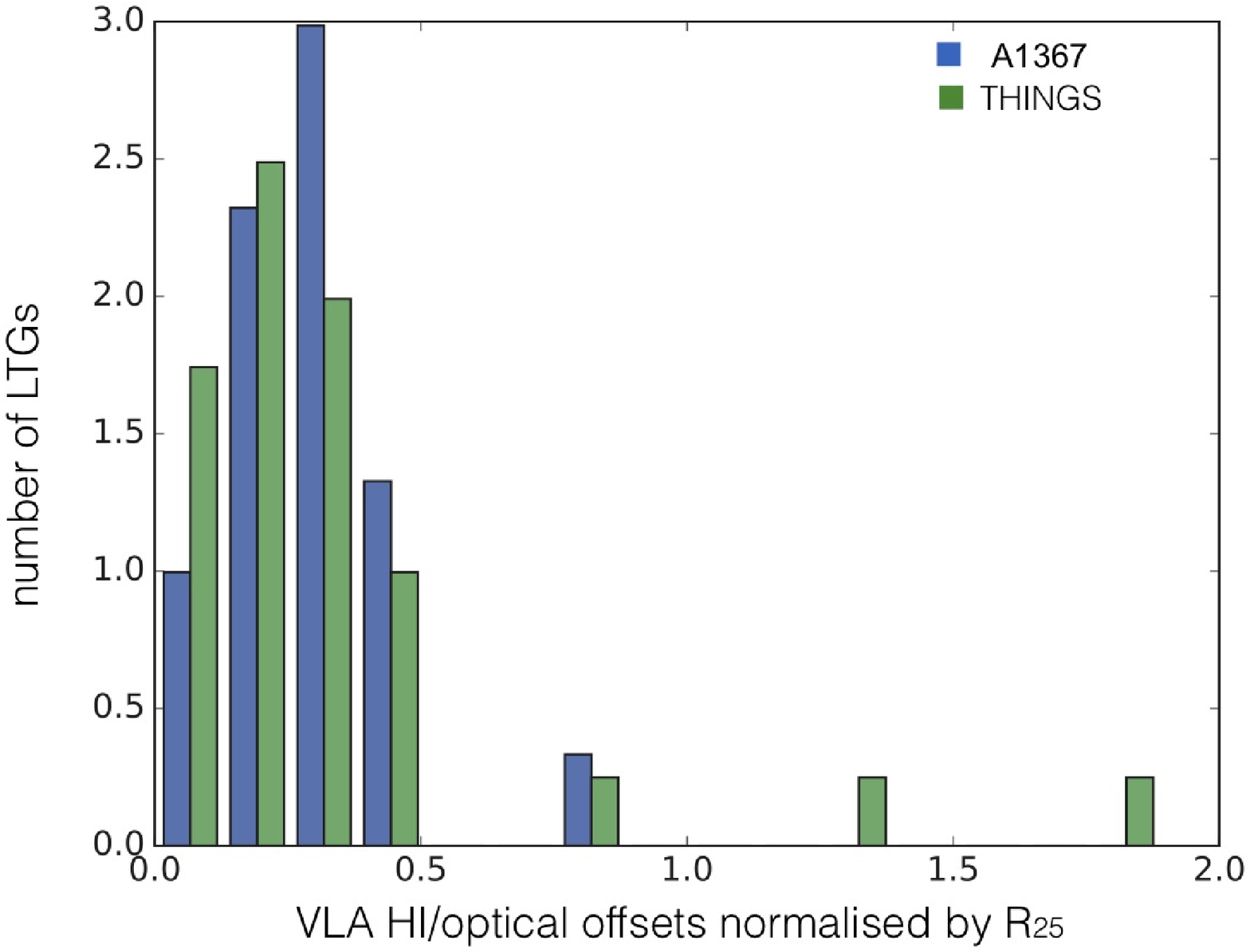}
\vspace{1cm}
\caption[ ]{\textcolor{black}{Comparative distribution of the \textcolor{black}{the number of LTGs (vertical axis) versus their  \hi\ projected offset magnitudes, normalised by their $R_{25}$,   derived from resolved VLA  A\,1367  and THINGS \hi\ maps (horizontal axis)}.} }
\label{fig6nn}
\end{center}
\end{figure}

12 LTGs (40\%) of \textcolor{black}{the} 30 objects with resolved VLA \hi\ offsets show VLA \hi\ offsets greater than the 2 $\sigma$ VLA/optical pointing uncertainties. To determine whether  the magnitude of the A\,1367 resolved VLA \hi\ offsets were abnormally large we  compared  them  with the resolved \hi\ offsets from  32 high resolution VLA \hi\ maps of  nearby LTGs  available from  the   THINGS\footnote{The \hi\ Nearby Galaxy Survey survey  \citep{walter08}.  M\,81 was excluded from our THINGS sample because of incomplete continuum subtraction from its \hi\ map in the archive.  } on--line archive.   Our analysis of the THINGS \hi\  offsets was carried out after convolving the THINGS   {\sc robust = 0} \hi\ maps  with a 90 arcsec beam, using the AIPS task {\sc convl}, to  compensate for the greater distance at which  the A\, 1367 galaxies were  observed.  Figure \ref{fig6nn} shows  histograms of the projected \hi\ offset magnitudes (normalised by \textcolor{black}{$R_{25}$})  for the  A\,1367   and  THINGS  LTGs.    Table  \ref{table4a} shows the median and mean of these normalised \hi\ offset values.  The median  normalised resolved VLA \hi\ offsets for the  A\,1367 and THINGS galaxies are 0.27  and 0.23 with  mean values of 0.27$\pm$0.15 and 0.33$\pm$0.38 respectively. 

Inspection of the high resolution THINGS \hi\  images reveals they have widely diverse  morphologies, with the \hi\ intensity maxima variously associated  with \hi\ central concentrations, rings, spiral arms and tidal features in ways that do not follow an easily  predictable pattern. This is reflected in  the large standard deviation from the mean offset for the \textcolor{black}{$R_{25}$} normalised  resolved \hi\   for the THINGS galaxies.  Within the limitations of the small sample sizes and large variation in  \hi\ offset magnitudes, the mean \textcolor{black}{$R_{25}$} normalised resolved VLA \hi\  offset in the A\,1367 galaxies cannot be \textcolor{black}{distinguished} from those in the THINGS galaxies  (i.e. LTGs near  the Virgo cluster outskirts). 

\begin{table}
\centering
\begin{minipage}{180mm}
\caption[]{Magnitude \hi\ offsets normalised by optical \textcolor{black}{$R_{25}$}}
\label{table4a}
\begin{tabular}{@{}llllll@{}}
\hline
\hi\ maxima &Survey&	Median	&	Mean	&	n\footnote{number of \hi\ detections with optical counterparts.}	\\
\hline
Resolved (VLA)&A\,1367\footnote{\textcolor{black}{$R_{25}$} is from,  in order of preference, Hyperleda, NED\\  and our own estimate from SDSS images. }   &0.27&	0.27$\pm$ 0.15	&	29 \\
Resolved (VLA) &THINGS\footnote{\textcolor{black}{$R_{25}$} is from \textcolor{black}{$D_{25}$/2 from} \cite{walter08}. }  &	0.23 	&	0.33$\pm$ 0.38	&	32 \\
\hline
Unresolved (Arecibo)&AGES  \ &0.52	&	0.79$\pm$ 0.74	&	43 \\
 \hline
\end{tabular}
\end{minipage}
\end{table}

\subsubsection{ Unresolved (AGES) versus resolved VLA \hi\ offsets }
\label{offvla_ages}
For  21 galaxies in the A\,1367 volume we have both an AGES \hi\ position  and a resolved VLA \hi\ column density  maximum position  from the VLA maps. However, \textcolor{black}{ neither the magnitudes  of the offsets  nor the offset orientations from AGES and the VLA were statistically correlated. In both cases the null hypothesis that the variables are uncorrelated was accepted (one--tailed test) at  confidence levels of 99.95\%, with the Pearson \textit{r} and the \textit{p} values for offset magnitude being \textit{r}(\textcolor{black}{df =}19) = -0.123, p = 0.5964 and for orientation, r(\textcolor{black}{df =} 19) = 0.268, p = 0.240.}  Figure \ref{fig_5nn} shows the projected magnitude of  the AGES and resolved VLA \hi\ offsets in arcsec  for the 21  galaxies. The mean AGES \hi\ offset (22 $\pm$ 13 arcsec)  is  twice the  mean resolved VLA  \hi\ offset (11 $\pm$ 6 arcsec). 
 
\begin{figure*}
\begin{center}
\includegraphics[ angle=0,scale=.5] {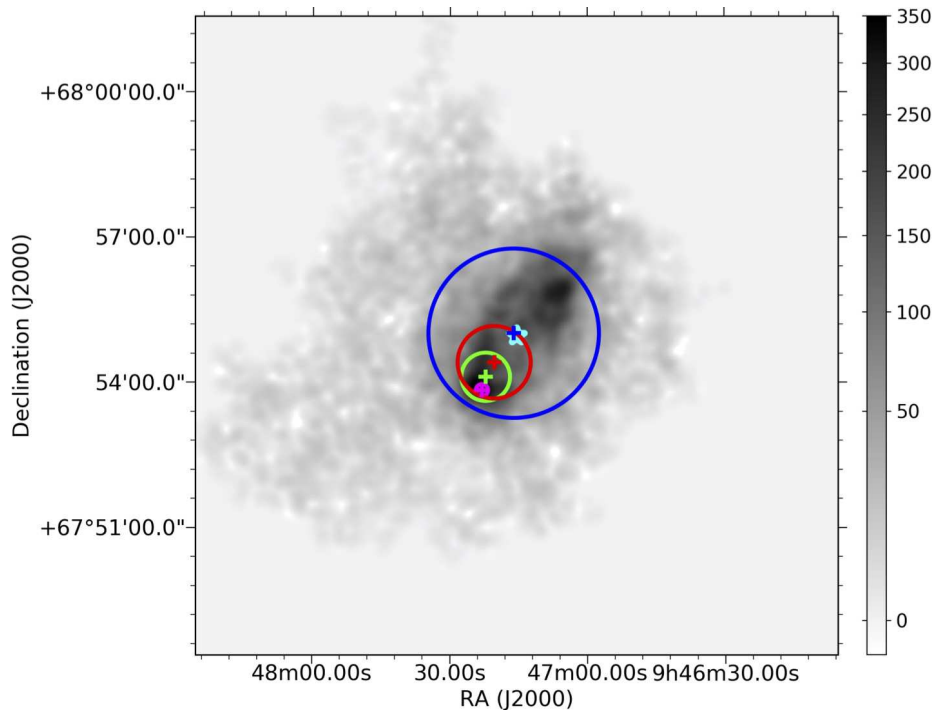}
\vspace{1cm}
\caption[ ]{ NGC 2976 (V$_{\mathrm{hel}}$ = 3\,km\,s$^{-1}$) THINGS \hi\ map showing the impact of resolution on the position of the \hi\ intensity maximum. The crosses mark the positions of the \hi\ intensity maxima from maps smoothed with the AIPS tasks {\sc convl} using beam sizes of 15 (magenta), 60 (green), 90 (red)  and 210 (blue)  arcsec.  The circles of the same color indicate the {\sc convl} beam size,  with the 210 arcsec beam approximately equal to the Arecibo 3.5 arcmin FWHP beam. The cyan star marks the optical centre of the galaxy. \textcolor{black}{The greyscale scale shows the \hi\ column density in the figure in units of 10$^{19}$ atoms cm$^2$. }  } 
\label{fig7n}
\end{center}
\end{figure*}

As part of trying to understand the reason for this difference we investigated the impact of spatial resolution on the position of the \hi\  maxima.  We used the AIPS task  {\sc convl} to progressively  smooth the highest resolution \hi\ images available, i.e.,  THINGS \hi\ maps of nearby galaxies.  Progressively increased  smoothing of the THINGS maps in almost all cases significantly changed  the position of the  \hi\  maximum intensity away from the highest column density clumps resolvable in the  THINGS images toward  the flux weighted mean position of the \hi\ in the whole galaxy. Figure \ref{fig7n} illustrates the effect of resolution on the position of \hi\  maximum intensity. For the THINGS galaxies the mean shift in \hi\  maximum intensity  between the images smoothed with  15 arcsec and 210 arcsec beams was 94 arcsec.   \textcolor{black}{From this we conclude the unresolved AGES A\,1367 \hi\ positions reflect the flux weighted mean position of each map's \hi, whereas its resolved VLA \hi\ column density maximum locates the highest density \hi\ within the resolved map's resolution limit. Moreover, our analysis in section  \ref{offvla} indicates that for highly resolved maps the position of the \hi\ column density maximum offset   has a large natural variation. The THINGS  results explain why for the A\,1367 LTGs the VLA \hi\ offsets are not correlated with the AGES \hi\ offsets or other lower density resolved \hi\ morphology or kinematic perturbation signatures, even in  cases where  the resolved maps show only some moderate \hi\ perturbation.}

\textcolor{black}{On the other hand, the THINGS results suggest that the unresolved AGES \hi\ offsets can reflect the effect of interactions which have  asymmetrically displaced significant masses of lower density \hi\ with  little or no impact on any offset derived on the basis of the resolved VLA \hi\ maps. This naturally explains why all of the LTGs with clearly identifiable diffuse one--sided  \hi\ tails (e.g., FGC\,1287 and \cg087; figure \ref{97087}),  shown  with  red squares in Figure \ref{fig_5nn},  have larger AGES \hi\ offsets than  VLA \hi\ offsets.  A good example is \cg062, a  galaxy with both optical and \hi\ one sided SE oriented tails. Table \textcolor{black}{\ref{table5a}} reveals it has an  AGES \hi\ offset  $>$ 2 $\sigma$ = 39.6$\pm$15.2 (16 kpc) but its  \hi\ column density maximum in the VLA C--array map is projected at the optical centre, as seen in Figure \ref{97062} in Appendix \ref{appenda}. Other examples  showing a low density one--sided \hi\ tail, but no significant resolved VLA \hi\ offset,  are \cg121 and \cg102 (see Appendix \textcolor{black}{\ref{appenda}}, including Figures \ref{97121}  and \ref{97102}).}

\subsection{\textcolor{black}{ \af\ distribution: comparison with the Virgo cluster} }
\label{iresoled-vla}
\textcolor{black}{In an effort to understand whether the \hi\ in A\,1367 LTGs  is more  perturbed than in other nearby clusters we  defined a  cluster volume for Virgo (RA [11:55:22, 13:06:10], DEC \textcolor{black}{[3:28:12, 21:12:00]}, V[-1000, 3000]). This cluster volume was calculated by scaling the Virgo  cluster volume relative to the dimensions of the A\,1367--AGES volume, i.e.,  1.64 $\times$ $R_{200}$  and a velocity range $\sim$ 4  $\times$ the cluster's velocity dispersion. A substantial fraction of the Virgo volume defined in this way is included within the ALFALFA blind \hi\ surveys \citep{giovan2007,kent08} and  like the AGES survey for A\,1367, these surveys were carried out with the ALFA receiver and  the  Arecibo 305m  telescope. From here on we refer to the ALFALFA surveyed portion of the Virgo volume  as the Virgo--ALFALFA volume.  } 

\textcolor{black}{ We compiled a sample of 220 spectra for \hi\ detections within the  Virgo--ALFALFA volume which have optical counterparts and  spectra of high quality (quality parameter = 1) in \cite{giovan2007} and \cite{kent08}. After  reviewing the notes for these ALFALFA surveys we excluded 11 of these spectra because they were were noted as being contaminated by RFI or confused with another sources. We retrieved  the remaining sample spectra, with the exception of two spectra that were unavailable, from the NED ALFALFA data archive \citep{hayn11}, i.e., a final sample of  207 spectra.  Although using observations from the same telescope aids comparability of the spectra, because Virgo is closer than A\,1367 the Virgo-ALFALFA sample has a lower \hi\ mass detection threshold.  To ensure a proper comparison  between  the distribution of \af\ in Virgo and  A\,1367  we selected only Virgo--ALFALFA spectra  for galaxies with \mhi\ $ >$ 3 $\times$ 10$^8$ \msolar, i.e., the lowest \hi\ mass LTG in our A\,1367--AGES sample; 90 of the 207  Virgo--ALFALFA  sample galaxies met this criteria  }

\textcolor{black}{We determined \af\ for the 90 Virgo--ALFALFA spectra and the distribution of their \af\ ratios is shown in Figure \ref{fig2} (lower panel)  compared to the half  Gaussian fit to the  \af\ distribution from  the AMIGA sample of isolated galaxies \citep{espada11}.  \textcolor{black}{14 (16\%)} of the Virgo--ALFALFA   galaxies (\mhi\ $ >$ 3 $\times$ 13$^8$ \msolar) have \af\ $>$ 3 $\sigma$ (1.39) in the AMIGA sample. This compares to  26\% for A\,1367. The mean \af\ uncertainty for the Virgo--ALFALFA  and A\,1367--AGES  samples are 0.07 and  0.05, respectively.   We also examined the distribution of \af\ ratios for the 117 Virgo--ALFALFA galaxies  excluded from the sample of \textcolor{black}{90} galaxies solely because their \mhi\ was $ \leq$ 3 $\times$ 10$^8$ \msolar.  49 (42\%) of these galaxies have \af\ $>$ 3 $\sigma$ (1.39) in the AMIGA sample. }

\textcolor{black}{The VIVA\footnote{VLA Imaging of Virgo galaxies in Atomic gas} VLA \hi\ imaging survey of Virgo galaxies revealed five galaxies and an interacting pair with long  one sided \hi\ tails \citep{chung07}. \textcolor{black}{All of these galaxies except NGC\,4424 have an \hi\ mass $>$ 3 $\times$ 10$^8$ \msolar.}  Three of the five galaxies (NGC\,4654, NGC\,4302 and NGC\, 4330) had \af\ from  their ALFALFA spectra $>$ 2 $\sigma$   from the AMIGA sample distribution (1.26). For  the pair (NGC\,4294/4299), which have a projected separation of  5.6 arcmin, only the higher inclination  member, NGC 4294,  has an ALFALFA \af \ $>$ 1.26. We conclude that   \af\  $>$ 1.26 is associated with a majority of VIVA galaxies with long \hi\ tails, but it  is not the case for all of them. }

\textcolor{black}{For the Virgo and A\,1367 samples uncertainties about the fraction of galaxies with  \af\ $>$ 1.39 are introduced by the process of selecting eligible spectra, e.g., because of Arecibo's fixed FWHP beam size and the greater distance to A\,1367, \hi\ detections are more likely to be confused than those in the nearer isolated and Virgo samples. For the parts of the A\,1367 volume where we have VLA \hi\ data we were able to correct for this by substituting unconfused  VLA \af\ values (\cg068 and \cg125), but in the rest of the A\,1367 volume we cannot rule out confusion.    }

\textcolor{black}{Even allowing for possible confusion, the comparison between distribution of \af\ values  from the A\,1367 and Virgo samples  with \mhi\ $ >$ 3 $\times$ 10$^8$ \msolar\ provides clear evidence  of a higher frequency of strong  \hi\ \textcolor{black}{profile} perturbations   in A\,1367 LTGs.  Also, the higher  fraction of  Virgo galaxies with   \af\ above the AMIGA 3 $\sigma$ value in the  lower \hi\ mass sample compared to the fraction from the higher \hi\ mass sample,  suggests the lower \hi\ mass galaxies are preferentially  perturbed.   However, this higher fraction \textcolor{black}{of  lower \hi\ mass galaxies in Virgo}  is also, to some extent, probably due to the presence of perturbed  residual \hi\ in high stellar mass galaxies at a late stage of gas stripping. }

\subsection{ICM and Galaxy density}
\label{icm}

\begin{figure*}
\centering
\includegraphics[scale=0.62, angle=0]{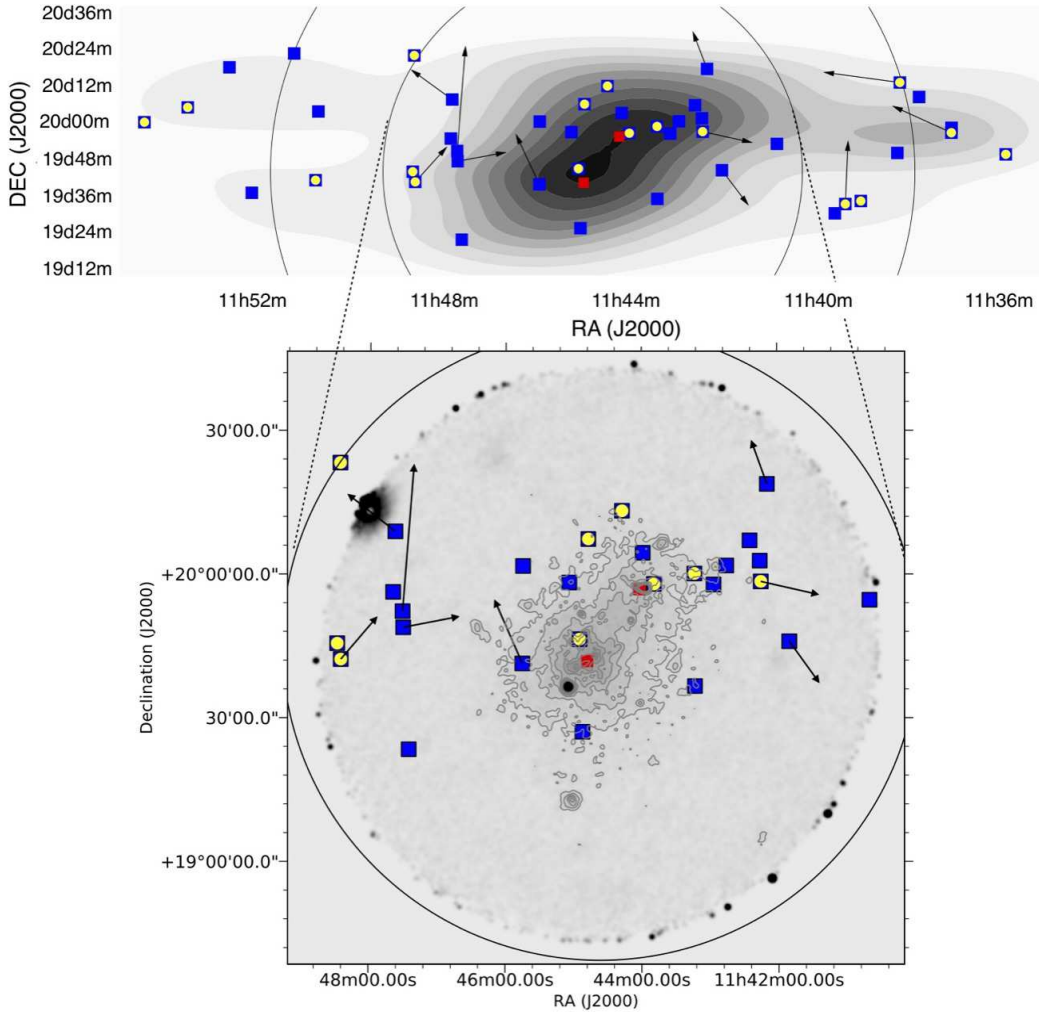}  
\caption{ \textbf{A\,1367  positions of galaxies with \textcolor{black}{\af}$>$ 2 $\sigma$ (yellow dots), projected offset magnitudes (x 20) and  orientations of the  AGES \hi\ offsets $>$ 2 $\sigma$ (black arrows) and the 43  AGES \hi\ detections (blue squares)}: \textit{Top panel:}. Positions projected on grayscale contours from a density map \textcolor{black}{for} all galaxies with SDSS spectroscopic redshifts in the A\,1367 volume. \textit{Lower panel: }Zoom in of the data  projected on a \textit{ROSAT} X--ray image with gray contours.  The red squares indicate the location of the SE and NW subcluster cores \citep{don98}. The outer larger circle is the virial radius and the inner larger circle is the \textcolor{black}{$R_{200}$} radius. } 
\label{fig8}
\end{figure*}
Modelling \cite[e.g.,][]{tonn07} supports the general proposition that   ram pressure stripping is  the principal mechanism for gas  loss in nearby cluster LTGs. But in A\,1367,  ram pressure stripping modelling \textcolor{black}{reported} in Paper I  for velocities approximately equal to the cluster velocity dispersion, suggests  most of its LTGs are only subject to moderate ram pressure, except within $\sim$ 200 kpc of the subcluster cores.  Clumpy ICM, shocks and bulk ICM motions could potentially enhance ram pressure  by up to an order of magnitude \citep{ken04}.  However, we see no evidence of  ICM clumps at the sensitivity and resolution of  available  \textit{ROSAT} and  \textit{XMM Newton} X--ray data. Projection  effects may mask a preferred direction for AGES \hi\ offsets. However,  if the AGES \hi\ offsets are  due to a systematic, as opposed to \textcolor{black}{a} sudden increase in  ram pressure stripping during their infall,  we would expect the projected magnitude of the \hi\ offsets to increase with proximity to the cluster centre. \textcolor{black}{However,}  we do not see evidence of this in Figure \ref{fig8}.

Figure \ref{fig8}  shows the AGES \hi\ offsets $>$ 2 $\sigma$ (black arrows) and \textcolor{black}{\af}  $>$ 2 $\sigma$ (yellow filled circles) projected against galaxy density and ICM X--ray emission (\textcolor{black}{\textit{ROSAT}}) as well as the 43 AGES \hi\ detections (blue squares). The LTGs with \textcolor{black}{\af}  $>$ 2 $\sigma$ are projected more or less randomly within the regions of the in the A\,1367 volume \textcolor{black}{enclosed by the lowest galaxy density contour}.  On the other hand the AGES \hi\ offsets $>$ 2 $\sigma$ appear preferentially projected at clustercentric  radii  between half of the \textcolor{black}{$R_{200}$} and the viral radius. We do not see any clear correlation between either of the AGES \hi\ \textcolor{black}{offsets}  or \textcolor{black}{\af}   $>$ 2 $\sigma$  and either galaxy density or ICM density traced by X--ray emission. \textcolor{black}{ It is important to not confuse the lack of correlation with projected position  for LTGs having \af\ $>$ 2 $\sigma$ in Figure \ref{fig8} with the  correlation between cluster centric distance and the magnitude of  \af\ $>$ 2 $\sigma$  shown in Figure \ref{Fig_3nn}.}

\section{Concluding remarks}
\label{concl}
Using the single dish and resolved \hi\ data available to us for A\,1367 LTGs, including VLA C--configuration data for the 16 objects  presented in this paper, we analysed \textcolor{black}{four} types of  \hi\ perturbation signature (\hi\ profile asymmetries\textcolor{black}{, \hi/optical velocity offsets} as well as projected resolved and unresolved \hi\ position versus optical position offsets). 
\begin{itemize}
\item
26\% of A\,1367 LTGs have asymmetric AGES  \hi\ profiles  with an \textcolor{black}{\af} ratio  greater than the \textcolor{black}{3 $\times$ the 1 $\sigma$ dispersion in the ratio (3 \textcolor{black}{$\sigma$} for short)} from a sample of isolated galaxies. Only  $\sim$ 2\% of the isolated sample have an \textcolor{black}{\af} ratio  \textcolor{black}{greater than its 3 $\sigma$ value } and for other samples \textcolor{black}{probing}  \textcolor{black}{denser} environments the fractions are between 10\% to 20 \%. 
\item 
For the A\,1367 LTGs there is a statistically significant correlation between  their  \textcolor{black}{\af}  $>$ 2 $\sigma$ from AGES and both their proximity to the subcluster centres and their \hi\ deficiency. This indicates \textcolor{black}{\af}  $>$ 2 $\sigma$ from AGES can be a signature of recent or ongoing \hi\ stripping; 56\% of the A\,1367 LTGs, which are members of groups or pairs, have an \textcolor{black}{\af} ratio $>$ 2 $\sigma$. These groups and pairs \textcolor{black}{of}  LTGs  make up 31\% of the A\,1367 LTGs with \textcolor{black}{an} \af ratio  $>$ 2 $\sigma$. This suggests  interactions between group or  pair members \textcolor{black}{may  significantly contribute} to the number of A\,1367 galaxies  displaying an  \textcolor{black}{\af} ratio $>$ 2 $\sigma$.
\item
\textcolor{black}{For A\,1367 LTGs } we did not find a statistically significant correlation between   \af\  \textcolor{black} {magnitudes}  $>$ 2 $\sigma$ and \textcolor{black} {the magnitudes of}  AGES \hi\ offsets $>$ \textcolor{black}{the} 2 $\sigma$ pointing error. However,  in cases where both of these signatures are present and the  resolved VLA \hi\ mapping is available, the resolved \hi\ morphology and kinematics confirm  a strongly perturbed \hi\ disk.
\item
\textcolor{black}{For \textcolor{black}{the} 19 A\,1367 LTGs \textcolor{black}{for which we have both AGES offsets and resolved  VLA \hi\  offsets. No statistical correlation was found \textcolor{black}{for either the \hi\ offset magnitudes or orientations from the two telescopes. However,} the  mean resolved VLA \hi\ offset magnitude is only about half that} of the mean resolved AGES \hi\ offset. The LTGs  with the most extreme  differences  between their  AGES and  VLA  \hi\ offset \textcolor{black}{magnitudes  are those where the VLA \hi\ maps reveal low density } one sided \hi\ tails, \textcolor{black}{but with little or no evidence of an impact on the highest density \hi\  from the  the resolved \hi\ offsets.} This suggests that the AGES \hi\ offsets $>$ 2 $\sigma$ pointing uncertainties are sensitive to  interactions which asymmetrically displace  significant  masses of  \textcolor{black}{low density \hi.  }}
\item
\textcolor{black}{The 26\% of A\,1367 LTGs  with  \af\ $>$ 3 $\sigma$ in an isolated sample,  compares to 16\% from a comparable  sample of Virgo galaxies drawn from the ALFALFA blind \hi\ surveys. These fractions indicate a higher frequency of strong \hi\  perturbations in A\,1367 LTGs in comparison to  Virgo.} 
\item
\textcolor{black}{Within the A\,1367 volume, the projected distribution \textcolor{black}{of} neither the LTGs with  unresolved AGES \hi\ offsets, $>$ 2$\sigma$ pointing uncertainties,  nor those with an  \textcolor{black}{\af} ratio $>$ 2$\sigma$   display a clear correlation with distance from the cluster centre, galaxy density or the cluster's ICM X--ray emission (\textcolor{black}{\textit{ROSAT}}).}

\end{itemize}
This study confirms that the A\,1367 LTGs are suffering ongoing strong and widespread interactions which are significantly perturbing their \hi. While the evidence so far reveals examples which are consistent with ram pressures stripping and/or tidal interactions,  determining (with greater certainty) the dominant  interaction mechanism  for each galaxy and for the cluster as whole will require further detailed multi--wavelength studies and modelling. 

\section*{Acknowledgements}
\textcolor{black}{We are grateful to the anonymous referee for their helpful comments that have significantly improved the paper.} This work was supported by Funda\c{c}\~{a}o para a Ci\^{e}ncia e a Tecnologia (FCT) through national funds (UID/FIS/04434/2013) and by FEDER through COMPETE2020 (POCI-01-0145-FEDER-007672). TS acknowledges the support by the fellowship SFRH/BPD/103385/2014 funded by FCT (Portugal) and POPH/FSE (EC).
 \textcolor{black}{EB acknowledges support from the UK Science and Technology Facilities Council [grant number ST/M001008/1].} \textcolor{black}{We would like thank Dr Dana Ficut--Vicas for \textcolor{black}{her role in developing} the AIPS scripts for the reduction of data taken during the VLA--EVLA transition.}

This research has made use of the NASA/IPAC Extragalactic Database (NED) which is operated by the Jet Propulsion Laboratory, California Institute of Technology, under contract with the National Aeronautics and Space Administration.

This research has made use of the Sloan Digital Sky Survey (SDSS). Funding for the SDSS and SDSS-II has been provided by the Alfred P. Sloan Foundation, the Participating Institutions, the National Science Foundation, the U.S. Department of Energy, the National Aeronautics and Space Administration, the Japanese Monbukagakusho, the Max Planck Society, and the Higher Education Funding Council for England. The SDSS Web Site is http://www.sdss.org/.

\textcolor{black}{This research made use of APLpy, an open--source plotting package for Python \citep{Robitaille2012}.}

\bibliographystyle{mnras}
\bibliography{cluster}

\begin{thebibliography}{}
\makeatletter
\relax
\def\mn@urlcharsother{\let\do\@makeother \do\$\do\&\do\#\do\^\do\_\do\%\do\~}
\def\mn@doi{\begingroup\mn@urlcharsother \@ifnextchar [ {\mn@doi@}
  {\mn@doi@[]}}
\def\mn@doi@[#1]#2{\def\@tempa{#1}\ifx\@tempa\@empty \href
  {http://dx.doi.org/#2} {doi:#2}\else \href {http://dx.doi.org/#2} {#1}\fi
  \endgroup}
\def\mn@eprint#1#2{\mn@eprint@#1:#2::\@nil}
\def\mn@eprint@arXiv#1{\href {http://arxiv.org/abs/#1} {{\tt arXiv:#1}}}
\def\mn@eprint@dblp#1{\href {http://dblp.uni-trier.de/rec/bibtex/#1.xml}
  {dblp:#1}}
\def\mn@eprint@#1:#2:#3:#4\@nil{\def\@tempa {#1}\def\@tempb {#2}\def\@tempc
  {#3}\ifx \@tempc \@empty \let \@tempc \@tempb \let \@tempb \@tempa \fi \ifx
  \@tempb \@empty \def\@tempb {arXiv}\fi \@ifundefined
  {mn@eprint@\@tempb}{\@tempb:\@tempc}{\expandafter \expandafter \csname
  mn@eprint@\@tempb\endcsname \expandafter{\@tempc}}}

\bibitem[\protect\citeauthoryear{{Amram}, {Gavazzi}, {Marcelin}, {Boselli},
  {V{\'{\i}}lchez}, {Iglesias-Paramo}  \& {Tarenghi}}{{Amram}
  et~al.}{2002}]{amram02}
{Amram} P.,  {Gavazzi} G.,  {Marcelin} M.,  {Boselli} A.,  {V{\'{\i}}lchez}
  J.~M.,  {Iglesias-Paramo} J.,   {Tarenghi} M.,  2002, \mn@doi [Ap\&SS]
  {10.1023/A:1019502105186}, \href
  {http://adsabs.harvard.edu/abs/2002Ap%26SS.281..401A} {281, 401}

\bibitem[\protect\citeauthoryear{{Bekki}}{{Bekki}}{2014}]{bekki14}
{Bekki} K.,  2014, \mn@doi [MNRAS] {10.1093/mnras/stt2216}, \href
  {http://adsabs.harvard.edu/abs/2014MNRAS.438..444B} {438, 444}

\bibitem[\protect\citeauthoryear{{Boselli} \& {Gavazzi}}{{Boselli} \&
  {Gavazzi}}{2006}]{bosel06a}
{Boselli} A.,  {Gavazzi} G.,  2006, \mn@doi [PASP] {10.1086/500691}, \href
  {http://adsabs.harvard.edu/abs/2006PASP..118..517B} {118, 517 }

\bibitem[\protect\citeauthoryear{{Boselli}, {Gavazzi}, {Combes}  \&
  {Lequeux}}{{Boselli} et~al.}{1994}]{bosel94}
{Boselli} A.,  {Gavazzi} G.,  {Combes} F.,   {Lequeux} J.,  1994, A\&A, \href
  {http://adsabs.harvard.edu/abs/1994A%26A...285...69B} {285, 69}

\bibitem[\protect\citeauthoryear{{Boselli}, {Boissier}, {Cortese}, {Gil de
  Paz}, {Seibert}, {Madore}, {Buat}  \& {Martin}}{{Boselli}
  et~al.}{2006}]{bosel06b}
{Boselli} A.,  {Boissier} S.,  {Cortese} L.,  {Gil de Paz} A.,  {Seibert} M.,
  {Madore} B.~F.,  {Buat} V.,   {Martin} D.~C.,  2006, \mn@doi [ApJ]
  {10.1086/507766}, \href {http://adsabs.harvard.edu/abs/2006ApJ...651..811B}
  {651, 811}

\bibitem[\protect\citeauthoryear{{Boselli}, {Cortese}, {Boquien}, {Boissier},
  {Catinella}, {Gavazzi}, {Lagos}  \& {Saintonge}}{{Boselli}
  et~al.}{2014}]{bosel14}
{Boselli} A.,  {Cortese} L.,  {Boquien} M.,  {Boissier} S.,  {Catinella} B.,
  {Gavazzi} G.,  {Lagos} C.,   {Saintonge} A.,  2014, \mn@doi [A\&A]
  {10.1051/0004-6361/201322313}, \href
  {http://adsabs.harvard.edu/abs/2014A%26A...564A..67B} {564, A67}

\bibitem[\protect\citeauthoryear{{Boselli} et~al.,}{{Boselli}
  et~al.}{2016}]{bosel16}
{Boselli} A.,  et~al., 2016, \mn@doi [A\&A] {10.1051/0004-6361/201527795},
  \href {http://esoads.eso.org/abs/2016A%26A...587A..68B} {587, A68}

\bibitem[\protect\citeauthoryear{{Bravo-Alfaro}, {Cayatte}, {van Gorkom}  \&
  {Balkowski}}{{Bravo-Alfaro} et~al.}{2000}]{bravo00}
{Bravo-Alfaro} H.,  {Cayatte} V.,  {van Gorkom} J.~H.,   {Balkowski} C.,  2000,
  \mn@doi [AJ] {10.1086/301194}, \href
  {http://adsabs.harvard.edu/abs/2000AJ....119..580B} {119, 580}

\bibitem[\protect\citeauthoryear{{Briggs}}{{Briggs}}{1995}]{briggs95}
{Briggs} D.~S.,  1995, BAAS, \href
  {http://adsabs.harvard.edu/abs/1995AAS...18711202B} {27, 1444}

\bibitem[\protect\citeauthoryear{{Brown} et~al.,}{{Brown}
  et~al.}{2017}]{brown17}
{Brown} T.,  et~al., 2017, \mn@doi [MNRAS] {10.1093/mnras/stw2991}, \href
  {http://esoads.eso.org/abs/2017MNRAS.466.1275B} {466, 1275}

\bibitem[\protect\citeauthoryear{{Chung}, {van Gorkom}, {Kenney}  \&
  {Vollmer}}{{Chung} et~al.}{2007}]{chung07}
{Chung} A.,  {van Gorkom} J.~H.,  {Kenney} J.~D.~P.,   {Vollmer} B.,  2007,
  \mn@doi [ApJL] {10.1086/518034}, \href
  {http://adsabs.harvard.edu/abs/2007ApJ...659L.115C} {659, L115}

\bibitem[\protect\citeauthoryear{{Chung}, {van Gorkom}, {Kenney}, {Crowl}  \&
  {Vollmer}}{{Chung} et~al.}{2009}]{chung09}
{Chung} A.,  {van Gorkom} J.~H.,  {Kenney} J.~D.~P.,  {Crowl} H.,   {Vollmer}
  B.,  2009, \mn@doi [AJ] {10.1088/0004-6256/138/6/1741}, \href
  {http://adsabs.harvard.edu/abs/2009AJ....138.1741C} {138, 1741}

\bibitem[\protect\citeauthoryear{{Conselice}}{{Conselice}}{2003}]{consel03}
{Conselice} C.~J.,  2003, \mn@doi [ApJS] {10.1086/375001}, \href
  {http://adsabs.harvard.edu/abs/2003ApJS..147....1C} {147, 1}

\bibitem[\protect\citeauthoryear{{Consolandi}, {Gavazzi}, {Fossati},
  {Fumagalli}, {Boselli}, {Yagi}  \& {Yoshida}}{{Consolandi}
  et~al.}{2017}]{cosolandi17}
{Consolandi} G.,  {Gavazzi} G.,  {Fossati} M.,  {Fumagalli} M.,  {Boselli} A.,
  {Yagi} M.,   {Yoshida} M.,  2017, \mn@doi [\aap]
  {10.1051/0004-6361/201731218}, \href
  {http://esoads.eso.org/abs/2017A%26A...606A..83C} {606, A83}

\bibitem[\protect\citeauthoryear{{Cortese}, {Gavazzi}, {Boselli},
  {Iglesias-Paramo}  \& {Carrasco}}{{Cortese} et~al.}{2004}]{cort04}
{Cortese} L.,  {Gavazzi} G.,  {Boselli} A.,  {Iglesias-Paramo} J.,   {Carrasco}
  L.,  2004, \mn@doi [A\&A] {10.1051/0004-6361:20040381}, \href
  {http://adsabs.harvard.edu/abs/2004A%26A...425..429C} {425, 429 }

\bibitem[\protect\citeauthoryear{{Cortese}, {Gavazzi}, {Boselli}, {Franzetti},
  {Kennicutt}, {O'Neil}  \& {Sakai}}{{Cortese} et~al.}{2006}]{cort06}
{Cortese} L.,  {Gavazzi} G.,  {Boselli} A.,  {Franzetti} P.,  {Kennicutt}
  R.~C.,  {O'Neil} K.,   {Sakai} S.,  2006, \mn@doi [A\&A]
  {10.1051/0004-6361:20064873}, \href
  {http://adsabs.harvard.edu/abs/2006A%26A...453..847C} {453, 847}

\bibitem[\protect\citeauthoryear{{Cortese} et~al.,}{{Cortese}
  et~al.}{2008}]{cort08}
{Cortese} L.,  et~al., 2008, \mn@doi [MNRAS]
  {10.1111/j.1365-2966.2007.12664.x}, \href
  {http://adsabs.harvard.edu/abs/2008MNRAS.383.1519C} {383, 1519}

\bibitem[\protect\citeauthoryear{{Cortese} et~al.,}{{Cortese}
  et~al.}{2012}]{cort12}
{Cortese} L.,  et~al., 2012, \mn@doi [A\&A] {10.1051/0004-6361/201219312},
  \href {http://adsabs.harvard.edu/abs/2012A%26A...544A.101C} {544, A101}

\bibitem[\protect\citeauthoryear{{Di Teodoro} \& {Fraternali}}{{Di Teodoro} \&
  {Fraternali}}{2015}]{diTeodoro2015}
{Di Teodoro} E.~M.,  {Fraternali} F.,  2015, \mn@doi [MNRAS]
  {10.1093/mnras/stv1213}, \href
  {http://adsabs.harvard.edu/abs/2015MNRAS.451.3021D} {451, 3021}

\bibitem[\protect\citeauthoryear{{Dickey} \& {Gavazzi}}{{Dickey} \&
  {Gavazzi}}{1991}]{dick91}
{Dickey} J.~M.,  {Gavazzi} G.,  1991, \mn@doi [ApJ] {10.1086/170056}, \href
  {http://adsabs.harvard.edu/abs/1991ApJ...373..347D} {373, 347}

\bibitem[\protect\citeauthoryear{{Dom{\'{\i}}nguez}, {Muriel}  \&
  {Lambas}}{{Dom{\'{\i}}nguez} et~al.}{2001}]{doming01}
{Dom{\'{\i}}nguez} M.,  {Muriel} H.,   {Lambas} D.~G.,  2001, \mn@doi [AJ]
  {10.1086/319405}, \href {http://adsabs.harvard.edu/abs/2001AJ....121.1266D}
  {121, 1266}

\bibitem[\protect\citeauthoryear{{Donnelly}, {Markevitch}, {Forman}, {Jones},
  {David}, {Churazov}  \& {Gilfanov}}{{Donnelly} et~al.}{1998}]{don98}
{Donnelly} R.~H.,  {Markevitch} M.,  {Forman} W.,  {Jones} C.,  {David} L.~P.,
  {Churazov} E.,   {Gilfanov} M.,  1998, \mn@doi [ApJ] {10.1086/305719}, \href
  {http://adsabs.harvard.edu/abs/1998ApJ...500..138D} {500, 138}

\bibitem[\protect\citeauthoryear{{Dressler}}{{Dressler}}{2004}]{dress04}
{Dressler} A.,  2004, in {Mulchaey} J.~S.,  {Dressler} A.,   {Oemler} A.,  eds,
  Clusters of Galaxies: Probes of Cosmological Structure and Galaxy Evolution.
  p.~206

\bibitem[\protect\citeauthoryear{{Espada}, {Verdes-Montenegro}, {Huchtmeier},
  {Sulentic}, {Verley}, {Leon}  \& {Sabater}}{{Espada} et~al.}{2011}]{espada11}
{Espada} D.,  {Verdes-Montenegro} L.,  {Huchtmeier} W.~K.,  {Sulentic} J.,
  {Verley} S.,  {Leon} S.,   {Sabater} J.,  2011, \mn@doi [A\&A]
  {10.1051/0004-6361/201016117}, \href
  {http://adsabs.harvard.edu/abs/2011A%26A...532A.117E} {532, A117}

\bibitem[\protect\citeauthoryear{{Fasano}, {Poggianti}, {Couch}, {Bettoni},
  {Kj{\ae}rgaard}  \& {Moles}}{{Fasano} et~al.}{2000}]{fasa00}
{Fasano} G.,  {Poggianti} B.~M.,  {Couch} W.~J.,  {Bettoni} D.,
  {Kj{\ae}rgaard} P.,   {Moles} M.,  2000, \mn@doi [ApJ] {10.1086/317047},
  \href {http://adsabs.harvard.edu/abs/2000ApJ...542..673F} {542, 673}

\bibitem[\protect\citeauthoryear{{Fossati}, {Fumagalli}, {Boselli}, {Gavazzi},
  {Sun}  \& {Wilman}}{{Fossati} et~al.}{2016}]{fossati16}
{Fossati} M.,  {Fumagalli} M.,  {Boselli} A.,  {Gavazzi} G.,  {Sun} M.,
  {Wilman} D.~J.,  2016, \mn@doi [MNRAS] {10.1093/mnras/stv2400}, \href
  {http://esoads.eso.org/abs/2016MNRAS.455.2028F} {455, 2028}

\bibitem[\protect\citeauthoryear{{Gavazzi}}{{Gavazzi}}{1989}]{gava89}
{Gavazzi} G.,  1989, \mn@doi [ApJ] {10.1086/167985}, \href
  {http://adsabs.harvard.edu/abs/1989ApJ...346...59G} {346, 59}

\bibitem[\protect\citeauthoryear{{Gavazzi} \& {Jaffe}}{{Gavazzi} \&
  {Jaffe}}{1987}]{gava87}
{Gavazzi} G.,  {Jaffe} W.,  1987, A\&A, \href
  {http://adsabs.harvard.edu/abs/1987A%26A...186L...1G} {186, L1}

\bibitem[\protect\citeauthoryear{{Gavazzi}, {Contursi}, {Carrasco}, {Boselli},
  {Kennicutt}, {Scodeggio}  \& {Jaffe}}{{Gavazzi} et~al.}{1995}]{gava95}
{Gavazzi} G.,  {Contursi} A.,  {Carrasco} L.,  {Boselli} A.,  {Kennicutt} R.,
  {Scodeggio} M.,   {Jaffe} W.,  1995, A\&A, \href
  {http://adsabs.harvard.edu/abs/1995A%26A...304..325G} {304, 325}

\bibitem[\protect\citeauthoryear{{Gavazzi}, {Marcelin}, {Boselli}, {Amram},
  {V{\'{\i}}lchez}, {Iglesias-Paramo}  \& {Tarenghi}}{{Gavazzi}
  et~al.}{2001a}]{gava01a}
{Gavazzi} G.,  {Marcelin} M.,  {Boselli} A.,  {Amram} P.,  {V{\'{\i}}lchez}
  J.~M.,  {Iglesias-Paramo} J.,   {Tarenghi} M.,  2001a, \mn@doi [A\&A]
  {10.1051/0004-6361:20010991}, \href
  {http://adsabs.harvard.edu/abs/2001A%26A...377..745G} {377, 745}

\bibitem[\protect\citeauthoryear{{Gavazzi}, {Boselli}, {Mayer},
  {Iglesias-Paramo}, {V{\'{\i}}lchez}  \& {Carrasco}}{{Gavazzi}
  et~al.}{2001b}]{gava01b}
{Gavazzi} G.,  {Boselli} A.,  {Mayer} L.,  {Iglesias-Paramo} J.,
  {V{\'{\i}}lchez} J.~M.,   {Carrasco} L.,  2001b, \mn@doi [ApJL]
  {10.1086/338389}, \href {http://adsabs.harvard.edu/abs/2001ApJ...563L..23G}
  {563, L23}

\bibitem[\protect\citeauthoryear{{Gavazzi}, {Boselli}, {Donati}, {Franzetti}
  \& {Scodeggio}}{{Gavazzi} et~al.}{2003}]{gava03b}
{Gavazzi} G.,  {Boselli} A.,  {Donati} A.,  {Franzetti} P.,   {Scodeggio} M.,
  2003, \mn@doi [A\&A] {10.1051/0004-6361:20030026}, \href
  {http://adsabs.harvard.edu/abs/2003A%26A...400..451G} {400, 451}

\bibitem[\protect\citeauthoryear{{Giovanelli} et~al.,}{{Giovanelli}
  et~al.}{2005}]{giovanelli05a}
{Giovanelli} R.,  et~al., 2005, \mn@doi [AJ] {10.1086/497431}, \href
  {http://esoads.eso.org/abs/2005AJ....130.2598G} {130, 2598}

\bibitem[\protect\citeauthoryear{{Giovanelli}, {Haynes}, {Kent}, {Saintonge},
  {Stierwalt}  \& {Altaf}}{{Giovanelli} et~al.}{2007}]{giovan2007}
{Giovanelli} R.,  {Haynes} M.~P.,  {Kent} B.~R.,  {Saintonge} A.,  {Stierwalt}
  S.,   {Altaf} A.,  2007, \mn@doi [AJ] {10.1086/516635}, \href
  {http://adsabs.harvard.edu/abs/2007AJ....133.2569G} {133, 2569}

\bibitem[\protect\citeauthoryear{{Goto}, {Yamauchi}, {Fujita}, {Okamura},
  {Sekiguchi}, {Smail}, {Bernardi}  \& {Gomez}}{{Goto} et~al.}{2003}]{goto03}
{Goto} T.,  {Yamauchi} C.,  {Fujita} Y.,  {Okamura} S.,  {Sekiguchi} M.,
  {Smail} I.,  {Bernardi} M.,   {Gomez} P.~L.,  2003, \mn@doi [MNRAS]
  {10.1046/j.1365-2966.2003.07114.x}, \href
  {http://adsabs.harvard.edu/abs/2003MNRAS.346..601G} {346, 601}

\bibitem[\protect\citeauthoryear{{Gunn} \& {Gott}}{{Gunn} \&
  {Gott}}{1972}]{gun72}
{Gunn} J.~E.,  {Gott} J.~R.~I.,  1972, ApJ, \href
  {http://adsabs.harvard.edu/abs/1972ApJ...176....1G} {176, 1}

\bibitem[\protect\citeauthoryear{{Haynes} \& {Giovanelli}}{{Haynes} \&
  {Giovanelli}}{1984}]{hayn84}
{Haynes} M.~P.,  {Giovanelli} R.,  1984, \mn@doi [AJ] {10.1086/113573}, \href
  {http://adsabs.harvard.edu/abs/1984AJ.....89..758H} {89, 758}

\bibitem[\protect\citeauthoryear{{Haynes}, {Giovanelli}  \& {Kent}}{{Haynes}
  et~al.}{2007}]{hayn07}
{Haynes} M.~P.,  {Giovanelli} R.,   {Kent} B.~R.,  2007, \mn@doi [ApJL]
  {10.1086/521188}, \href {http://esoads.eso.org/abs/2007ApJ...665L..19H} {665,
  L19}

\bibitem[\protect\citeauthoryear{{Haynes} et~al.,}{{Haynes}
  et~al.}{2011}]{hayn11}
{Haynes} M.~P.,  et~al., 2011, \mn@doi [AJ] {10.1088/0004-6256/142/5/170},
  \href {http://adsabs.harvard.edu/abs/2011AJ....142..170H} {142, 170}

\bibitem[\protect\citeauthoryear{{Hess} \& {Wilcots}}{{Hess} \&
  {Wilcots}}{2013}]{hess2013}
{Hess} K.~M.,  {Wilcots} E.~M.,  2013, \mn@doi [AJ]
  {10.1088/0004-6256/146/5/124}, \href
  {http://esoads.eso.org/abs/2013AJ....146..124H} {146, 124}

\bibitem[\protect\citeauthoryear{{Jaff{\'e}}, {Smith}, {Candlish}, {Poggianti},
  {Sheen}  \& {Verheijen}}{{Jaff{\'e}} et~al.}{2015}]{jaffe15}
{Jaff{\'e}} Y.~L.,  {Smith} R.,  {Candlish} G.~N.,  {Poggianti} B.~M.,  {Sheen}
  Y.-K.,   {Verheijen} M.~A.~W.,  2015, \mn@doi [MNRAS] {10.1093/mnras/stv100},
  \href {http://esoads.eso.org/abs/2015MNRAS.448.1715J} {448, 1715}

\bibitem[\protect\citeauthoryear{{Kenney}, {van Gorkom}  \& {Vollmer}}{{Kenney}
  et~al.}{2004}]{ken04}
{Kenney} J.~D.~P.,  {van Gorkom} J.~H.,   {Vollmer} B.,  2004, \mn@doi [AJ]
  {10.1086/420805}, \href {http://adsabs.harvard.edu/abs/2004AJ....127.3361K}
  {127, 3361}

\bibitem[\protect\citeauthoryear{{Kent} et~al.,}{{Kent} et~al.}{2008}]{kent08}
{Kent} B.~R.,  et~al., 2008, \mn@doi [AJ] {10.1088/0004-6256/136/2/713}, \href
  {http://adsabs.harvard.edu/abs/2008AJ....136..713K} {136, 713}

\bibitem[\protect\citeauthoryear{{Koopmann} \& {Kenney}}{{Koopmann} \&
  {Kenney}}{2004}]{koop04}
{Koopmann} R.~A.,  {Kenney} J.~D.~P.,  2004, \mn@doi [ApJ] {10.1086/423190},
  \href {http://adsabs.harvard.edu/abs/2004ApJ...613..851K} {613, 851}

\bibitem[\protect\citeauthoryear{{Makarov}, {Prugniel}, {Terekhova}, {Courtois}
   \& {Vauglin}}{{Makarov} et~al.}{2014}]{makarov14}
{Makarov} D.,  {Prugniel} P.,  {Terekhova} N.,  {Courtois} H.,   {Vauglin} I.,
  2014, \mn@doi [A\&A] {10.1051/0004-6361/201423496}, \href
  {http://adsabs.harvard.edu/abs/2014A%26A...570A..13M} {570, A13}

\bibitem[\protect\citeauthoryear{{Montes} \& {Trujillo}}{{Montes} \&
  {Trujillo}}{2014}]{montes2014}
{Montes} M.,  {Trujillo} I.,  2014, \mn@doi [ApJ]
  {10.1088/0004-637X/794/2/137}, \href
  {http://adsabs.harvard.edu/abs/2014ApJ...794..137M} {794, 137}

\bibitem[\protect\citeauthoryear{{Moore}, {Katz}, {Lake}, {Dressler}  \&
  {Oemler}}{{Moore} et~al.}{1996}]{moore96}
{Moore} B.,  {Katz} N.,  {Lake} G.,  {Dressler} A.,   {Oemler} A.,  1996,
  \mn@doi [Nature] {10.1038/379613a0}, \href
  {http://adsabs.harvard.edu/abs/1996Natur.379..613M} {379, 613}

\bibitem[\protect\citeauthoryear{{Moss}}{{Moss}}{2006}]{moss06}
{Moss} C.,  2006, \mn@doi [MNRAS] {10.1111/j.1365-2966.2006.11000.x}, \href
  {http://adsabs.harvard.edu/abs/2006MNRAS.373..167M} {373, 167}

\bibitem[\protect\citeauthoryear{{Plionis}, {Tovmassian}  \&
  {Andernach}}{{Plionis} et~al.}{2009}]{plionis09}
{Plionis} M.,  {Tovmassian} H.~M.,   {Andernach} H.,  2009, \mn@doi [MNRAS]
  {10.1111/j.1365-2966.2009.14507.x}, \href
  {http://adsabs.harvard.edu/abs/2009MNRAS.395....2P} {395, 2}

\bibitem[\protect\citeauthoryear{{Poggianti}, {Arag{\'o}n-Salamanca},
  {Zaritsky}, {De Lucia}, {Milvang-Jensen}, {Desai}  \& {Jablonka}}{{Poggianti}
  et~al.}{2009}]{poggianti09}
{Poggianti} B.~M.,  {Arag{\'o}n-Salamanca} A.,  {Zaritsky} D.,  {De Lucia} G.,
  {Milvang-Jensen} B.,  {Desai} V.,   {Jablonka} P.,  2009, \mn@doi [ApJ]
  {10.1088/0004-637X/693/1/112}, \href
  {http://esoads.eso.org/abs/2009ApJ...693..112P} {693, 112}

\bibitem[\protect\citeauthoryear{{Robitaille} \& {Bressert}}{{Robitaille} \&
  {Bressert}}{2012}]{Robitaille2012}
{Robitaille} T.,  {Bressert} E.,  2012, {APLpy: Astronomical Plotting Library
  in Python}, Astrophysics Source Code Library (\mn@eprint {ascl} {1208.017})

\bibitem[\protect\citeauthoryear{{Roediger} \& {Hensler}}{{Roediger} \&
  {Hensler}}{2005}]{roed05}
{Roediger} E.,  {Hensler} G.,  2005, \mn@doi [A\&A]
  {10.1051/0004-6361:20042131}, \href
  {http://adsabs.harvard.edu/abs/2005A%26A...433..875R} {433, 875}

\bibitem[\protect\citeauthoryear{{Sakai}, {Kennicutt}, {van der Hulst}  \&
  {Moss}}{{Sakai} et~al.}{2002}]{sakai02}
{Sakai} S.,  {Kennicutt} Jr. R.~C.,  {van der Hulst} J.~M.,   {Moss} C.,  2002,
  \mn@doi [ApJ] {10.1086/342622}, \href
  {http://adsabs.harvard.edu/abs/2002ApJ...578..842S} {578, 842}

\bibitem[\protect\citeauthoryear{{Scott} et~al.,}{{Scott}
  et~al.}{2010}]{scott10}
{Scott} T.~C.,  et~al., 2010, \mn@doi [MNRAS]
  {10.1111/j.1365-2966.2009.16204.x}, \href
  {http://adsabs.harvard.edu/abs/2010MNRAS.403.1175S} {403, 1175 (Paper I)}

\bibitem[\protect\citeauthoryear{{Scott}, {Cortese}, {Brinks}, {Bravo-Alfaro},
  {Auld}  \& {Minchin}}{{Scott} et~al.}{2012}]{scott12}
{Scott} T.~C.,  {Cortese} L.,  {Brinks} E.,  {Bravo-Alfaro} H.,  {Auld} R.,
  {Minchin} R.,  2012, \mn@doi [MNRAS] {10.1111/j.1745-3933.2011.01169.x},
  \href {http://adsabs.harvard.edu/abs/2012MNRAS.419L..19S} {419, L19 (Paper
  II)}

\bibitem[\protect\citeauthoryear{{Scott}, {Usero}, {Brinks}, {Boselli},
  {Cortese}  \& {Bravo-Alfaro}}{{Scott} et~al.}{2013}]{scott13}
{Scott} T.~C.,  {Usero} A.,  {Brinks} E.,  {Boselli} A.,  {Cortese} L.,
  {Bravo-Alfaro} H.,  2013, \mn@doi [MNRAS] {10.1093/mnras/sts328}, \href
  {http://adsabs.harvard.edu/abs/2013MNRAS.429..221S} {429, 221 (Paper III)}

\bibitem[\protect\citeauthoryear{{Scott}, {Usero}, {Brinks}, {Bravo-Alfaro},
  {Cortese}, {Boselli}  \& {Argudo-Fern{\'a}ndez}}{{Scott}
  et~al.}{2015}]{scott15}
{Scott} T.~C.,  {Usero} A.,  {Brinks} E.,  {Bravo-Alfaro} H.,  {Cortese} L.,
  {Boselli} A.,   {Argudo-Fern{\'a}ndez} M.,  2015, \mn@doi [MNRAS]
  {10.1093/mnras/stv1592}, \href
  {http://adsabs.harvard.edu/abs/2015MNRAS.453..328S} {453, 328 (Paper IV)}

\bibitem[\protect\citeauthoryear{{Sengupta}, {Dwarakanath}, {Saikia}  \&
  {Scott}}{{Sengupta} et~al.}{2013}]{sengupta2013}
{Sengupta} C.,  {Dwarakanath} K.~S.,  {Saikia} D.~J.,   {Scott} T.~C.,  2013,
  \mn@doi [MNRAS] {10.1093/mnrasl/sls039}, \href
  {http://esoads.eso.org/abs/2013MNRAS.431L...1S} {431, L1}

\bibitem[\protect\citeauthoryear{{Sengupta}, {Scott}, {Paudel}, {Saikia},
  {Dwarakanath}  \& {Sohn}}{{Sengupta} et~al.}{2015}]{sengupta2015}
{Sengupta} C.,  {Scott} T.~C.,  {Paudel} S.,  {Saikia} D.~J.,  {Dwarakanath}
  K.~S.,   {Sohn} B.~W.,  2015, \mn@doi [A\&A] {10.1051/0004-6361/201425149},
  \href {http://esoads.eso.org/abs/2015A%26A...584A.114S} {584, A114}

\bibitem[\protect\citeauthoryear{{Solanes}, {Manrique},
  {Garc{\'{\i}}a-G{\'o}mez}, {Gonz{\'a}lez-Casado}, {Giovanelli}  \&
  {Haynes}}{{Solanes} et~al.}{2001}]{sola01}
{Solanes} J.~M.,  {Manrique} A.,  {Garc{\'{\i}}a-G{\'o}mez} C.,
  {Gonz{\'a}lez-Casado} G.,  {Giovanelli} R.,   {Haynes} M.~P.,  2001, \mn@doi
  [ApJ] {10.1086/318672}, \href
  {http://adsabs.harvard.edu/abs/2001ApJ...548...97S} {548, 97}

\bibitem[\protect\citeauthoryear{{Spergel} et~al.,}{{Spergel}
  et~al.}{2007}]{sperg07}
{Spergel} D.~N.,  et~al., 2007, \mn@doi [ApJS] {10.1086/513700}, \href
  {http://adsabs.harvard.edu/abs/2007ApJS..170..377S} {170, 377}

\bibitem[\protect\citeauthoryear{{Tonnesen}}{{Tonnesen}}{2007}]{tonn07}
{Tonnesen} S.,  2007, \mn@doi [New Astronomy Review]
  {10.1016/j.newar.2006.11.055}, \href
  {http://adsabs.harvard.edu/abs/2007NewAR..51...80T} {51, 80}

\bibitem[\protect\citeauthoryear{{Venkatapathy} et~al.,}{{Venkatapathy}
  et~al.}{2017}]{venkata2017}
{Venkatapathy} Y.,  et~al., 2017, \mn@doi [\aj] {10.3847/1538-3881/aa8df8},
  \href {http://cdsads.u-strasbg.fr/abs/2017AJ....154..227V} {154, 227}

\bibitem[\protect\citeauthoryear{{Vollmer}, {Braine}, {Pappalardo}  \&
  {Hily-Blant}}{{Vollmer} et~al.}{2008}]{voll08}
{Vollmer} B.,  {Braine} J.,  {Pappalardo} C.,   {Hily-Blant} P.,  2008, \mn@doi
  [A\&A] {10.1051/0004-6361:200810432}, \href
  {http://adsabs.harvard.edu/abs/2008A%26A...491..455V} {491, 455}

\bibitem[\protect\citeauthoryear{{Walter}, {Brinks}, {de Blok}, {Bigiel},
  {Kennicutt}, {Thornley}  \& {Leroy}}{{Walter} et~al.}{2008}]{walter08}
{Walter} F.,  {Brinks} E.,  {de Blok} W.~J.~G.,  {Bigiel} F.,  {Kennicutt} Jr.
  R.~C.,  {Thornley} M.~D.,   {Leroy} A.,  2008, \mn@doi [AJ]
  {10.1088/0004-6256/136/6/2563}, \href
  {http://adsabs.harvard.edu/abs/2008AJ....136.2563W} {136, 2563}

\bibitem[\protect\citeauthoryear{{Yagi}, {Yoshida}, {Gavazzi}, {Komiyama},
  {Kashikawa}  \& {Okamura}}{{Yagi} et~al.}{2017}]{yagi17}
{Yagi} M.,  {Yoshida} M.,  {Gavazzi} G.,  {Komiyama} Y.,  {Kashikawa} N.,
  {Okamura} S.,  2017, \mn@doi [ApJ] {10.3847/1538-4357/aa68e3}, \href
  {http://esoads.eso.org/abs/2017ApJ...839...65Y} {839, 65}

\bibitem[\protect\citeauthoryear{{Yoon}, {Chung}, {Smith}  \&
  {Jaff{\'e}}}{{Yoon} et~al.}{2017}]{yoon2017}
{Yoon} H.,  {Chung} A.,  {Smith} R.,   {Jaff{\'e}} Y.~L.,  2017, \mn@doi [ApJ]
  {10.3847/1538-4357/aa6579}, \href
  {http://esoads.eso.org/abs/2017ApJ...838...81Y} {838, 81}

\bibitem[\protect\citeauthoryear{{Zwicky}, {Herzog}  \& {Wild}}{{Zwicky}
  et~al.}{1968}]{zwicky68}
{Zwicky} F.,  {Herzog} E.,   {Wild} P.,  1968, {Catalogue of galaxies and of
  clusters of galaxies}

\bibitem[\protect\citeauthoryear{{van Gorkom}}{{van Gorkom}}{2004}]{vgork04}
{van Gorkom} J.~H.,  2004, in {Mulchaey} J.~S.,  {Dressler} A.,   {Oemler} A.,
  eds, Clusters of Galaxies: Probes of Cosmological Structure and Galaxy
  Evolution. p.~305

\makeatother
\end{thebibliography}
\newpage
\begin{appendices}
	\begin{center}
		\textbf{\appendixpage}
	\end{center}
\appendix
\section{\textbf{VLA C--ARRAY \hi\  DETECTIONS}}
\label{appenda}
This appendix presents \textcolor{black}{the velocity integrated \hi\ maps and \hi\ velocity fields for the galaxies and extra--galactic \hii\ region listed in Table \ref{table2}.} A brief  analysis is also presented for each galaxy. In each case we comment, from the available data, on the evidence for or against a perturbed old stellar disk as an indication of a recent tidal interaction. Limitations on the use of NIR images for assessing perturbations  of  old stellar populations are considered in \textcolor{black}{Paper IV}\nocite{scott15}. \textcolor{black}{The galaxies are presented in ascending RA order}.

\begin{figure*}
\begin{center}
\includegraphics[scale=0.62] {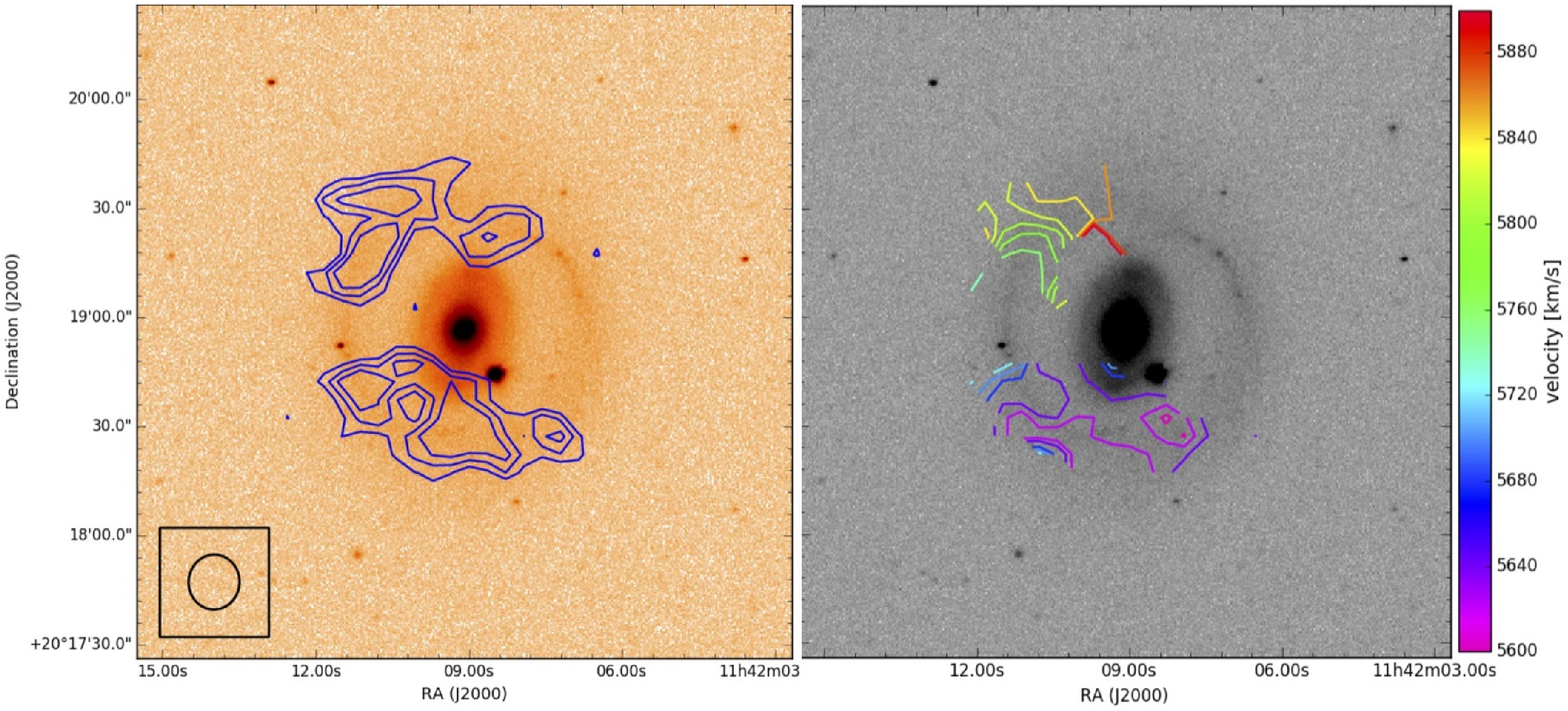}
\caption{CGCG 127--032  \textbf{Left:}  \hi\ integrated intensity contours from the PBCOR corrected map with contours at \textcolor{black}{1.6, 2.1 and 2.6 $\times$ 10$^{20}$ atoms cm$^{-2}$.}  \textcolor{black}{\textbf{Right:} \hi\ velocity field  with velocities of the contours corresponding to the colour scale at the right of the figure. The separation between contours, unless noted otherwise,  is 20 \km. The boxed ellipse  indicates} the size of the VLA synthesised C--array beam.  The background \textcolor{black}{for both maps is an  SDSS $g$ band} image.   }
\label{127032} 
\end{center}
\end{figure*}

\textbf{CGCG 127--032}  
Figure \ref{127032}  shows \hi\ detected in  two  elongated clumps S and NE of the optical centre.  This irregular  \hi\ morphology  contrasts strongly with the symmetric face--on optical spiral with two prominent spiral arms which encircle the galaxy in a rather symmetric way. CGCG\,127--032 has an AGES \hi\ offset of 29.5$\pm$14 arcsec which is consistent with the resolved VLA \hi\ morphology. The  maximum \hi\ column density in the map was \textcolor{black}{3.8} $\times$ 10$^{20}$ atoms cm$^{-2}$.    Unfortunately the best available NIR images for the galaxy are not deep enough to determine whether the old stellar disk edge is perturbed or not. 

\begin{figure*}
\begin{center}
\includegraphics[scale=0.62] {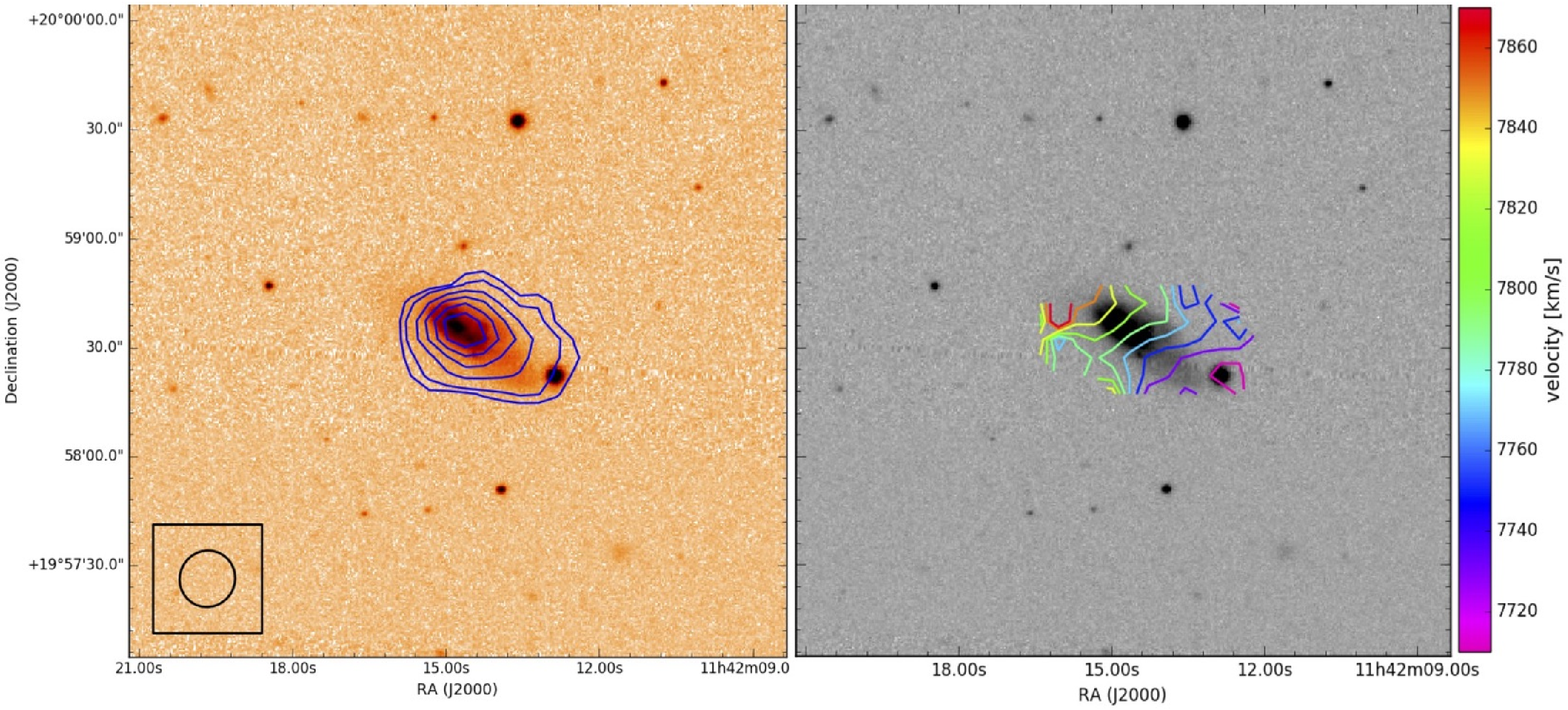}
\caption{\cg062  \textcolor{black}{\textbf{Left:}  \hi\ integrated intensity contours  at  \textcolor{black}{1.7, 2.9, 5.8, 8.7, 11.6 and 14.6 $\times$ 10$^{20}$ atoms cm$^{-2}$.} Other details  are per Figure \ref{127032}.} }
\label{97062} 
\end{center}
\end{figure*}

\textbf{\cg062}  This galaxy displays a cometary optical morphology with its tail oriented to the SE. As the  left panel of Figure  \ref{97062} shows, the  highest column density \hi\  (1.7 $\times$ 10$^{21}$ atoms cm$^{-2}$)  is projected at the optical centre  with a lower column  density \hi\ counterpart to the optical tail, extending to the star projected at the end of the optical tail. \textcolor{black}{See also section \ref{offvla_ages}.} The  \hi\ velocity field contours in the right panel of the figure show regular rotation across \textcolor{black}{the} brightest part of the optical disk, but signs of a  warp along the tail. \cg062 shows clear signs of perturbation of  the \hi\ in the outer disk most probably caused by the same mechanism that produced the optical tail. There are counterparts to the tail in the  $J, H$ and $K$ band  2MASS  images.  
\textcolor{black}{The \hi\ maps for this galaxy are based on  NRAO VLA archival data rather than our own observations (see Appendix B2). }

\begin{figure*}
\begin{center}
\includegraphics[scale=0.62] {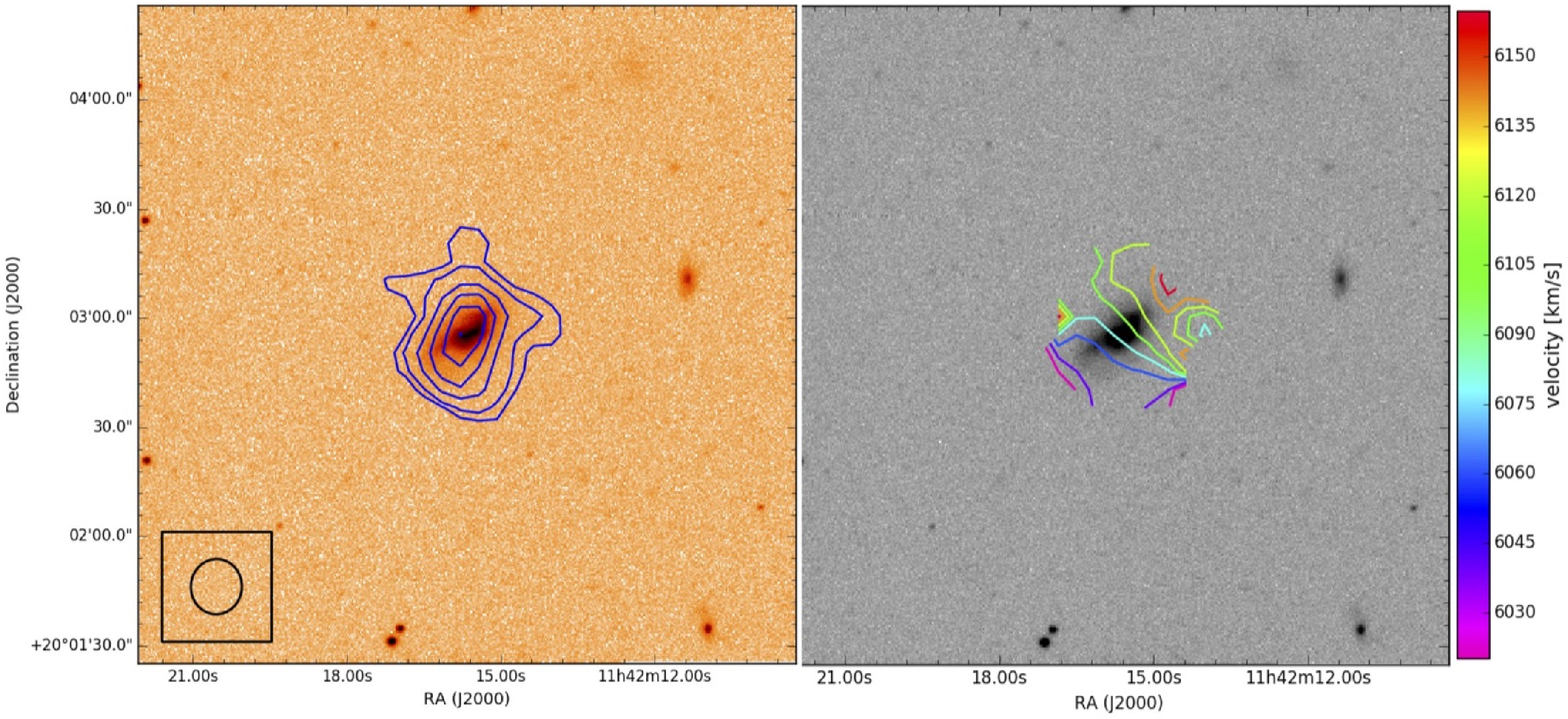}
\caption{\cg063  \textbf{Left:}  \hi\ integrated intensity contours  at \textcolor{black}{1.6, 2.6, 5.2, 7.8 and 10.4 $\times$ 10$^{20}$ atoms cm$^{-2}$.}  \textcolor{black}{ Other details  are per Figure \ref{127032}.}  }
\label{97063} 
\end{center}
\end{figure*}

\textbf{\cg063} Its column density maximum (\textcolor{black}{1.3 $\times$ 10$^{21}$} atoms cm$^{-2}$) is aligned, within the pointing uncertainties, with the optical centre 
(Figure \ref{97063}, left panel). The SE disk edge in a  smoothed SDSS \textcolor{black}{$z$} band image is  elongated compared to the NW and curves sharply \textcolor{black}{south}  toward the disk edge.  \textcolor{black}{The  velocity field  (Figure \ref{97063}, right panel)  isovelocity contours are perpendicular to the  optical disk major axis with a slight} hint of perturbation at the SE optical disk edge.  

\begin{figure*}
\begin{center}
\includegraphics[scale=0.62] {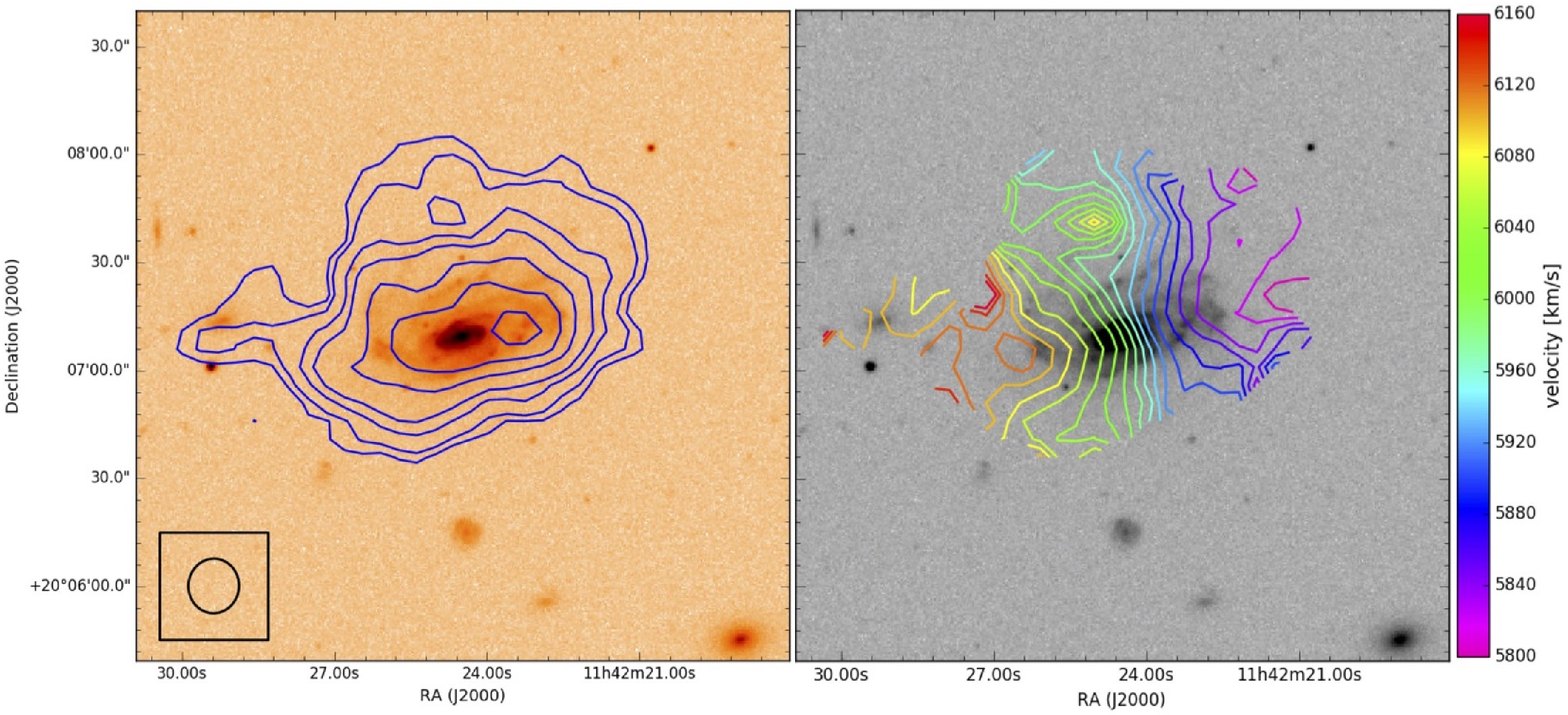}
\caption{\cg068 \textcolor{black}{\textbf{Left:}  \hi\ integrated intensity contours at  \textcolor{black}{1.6, 2.6, 5.2, 7.8, 12.9, 18.1 and 23.3 $\times$ 10$^{20}$ atoms cm$^{-2}$.}    The small galaxy within the eastern most \hi\ contour extension  is  J114229.18+200713.7.  Other details  are per Figure \ref{127032}.}   }
\label{97068} 
\end{center}
\end{figure*}

\textbf{\cg068}  This  gas rich   (\hi\ deficiency -0.3,  Table \ref{table2}) Sc galaxy shows several  \hi\ morphological and kinematic ongoing interaction signatures: i) in projection  the \hi\  3$\sigma$ contour in Figure  \ref{97068} (left panel)  extends  $\sim$ 58 arcsec (24 kpc)  \textcolor{black}{north}  of the optical center (perpendicular to the major axis), more than twice $\sim$ 26 arcsec (11 kpc)  it extends in the opposite direction. This confirms the earlier reports of \textcolor{black}{an} asymmetric \hi\ morphology in \cite{dick91} and  Paper I; ii)   \hi\ column densities rise much more rapidly   \textcolor{black}{south} of the optical centre than to the \textcolor{black}{north}  of it;   iii) An eastern \hi\ extension reaches the projected position of a smaller  Irr cluster galaxy, SDSS J114229.18+200713.7; iv) In the \hi\ velocity field (Figure  \ref{97068}, right panel) there is  a  systematic change in the \textcolor{black}{of the} angle \hi\ isovelocity lines, relative to the major optical axis  indicating stronger warping of the \hi\  disk at its eastern edge compared  to the western edge; v) FUV and NUV \textit{(GALEX)} images show \cg068 to have an extended  UV disk (XUV disk) with  a similar extent  to the \hi\ disk, except in the eastern \hi\ extension.   The column density maximum (\textcolor{black}{2.6} $\times$ 10$^{21}$ atoms cm$^{-2}$) is offset 18.6 $\pm$ 2.9 arcsec (7.7 kpc)  W of  the optical centre. Smoothed  \textit{Spitzer } 3.6 $\mu$m  and 4.5 $\mu$m images suggests an extension of diffuse NIR emission at the NE disk edge,  possibly due to an interaction with SDSS J114229.18+200713.7. This small companion only has  $\sim$ 1\%  \textcolor{black}{of}  \cg068's stellar mass  and  \textcolor{black}{an optical systemic velocity } $\sim$ 259 \km\ greater than \cg068.     An alternative interpretation is that the large scale \hi\ asymmetries in \cg068, including the warp, are caused by ram pressure as the galaxy's ISM interacts with the ICM on a south westerly  trajectory. In this scenario the eastern \hi\ extension would be explained as being similar to that seen in  the classic ram pressure stripping case NGC\,4522.   

\begin{figure*}
\begin{center}
\includegraphics[scale=0.62] {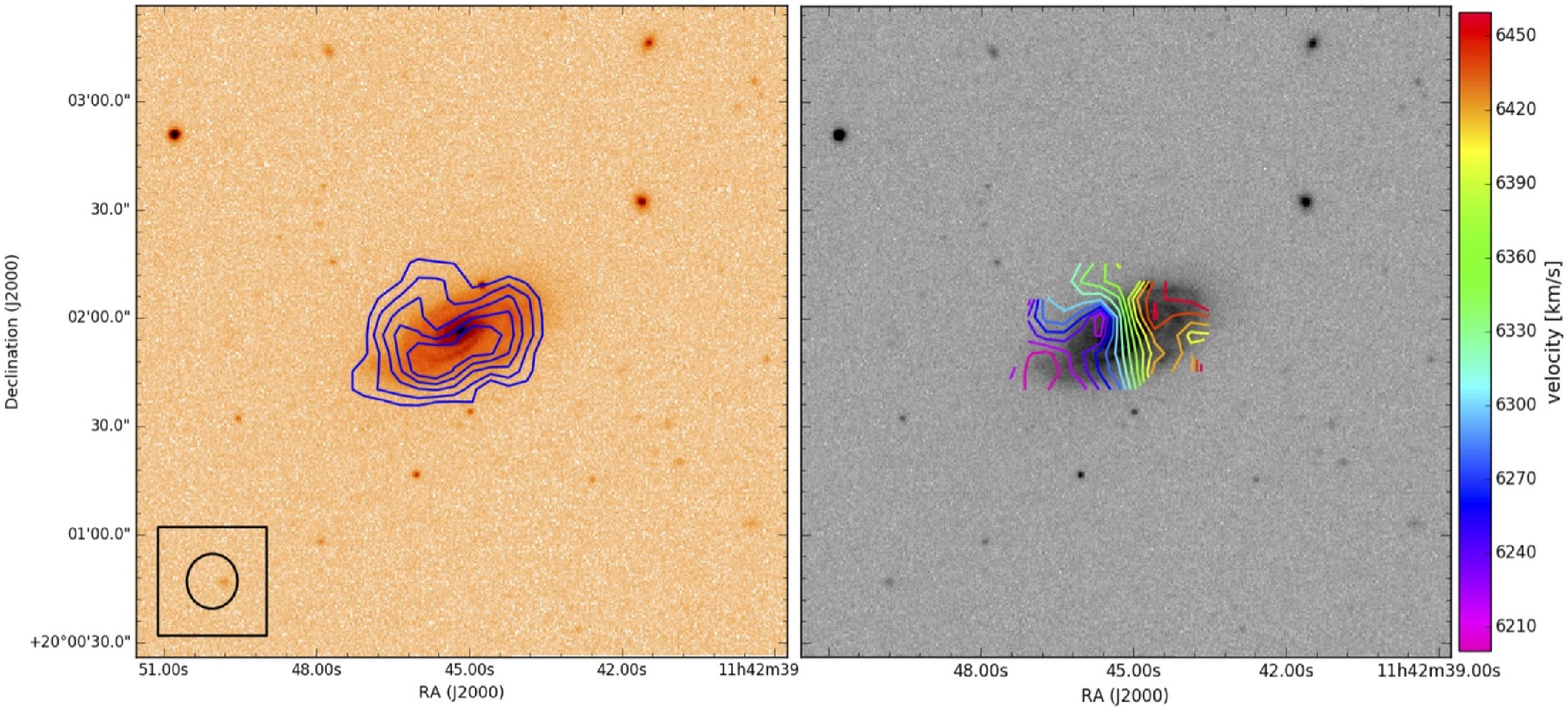}
\caption{\cg072 \textbf{Left:}  \hi\ integrated intensity contours  at  \textcolor{black}{1.6, 2.6, 3.9, 5.2 and 6.5 $\times$ 10$^{20}$ atoms cm$^{-2}$.}  \textcolor{black}{Other details  are per Figure \ref{127032}.}  }
\label{97072} 
\end{center}
\end{figure*}

\textbf{\cg072} The integrated \hi\ map for \cg072 (Figure \ref{97072}) shows the  \hi\ disk is truncated to approximately the radius of the optical disk    consistent with  its \hi\   deficiency of  0.5 (Table \ref{table2}), although  it has an  \hdos\ excess  of -0.20 (Paper III). Its \hi\ column density   maximum in the VLA C--array map (\textcolor{black}{8.2} $\times$ 10$^{20}$ atoms cm$^{-2}$) is offset 8.6$\pm$2.9 arcsec (3.6 kpc) SE of the optical centre.  An overall rotation pattern is seen in the velocity field in the right hand panel of the figure, but there is also evidence of  perturbed \hi\ kinematics and morphology \textcolor{black}{north}  of the \hi\ column density maximum.   Smoothed  \textit{Spitzer} 3.6 $\mu$m and 4.5 $\mu$m  band  images show  perturbation signatures   near the western edge of the disk,  which are possibly the result of   a tidal perturbation of  the outer disk within the relaxation time scale, which is assumed to be   approximately the time for a single disk rotation\footnote{ T$_{rot}$ [Gyr]= 6.1478 r/ \textcolor{black}{$V_{\mathrm {rot}}$}, where r = the optical radius [kpc] and \textcolor{black}{$V_{\mathrm {rot}}$}= 0.5 $\Delta$V  [\km]/ sin(i).}, i.e., $\sim$ 0.4 Gyr, based on  W$_{20}$   from Table \ref{table2}. 
 
\begin{figure*}
\begin{center}
\includegraphics[scale=0.62] {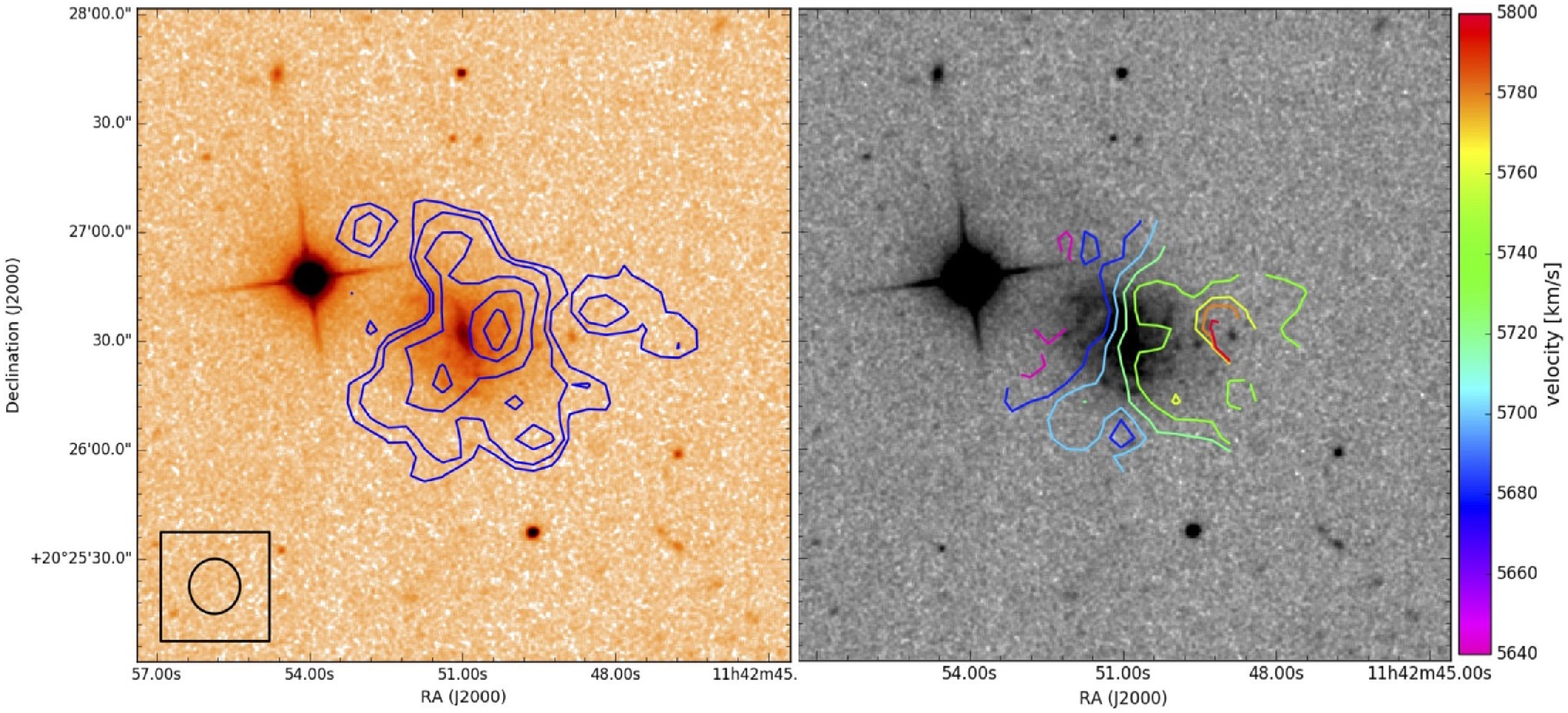}
\caption{SDSS J114250.97+202631.8 \textbf{Left:} 
 \hi\ integrated intensity contours  at \textcolor{black}{1.6, 2.6, 5.2, 7.8 and 10.4 $\times$ 10$^{20}$ atoms cm$^{-2}$.}   \textcolor{black}{Other details  are per Figure \ref{127032}}.}
\label{sdss1} 
\end{center}
\end{figure*}

\textbf{SDSS J114250.97+202631.8} 
Figure \ref{sdss1} shows  the integrated \hi\ map for this galaxy, which is projected $\sim$ 840 kpc from the NW subcluster core. SDSS J114250.97+202631.8 has its highest column density \hi\ (\textcolor{black}{1.2 $\times$ 10$^{21}$} atoms cm$^{-2}$) detected 10.6$\pm$2.9 arcsec (4 kpc) NW of the optical centre. As noted in section \ref{obs},  \hi\ in this galaxy was detected beyond the FWHP of the VLA primary beam and there is no AGES spectrum available for this galaxy so its \hi\ properties are more uncertain than for the other C--array LTGs.  
    
\begin{figure*}
\begin{center}
\includegraphics[scale=0.62] {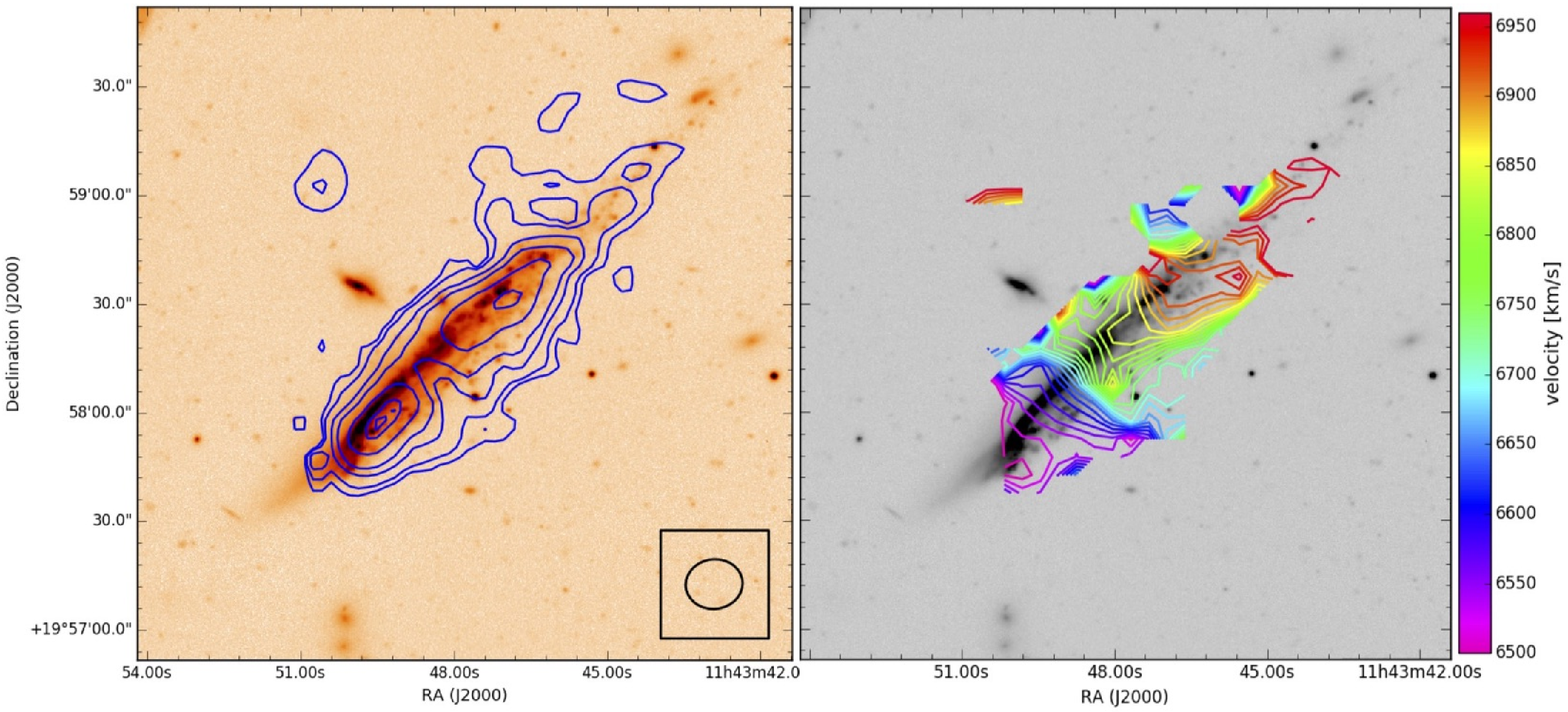}
\caption{CGCG 97-087  \textbf{Left:}  \hi\ integrated intensity contours  at \textcolor{black}{1.7, 2.9, 5.8, 8.7, 14.4, 20.2 and 21.7 $\times$ 10$^{20}$ atoms cm$^{-2}$.}  \textcolor{black}{ The background for both maps is an  Canada France Hawaii Telescope (CFHT) $u$ band  image. Other details  are per Figure \ref{127032}}. }
\label{97087} 
\end{center}
\end{figure*}

\begin{figure*}
\begin{center}
\includegraphics[scale=0.45] {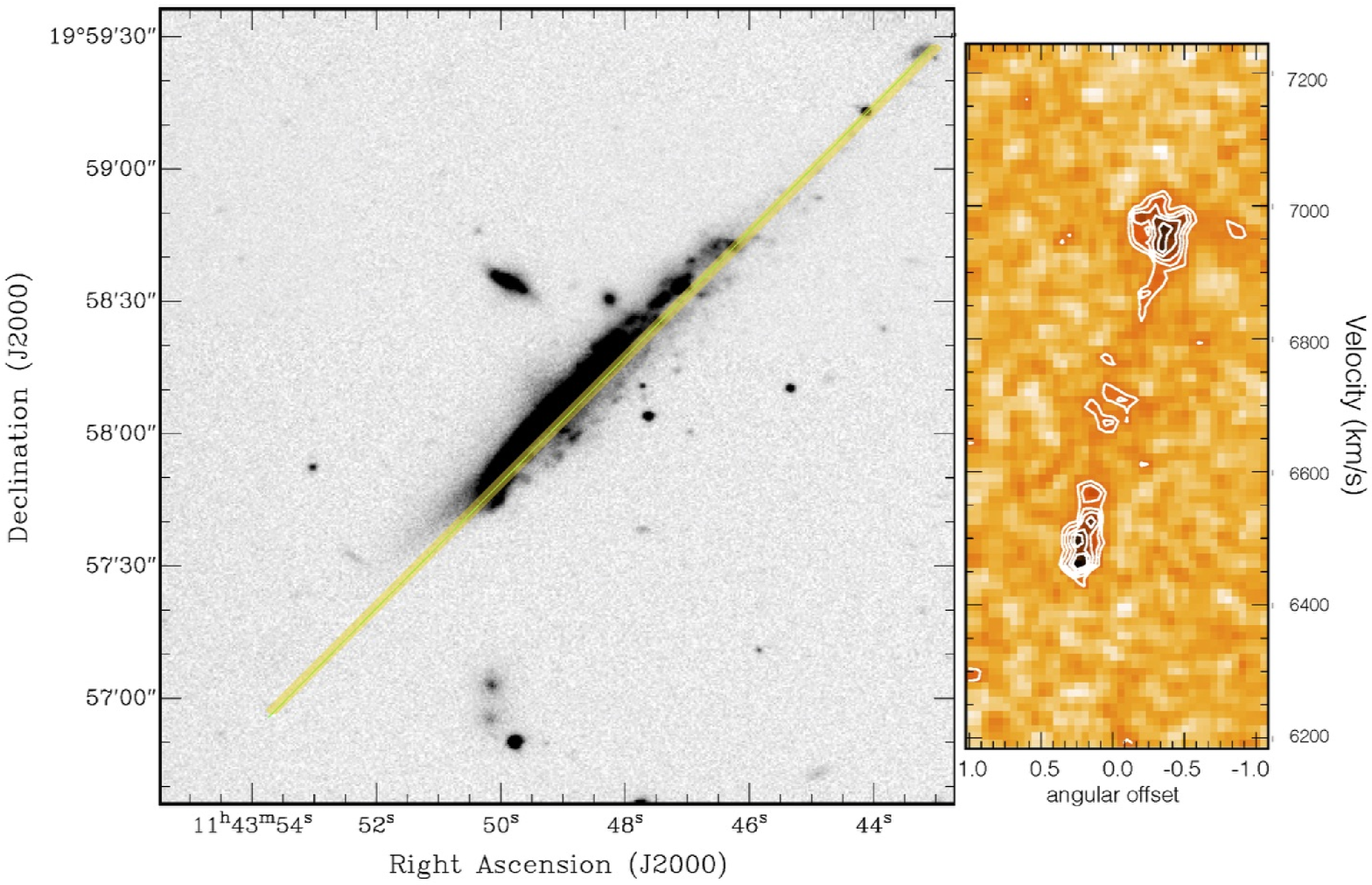}
\caption{CGCG 97-087  \textbf{Left:} \textcolor{black}{SDSS $g$ band image illustrating the direction along which the position--velocity (PV) diagram was taken (PA = 135 \degree).}  \textbf{Right:} \hi\ PV diagram PA = 135 \degree. Negative angular offsets [arcmin] in the PV diagram are north\textcolor{black}{west} of the optical centre and positive are to the south\textcolor{black}{east}.  }
\label{97087pv} 
\end{center}
\end{figure*}

\textbf{\cg087}
This is one of the best studied galaxies in the cluster, with our previous VLA D--array observations (Paper I) revealing a long diffuse $\sim$ 70 kpc \hi\ tail which has \halpha\ and radio continuum counterparts \textcolor{black}{\citep{bosel94,gava95,gava01a,yagi17}}.  As noted in Section \ref{results}  the VLA C--array  \textcolor{black}{(Figure  \ref{97087})} only maps the high  column density regions in the \textcolor{black}{VLA D--array map, but  resolves its \hi\ maxium}  into two distinct maxima SE and NW of the optical centre. The principal  \hi\ maximum ($\sim$ 11:43:49.6 + 19:57:58) is projected $\sim$10.9 arcsec (4.5 kpc)  SE of the optical centre and \textit{Spitzer} 4.5 $\mu$m maxima.   The  secondary \hi\ maximum ($\sim$ 11:43:47.4 +19:58:28) lies $\sim$ 71 arcsec  (29 kpc) NW of the optical  centre and  beyond the optical disk. There are two bright star forming clumps at  this  position, with  4.5 $\mu$m  and \halpha\ counterparts, possibly tracing in--situ star formation within the stripped gas.  \textcolor{black}{Both   C--array \hi\ maxima have counterparts  in  a VLT--MUSE \halpha\ image  \citep{cosolandi17}.  Kinematically the two maxima are quite  distinct. The \hi\ surronding the SE maximum ($V_{\mathrm {HI}}$ $\sim$ 6520 \km) shows a clear rotation signature across the optical disk with the changing angle of the isovelocity contours  indicating  the \hi\ disk is  warped. This contrasts with the \hi\ around the  NW \hi\ maximum, $V_{\mathrm {HI}}$ $\sim$ 6880 \km, which has a much shallower velocity gradient and less regular kinematics,  probably reflecting turbulence in the \hi\ displaced from the \hi\ disk.   Figure \ref{97087pv} shows the \hi\ position--velocity (PV) cut (PA = 135\degree) and the \hi\ PV diagram centred on the galaxy's optical centre. Immediately north of the optical centre \textcolor{black}{and between \hi\ maxima} the isovelocity contours are compressed  with the velocity rising $\sim$ 200 \km\  over a  projected distance of $\sim$ 10 arcsec (4 kpc). This is the position  where   \citep{gava01a} reported a $\sim$ 200 \km\  jump in \halpha\ velocities. The MUSE \halpha\ and [NII] emission maps show twin tails orginating from a  companion galaxy,  \cg087N (projected 30 arcsec to the NE), and extending into the \cg087 \hi\ velocity jump region. This and an asymmetrically  low \hi\ column density feature extending to the optical centre from the SW supports a senario in which \cg087N has made  a passage from the SW through \cg087s \hi\ disk as proposed in \citet{cosolandi17}. Unfortnately our \hi\ velocity coverage does not extend  to the velocity of \cg087N.  However, \cg087 is also underging strong ram pressure stripping so it  remains unclear to what extent the tidal interactions proposed in \cite{cosolandi17} or in \cite{gava01a} and  \cite{amram02}, account for the displacement of the  large mass of \hi\ now surrounding the NW \hi\ maximum and the \hi\ tail. Low velocity ($\lesssim300$ \km) tidal interactions are capable of displacing large fractions of an interacting pair's \hi\ beyond their optical disks \citep{sengupta2013,sengupta2015}, but in this case the companion is much less massive and the $\sim$800 \km\ velocity separation \citep{cosolandi17} suggests a high velocity low impact encounter.}

\begin{figure*}
\begin{center}
\includegraphics[scale=0.62] {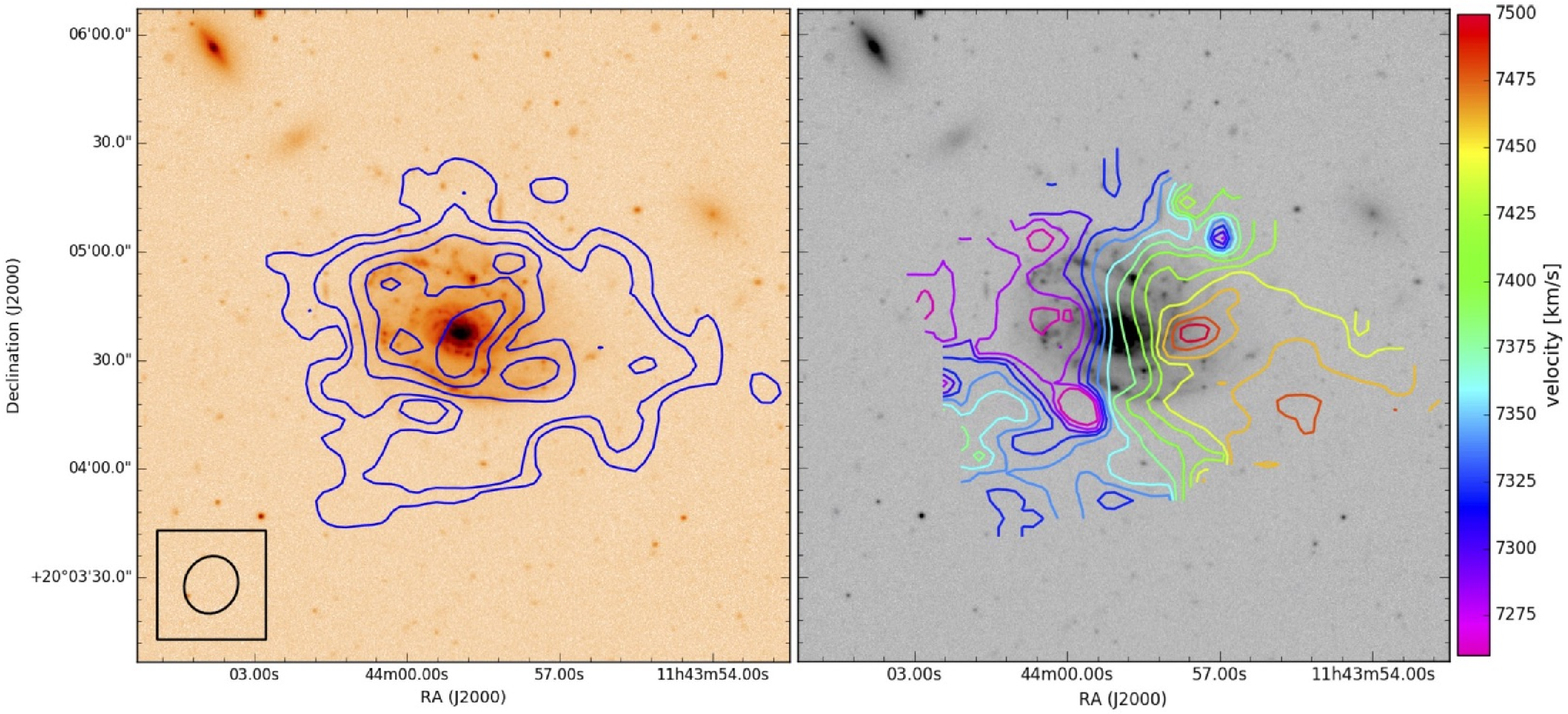}
\caption{CGCG 97-091  \textbf{Left:}  \hi\ integrated intensity contours  at   \textcolor{black}{1.6, 2.7, 5.4, 8.1 and 10.8 $\times$ 10$^{20}$ atoms cm$^{-2}$.} \textcolor{black}{Other details  are per Figure \ref{97087}}.  }
\label{97091} 
\end{center}
\end{figure*}

\textbf{\cg091}
To a first order \cg091 presents a symmetric \hi\ disk. The highest column density \hi\  (\textcolor{black}{13 $\times$ 10$^{20}$} atoms cm$^{-2}$)  is well aligned in projection with the optical centre,  although the  low \hi\ column density disk edge extends further  to the south and the east than north and west.   The closed isovelocity contours in \textcolor{black}{the} velocity field (Figure \ref{97091}-- right panel)  indicates the \hi\ disk is asymmetrically warped \textcolor{black}{E and W} of the optical centre.  Together the \hi\ morphology and kinematics suggest the \hi\ disk has suffered a recent weak perturbation. The  WISE 3.4 $\mu$m image is consistent with a symmetric unperturbed old stellar disk.

\begin{figure*}
\begin{center}
\includegraphics[scale=0.62] {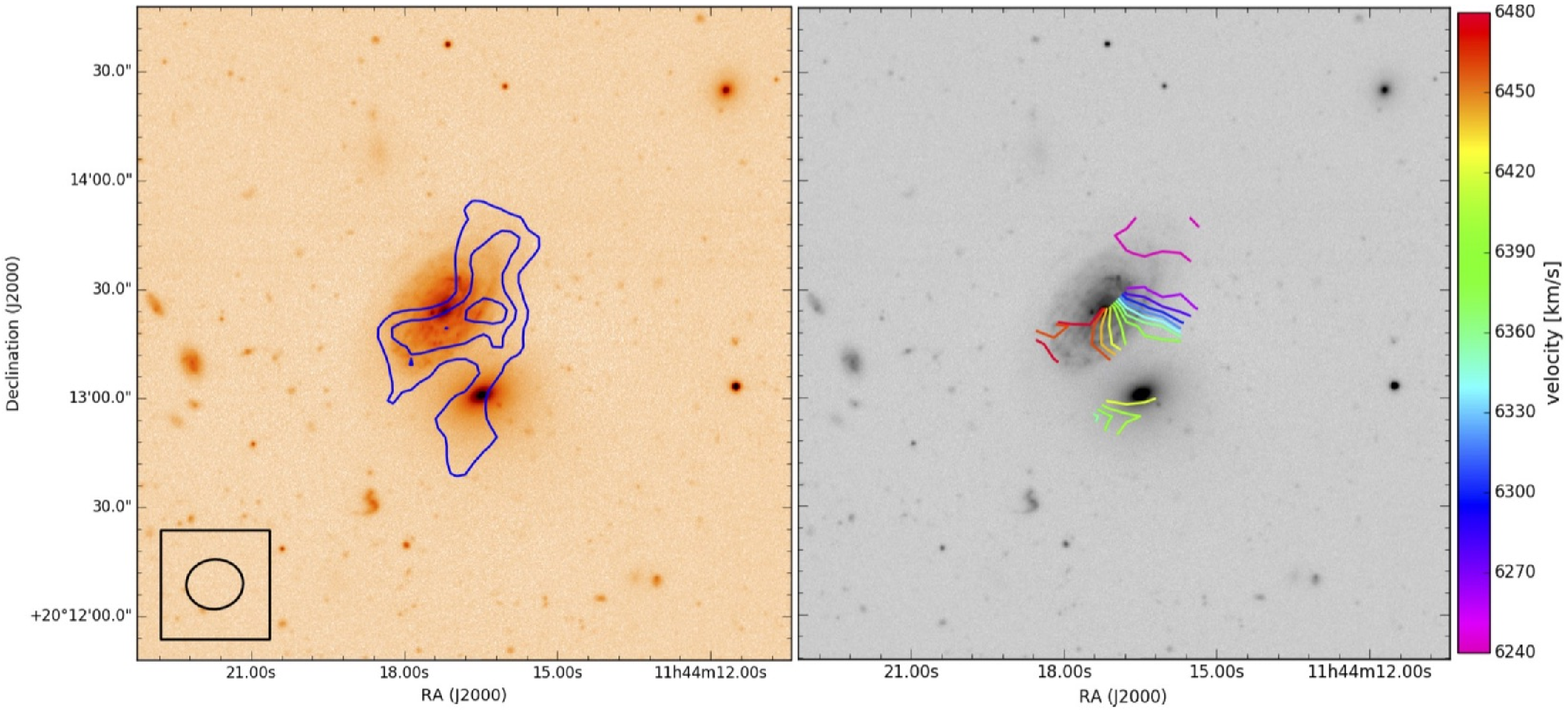}
\caption{CGCG 97-102N  \textbf{Left:}  \hi\ integrated intensity contours  at  \textcolor{black}{1.7, 2.9 and 4.0 $\times$ 10$^{20}$ atoms cm$^{-2}$.} \textcolor{black}{Other details  are per Figure \ref{97087}}. }
\label{97102} 
\end{center}
\end{figure*}

\textbf{\cg102}   \hi\  is projected against the optical disks of both members of this pair, \cg102N and \cg102S (Figure  \ref{97102}).  There is an  apparent \hi\ bridge/tail  joining the pair, although the \hi\ detection  in this feature where it is projected against \cg102S, is only at the  3 $\sigma$ level.  \cg102N displays a quite irregular  \hi\ morphology with the  column density maximum  (5.1 $\times$ 10$^{20}$ atoms cm$^{-2}$) offset 11.1$\pm$ 2.9 arcsec (4.6 kpc) W of the  optical centre and projected near the optical disk edge. 
\textcolor{black}{The} \hi\ morphology of \cg102N is  severely  skewed  to the S and W of the optical centre. The velocity field for \cg102N, right panel of the figure, shows an overall  rotating disk pattern.  But the changing angle of the isovelocity contours indicates \textcolor{black}{a strongly warped disk}, with the change in angle of the contours being greatest in the \hi\ velocity range of the \hi\ bridge/tail  projected against \cg102S, i.e., V$\sim$ \textcolor{black}{6380} \km\ to \textcolor{black}{6460} \km. \textcolor{black}{The \hi\ velocities of the} \hi\  projected against the \cg102S \textcolor{black}{optical} disk do not show a rotation pattern. This, the \af\ \textcolor{black}{for \cg102N}  and the classification of \cg102S as an elliptical (NED)  suggests the  \hi\ in the bridge projected against \cg102S  has been   tidally stripped from   \cg102N by an interaction between the pair.  A  tidal interaction scenario is supported by  evidence that the  \cg102N  molecular disk is \textcolor{black}{also} perturbed \nocite{scott13} \textcolor{black}{(Paper III)}.  The edges of the optical disks are projected against each other and the optical velocities of the pair from NED are \cg102N (6368$\pm$14 \km) and \cg102S (6364$\pm$9 \km) \textcolor{black}{give}  a difference of only 4 \km.  The CAS \citep{consel03} asymmetry parameter for \cg102N derived from an  SPM\footnote{Observatorio Astron\'omico Nacional, San Pedro M\'artir, Mexico. }  $J$ band image was 1.22 is  in the domain of  disturbed objects and provides evidence that the old stellar disk of \cg102N is perturbed in the SE, supporting  the  tidal interaction \textcolor{black}{scenario} \citep{venkata2017}.   

 
\begin{figure*}
\begin{center}
\includegraphics[scale=0.62] {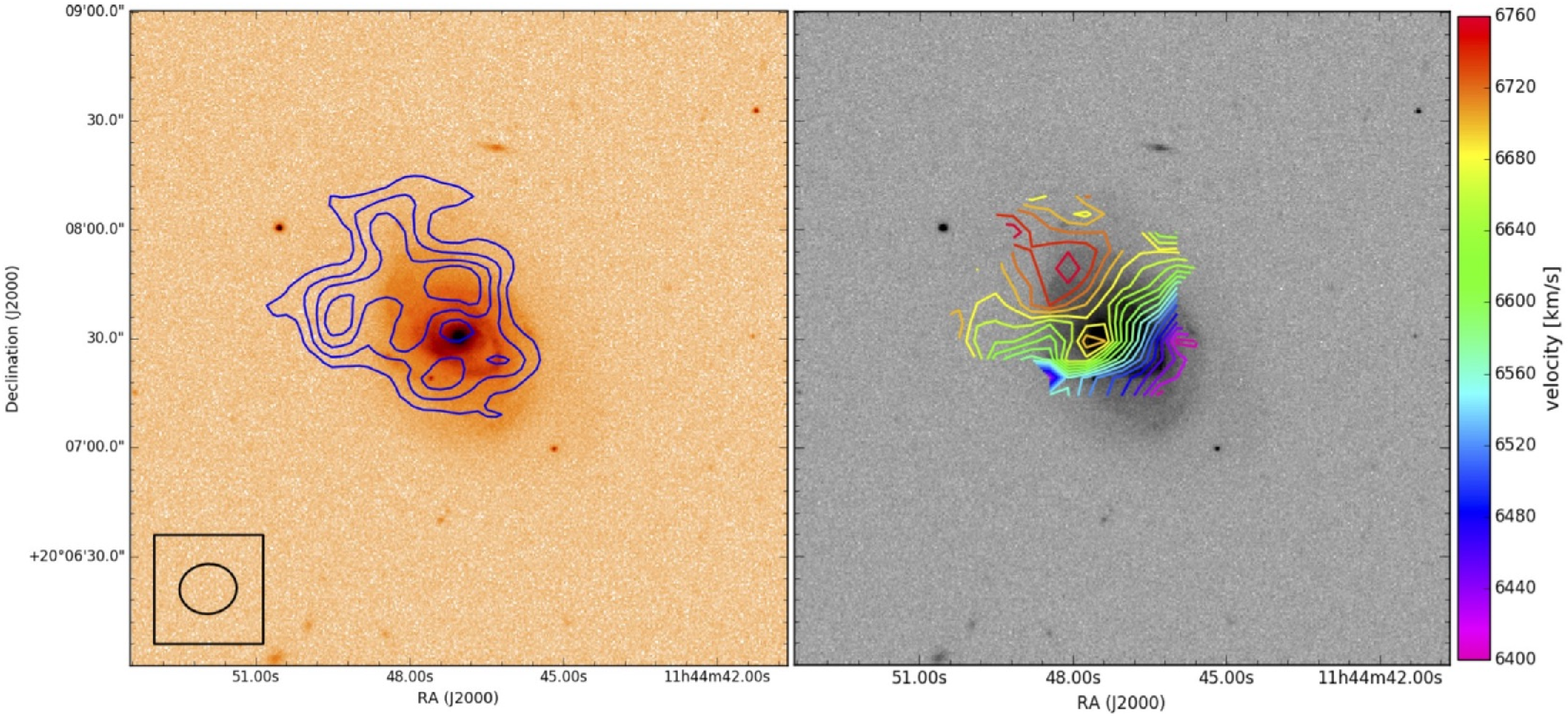}
\caption{CGCG 97-121 \textbf{Left:}  \hi\ integrated intensity contours  at   \textcolor{black}{1.7, 2.9, 4.0 and 5.2 $\times$ 10$^{20}$ atoms cm$^{-2}$.}    \textcolor{black}{Other details  are per Figure \ref{127032}}. }
\label{97121} 
\end{center}
\end{figure*}

\textbf{\cg121}
is an  Sab  galaxy  \textcolor{black}{projected} $\sim$  14.9 arcmin (370 kpc) from the NW subcluster core.  Although there is an unusual structure in the inner optical disk, smoothed  \textit{Spitzer } 3.6 $\mu$m and 4.5 $\mu$m  images \textcolor{black}{show} the morphology of the outer disk  to be rather symmetric, indicating the old stellar population there is unperturbed. This galaxy has \textcolor{black}{an} \hi\ deficiency of 0.4 (Table \ref{table2}) with the \hi\ in the SE truncated  $\sim$ 30 arcsec (12 kpc) inside the optical \textcolor{black}{disk. The} highest column density  \hi\ (\textcolor{black}{6.4} $\times$ 10$^{20}$ atoms cm$^{-2}$)  is projected $\sim$ 13.6  arcsec (6 kpc) \textcolor{black}{north} of the optical centre and near the northern edge of  the optical disk (Figure  \ref{97121}).  The \hi\ velocity field (Figure  \ref{97121}, right panel) and  \textcolor{black}{\af} = 1.92 $\pm$0.08 show the  \hi\ kinematics are consistent with an ongoing  interaction which is  significantly perturbing the \hi\ disk.  The galaxy is also deficient in molecular gas, \hdos\ deficiency $\sim$ 0.48 (Paper III).  The FUV and NUV (\textit{GALEX}) images show a broad low surface brightness linear feature extending out to the northern  edge of the optical disk and the 3 $\sigma$ \hi\ contour. The unperturbed old stellar disk, highly perturbed \hi\ and gas deficiencies   are all signatures expected from ram pressure stripping. But as discussed in \nocite{scott15} \textcolor{black}{(Paper IV)} such a large offset of highest column density \hi\ in a spiral with its M*  mass is not expected from simple ram pressure stripping models, which predict only moderate P$_{ram}$  $\sim$ 8 $\times$ 10 $^{12}$ dyne cm$^{-2}$ at the projected position of the galaxy for a \textcolor{black}{$V_{\mathrm {rel}}$} of 1000 \km. 

\begin{figure*}
\begin{center}
\includegraphics[scale=0.62] {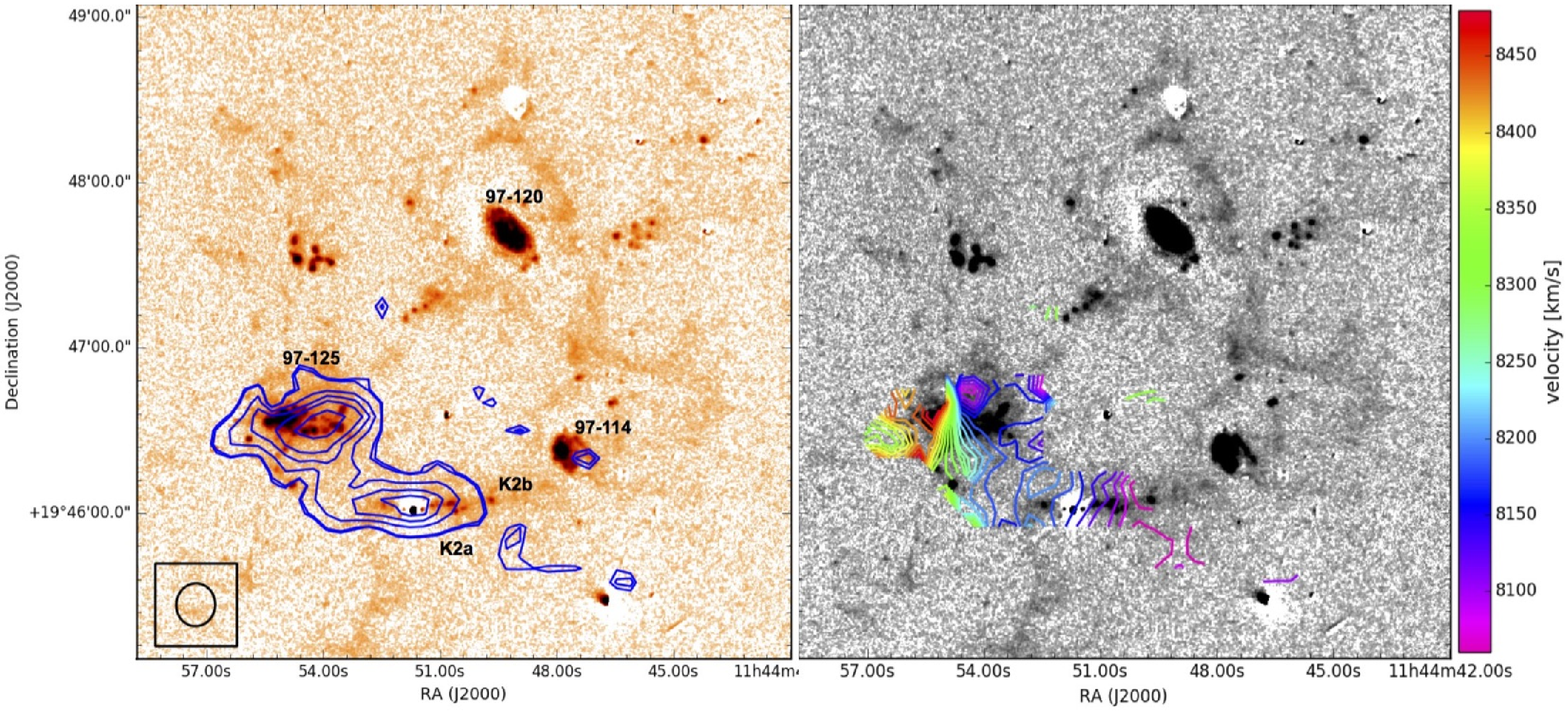}
\caption{Blue in--falling group (BIG). \hi\ integrated intensity contours on  deep \halpha\  image \citep{cort06}.   Contours are at   \textcolor{black}{1.8, 2.0, 4.0, 6.0, 8.0 and 12 $\times$ 10$^{20}$ atoms cm$^{-2}$.}  \textcolor{black}{The projected positions of CGCG galaxies and \hii\ regions (K2a and K2b) are also indicated on the integrated \hi\ map. Other details  are per Figure \ref{127032}}.  }
\label{big} 
\end{center}
\end{figure*}

\textbf{BIG compact Group}
BIG is projected near the centre of the SE subcluster but has \textcolor{black}{mean optical velocity, $<$V$_{opt}>$,} of  8230 \km, \textcolor{black}{which is} $\sim$ 1750 \km\ higher than the \textcolor{black}{$<$V$_{opt}>$} for A\,1367 \textcolor{black}{ \citep{sakai02,cort06}.} The \halpha\ image, Figure 2 in \cite{cort06}, shows spectacular \halpha\ emission  from tails tracing interaction debris within the group.   Figure \ref{big} shows the \hi\ in the easternmost  group member in the figure (\cg125) is highly perturbed with its \hi\ column density maximum offset 14.4 $\pm$2.9 arcsec (6 kpc)  E of the optical centre. Strikingly \cg125 has  a broad $\sim$ 80 arcsec (33 kpc) \textcolor{black}{long}  \hi\ tail curving \textcolor{black}{to the} SE. This tail contains a large  \hi\ mass (K2  in  the Cortese et al. 2006 figure), near the position of a bright foreground star. The Cortese  figure shows the eastern   \hi\ tail \textcolor{black}{\cg125} has an  \halpha\ counterpart which  extends beyond the \hi\ tail to the eastern side of the blue LTG, \cg114. The \hi\ velocity field shows the \hi\ velocity falling along  the \hi\ tail from 8277$\pm$18 \km\ at the \cg125 column density maximum to $\sim$\textcolor{black}{8100} \km\  at K2 (\textcolor{black}{$V_{\mathrm {opt}}$} $\sim$ 8075 \km\ at K2a). An  \hi\ clump is detected 10.4  arcsec (4 kpc) E of the \cg114 optical  centre with a \textcolor{black}{$V_{\mathrm {HI}}$}  = 8451$\pm$20 \km. \cg114 emits strongly in \halpha\ and has \textcolor{black}{$V_{\mathrm {opt}}$} = 8425 \km. Our \hi\ velocities for \cg114, \cg125 and K2 are in good agreement with those from the WSRT \citep{sakai02}.  This  and the   optical velocity of knot  K2b (8309 \km) suggests a tidal bridge  between \cg125 and \cg114 forms an arc in velocity space reaching its  minimum velocity at K2. BIG is projected $\sim$11 arcmin from the strong continuum source 3C264, making continuum subtraction in \textcolor{black}{this}  field particularly difficult. Incomplete continuum subtraction is the most likely reason for  the VLA C--array \hi\ flux for BIG being higher than from AGES, see Table \ref{table2}.   For this reason the first contour in the BIG \hi\ map  is drawn at 4.5 $\sigma$ rather than at 3  $\sigma$ as in the other \hi\ maps. \cite{yagi17} report an $\sim$ 300 kpc \halpha\ tail emanating from BIG, which suggests that gas debris from tidal preprocessing interactions \textcolor{black}{is} in the process of being removed from the compact group by \textcolor{black}{ram pressure stripping. }

\begin{figure*}
\begin{center}
\includegraphics[scale=0.62] {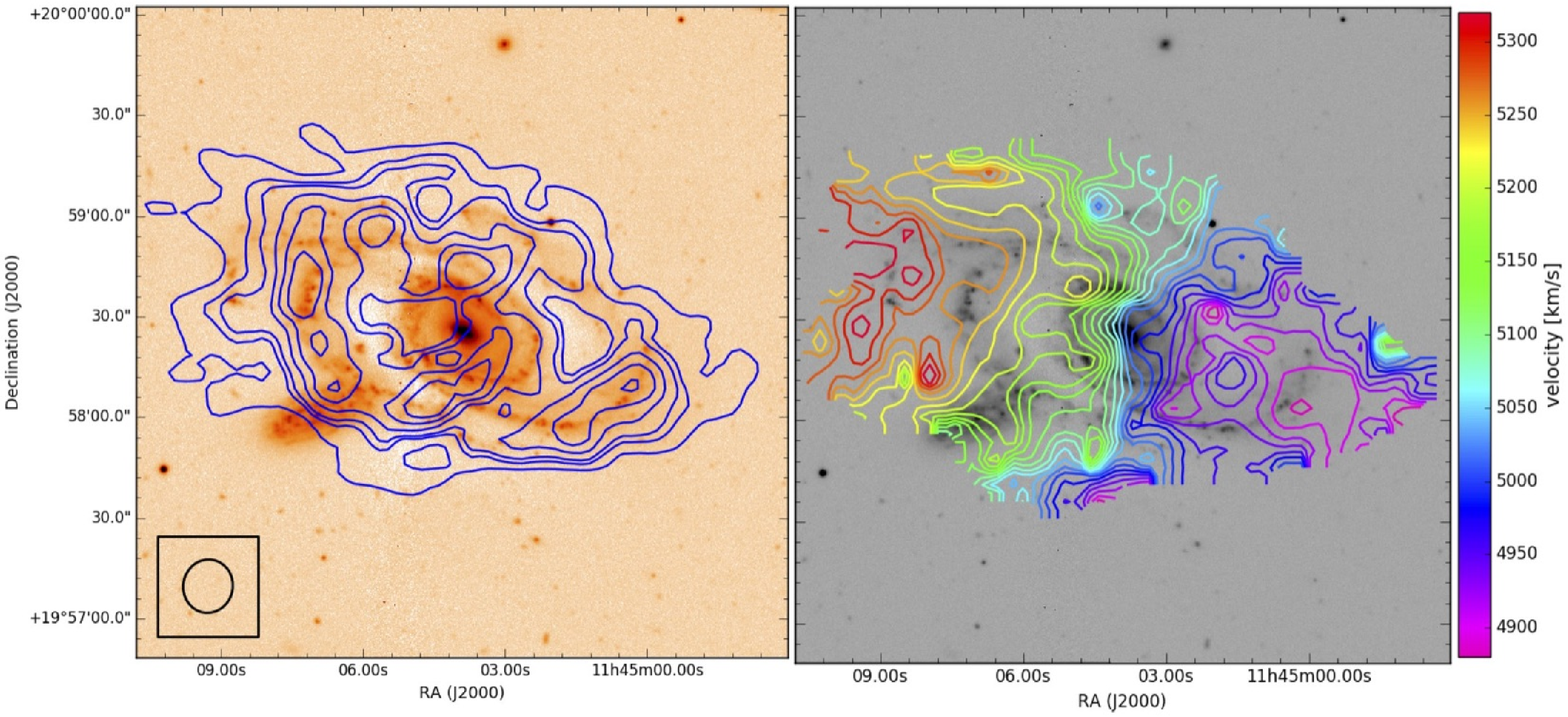}
\caption{CGCG 97-129  \textbf{Left:}  \hi\ integrated intensity contours  at \textcolor{black}{1.6, 4.1, 5.4, 8.2, 10.9 and 13.6 $\times$ 10$^{20}$ atoms cm$^{-2}$.} \textcolor{black}{Other details  are per Figure \ref{97087}}. }
\label{97129} 
\end{center}
\end{figure*}

\textbf{\cg0129W}
This large \textcolor{black}{barred} Sb spiral has a D$_{25}$ $\approx$ 2 arcmin (50 kpc). \textcolor{black}{ The  two impressive spiral arms   emanating from the ends of the bar encircle the galaxy} (Figure \ref{97129}). The \hi\  maximum (\textcolor{black}{1.7} $\times$ 10$^{21}$ atoms cm$^{-2}$)  is offset 32.4$\pm$2.9 arcsec (13 kpc) E of the optical centre. A smaller cluster member \cg129E is projected at the eastern edge of the optical disk but its  optical  velocity is  2455 \km\ higher than \cg129W. Smoothed \textit{Spitzer} 3.6$\mu$m and 4.5$\mu$m images suggest  the  old stellar disk of \cg129W, at  the radius of the encircling  optical arms, is to a first order symmetric. The  \hi\ disk  extends $\sim$ 30 arcsec (12 kpc) beyond the optical disk both  \textcolor{black}{to the} NE and SW. Overall the   \hi\ velocity field  (Figure \ref{97129}) reflects a rotating disk. However, the kinematics are quite perturbed  and the closed isovelocity contours \textcolor{black}{near the  disk  edges and changing isovelocity contour angles}  indicate an asymmetrically \textcolor{black}{warped \hi\  disk}.

\begin{figure*}
\begin{center}
\includegraphics[scale=0.62,angle=0] {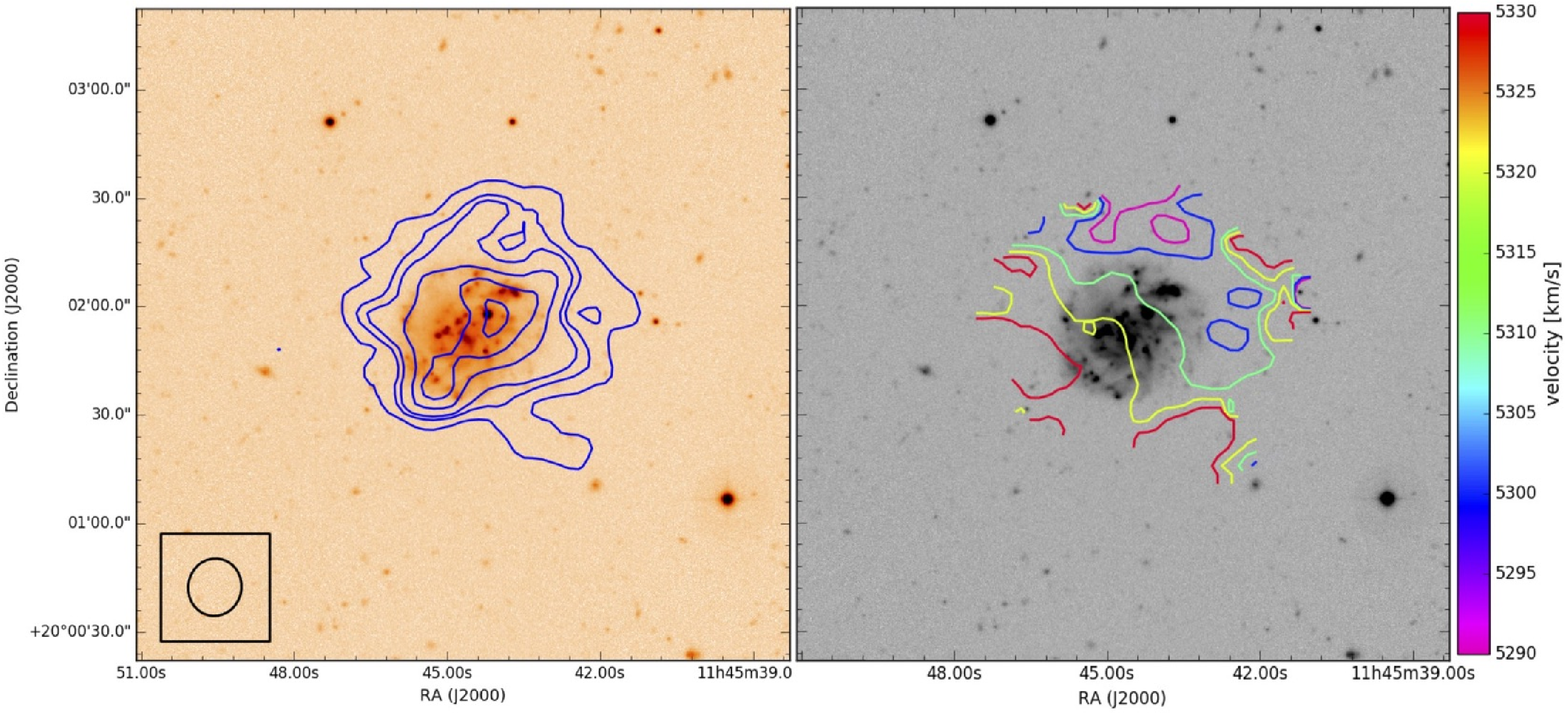}
\caption{CGCG 97-138   \textbf{Left:}  \hi\ integrated intensity contours  at   \textcolor{black}{1.6, 4.1, 5.4, 8.2, 10.9 and 13.6 $\times$ 10$^{20}$ atoms cm$^{-2}$.} \textcolor{black}{The separation between contours in the velocity field  is 20 \km. Other details  are per Figure \ref{97087}}. }
\label{97138} 
\end{center}
\end{figure*}

\textbf{\cg138}
This  blue (SDSS $g - i$  = 0.82)  galaxy  displays  a  rapid rise in \hi\ column density  in the SE where the edge of the \hi\ disk is truncated closer to the optical disk edge compared to  the \textcolor{black}{north} and \textcolor{black}{west}. The \hi\ intensity maximum (\textcolor{black}{1.4} $\times$ 10$^{21}$ atoms cm$^{-2}$) is located close to the NW optical disk edge.   The small \hi\ W$_{20}$ value of 64$\pm$3 \km\  and the velocity field isovelocity contours  (Figure \ref{97138}), which shows a systematic gradient  in the SE--NW direction, suggest we are viewing the \hi\ disk at low inclination angle (36.3\degree\ per Hyperleda for the optical disk). There are NUV \textit{(GALEX)} maxima at the optical centre and $\sim$ 10 arcsec (4 kpc) NW of  a bright m$_g$ = 16.4 star projected in front of the optical disk. Both NUV maxima have counterparts in the \halpha\ GOLDMine image. A smoothed H--band image (GoldMine) shows the disk edge to be quite irregular. It is not clear whether the $H$ band morphology  is the result of a recent interaction or the inherent morphology of the galaxy.   

 \section{\textbf{DATA REDUCTION: INDIVIDUAL FIELDS}}
 \label{vlafields}
\subsection{\bf Field B (day 5 observations)} 

This  field (Figure \ref{bigrosat}),  centred near the cluster centre,  has a significant spatial overlap with the SE of
Field D. Two discrete velocity \textcolor{black}{ranges} were observed using  IF1 and IF 2.  IF 1 covered a range of 7978 \,\km\ to 8511 \,\km\ with central velocity set to the  Blue Infalling Group's  \citep{cort06} mean velocity. The observations from this IF were combined in the \textit{uv} plane with C--array observations  from December 2002 at the same position and reported in Paper I. The combined data was used to produce an image cube with 48 channels covering the velocity range from 7978 \km\ to 8511 \km\ with a velocity width of 11 \km\
per channel.  \textcolor{black}{The} rms per channel of the combined cube was $\sim$ 0.26 mJy.   For the combined cube, the equivalent
\hi~mass detection threshold is \aprox\ \textcolor{black}{5.2} $\times$ 10$^{7}$ M$_{\odot}$ (corresponding to 3$\sigma$ in 2 consecutive 11 \km\  channels).  This is equivalent to \textcolor{black}{a} column density sensitivity for emission filling the beam of \textcolor{black}{1.7 $\times$ 10$^{19}$ cm$^{-2}$.}

IF 2 produced a cube with  a velocity range of 4836 \,\km\ to 5358 \,\km. For this cube, the equivalent \hi~mass detection threshold is \aprox\ \textcolor{black}{5.7} $\times$ 10$^{7}$ M$_{\odot}$ (corresponding to 3$\sigma$ in 2 consecutive 11 \km\  channels).  This is equivalent to a column density sensitivity for emission filling the beam of \textcolor{black}{1.9} $\times$ 10$^{19}$ cm$^{-2}$.

\subsection{\bf Field C (day 1 and 2 observations)}
\label{vlafieldc}
Field C (Figure \ref{bigrosat}) is centred   \aprox\ 25 arcmin NW of the NW cluster centre \citep{cort04}. It was planned that on day 1  two IFs  would cover two adjacent 500 \km\ velocity ranges with the similar set up on day 2 to extend the velocity range  by \textcolor{black}{a} further 1000 \km. The intention was to combine the cubes to produce a consolidated cube with a \textcolor{black}{contiguous} 2000 \km\  range. Unfortunately because of a telescope scheduling error, 7 hours of the 9 hours of planned day 2 observation time was applied to re--observing the day 1 velocity range. As a result  there were no \hi\ detections in the day 2 cube. However, the day 1 (lower velocity range) observations were correspondingly deeper than planned and combining IF1 and IF2  produced  an image cube with 100 channels covering the velocity range from 5504 \km\ to 6565 \km\ with a velocity width of 11 \km\ per channel.  The rms per channel of the combined cube was $\sim$ 0.22 mJy.   For the combined cube, the equivalent \hi~mass detection threshold is \aprox\ \textcolor{black}{4.4} $\times$ 10$^{7}$ M$_{\odot}$
(corresponding to 3$\sigma$ in 2 consecutive 11 \km\  channels).  This is equivalent to a column density sensitivity for emission filling the beam of \textcolor{black}{1.6} $\times$ 10$^{19}$ cm$^{-2}$.

The unsuccessful day 2 observation targets included \cg062. However, the  NRAO VLA archive contains VLA C--array \hi\ observations for this galaxy. We retrieved this archive data and reduced it with AIPS using the standard reduction procedure. For consistency the continuum subtraction technique applied for our own  C--array data was also applied to the \cg062 \hi\  data. We include \cg062 in our analysis of galaxies with resolved \hi.

\subsection{\bf  Field D (day 3 and 4 observations)} 
\label{vlafieldd}
Field D (Figure \ref{bigrosat}) is centred   \aprox\ 14 arcmin NE of the NW cluster centre \citep{cort04}. Cubes were produced from  observations of this field with the following discrete velocity ranges 7091 \km\ to 7463 \km\ (Day 3 IF1--52 channels),   4825 \km\ to 5369 \km\ (Day 3 IF2--52 channels) and 6194 \km\ to 7236 \km\ (Day 4 IF1 and IF2--98 channels).  The rms per channel of the  cubes were $\sim$ 0.34 mJy, 0.35 mJy and 0.25 mJy respectively, with their column density and mass sensitivity limits shown in Table \ref{table1}.

\end{appendices}

\end{document}